\documentstyle[pre,aps,epsfig]{revtex}  
  
\def\emline#1#2#3#4#5#6{%
       \put(#1,#2){\special{em:moveto}}%
       \put(#4,#5){\special{em:lineto}}}
\def\newpic#1{}

\def\eps{\varepsilon}  
\def\partt{\mbox{\boldmath $\partial$}}  
\def\const{{\rm const\,}}  
\def\Dm{\widetilde{\cal D}_{\mu}}  
\def\D{{\cal D}}  
\def\L{{\cal L}}  
\def\h{{\bf h}}  
\def\n{{\bf n}}  
\def\k{{\bf k}}  
\def\x{{\bf x}}  
\def\q{{\bf q}}  
\def\a{\bbox{a}}  
\def\r{{\bf r}}  
\def\bfv{{\bf v}}  
  
\def\gdot{\circle*{20}}  
\def\dhline#1#2#3#4#5#6{  {\countdef\nnn=255  
\dimendef\llx=0   \dimendef\lly=1  
\dimendef\dx=2   \dimendef\dy=3  
\llx=#1\unitlength  \lly=#2\unitlength  
\dx=#4\unitlength   \dy=#5\unitlength  \nnn=#6  
\divide\nnn by 2  
\advance\dx by-\llx \advance\dy by-\lly  
\div\nnn \div4 \lline \adv  
\multiply\dx by2 \multiply\dy by2  
\loop \adv \ifnum\nnn>1 \lline \adv \advance\nnn by-1  
\repeat \div2 \lline }}  
  
\def\div#1{ \divide\dx by#1  \divide\dy by#1 }  
\def\adv{ \advance\llx by\dx \advance\lly by\dy }  
  
\def\lline{ {  \divide\llx by\unitlength \divide\lly by\unitlength  
\divide\dx by\unitlength \divide\dy by\unitlength  
\advance\dx by\llx \advance\dy by\lly  
\emline{\number\llx}{\number\lly}{}{\number\dx}{\number\dy}{}}}  
  
\def\dA{  
\unitlength=0.04ex  
\special{em:linewidth 0.3pt}  
\begin{picture}(180,140)  
\emline{10}{10}{}{90}{130}{}  
\emline{90}{130}{}{170}{10}{}  
\dhline{20}{25}{}{160}{25}{10}  
\emline{38}{39}{}{26}{47}{}  
\emline{154}{47}{}{142}{39}{}  
\put(90,130){\gdot}  
\end{picture}}  
  
\def\dS{  
\unitlength=0.04ex  
\special{em:linewidth 0.3pt}  
\begin{picture}(180,40)  
\emline{10}{70}{}{170}{70}{}  
\emline{23}{63}{}{23}{77}{}  
\emline{130}{63}{}{130}{77}{}  
\dhline{36}{70}{}{45}{82}{2}  
\dhline{45}{82}{}{57}{92}{2}  
\dhline{57}{92}{}{70}{99}{2}  
\dhline{70}{99}{}{85}{103}{2}  
\dhline{85}{103}{}{101}{103}{2}  
\dhline{101}{103}{}{116}{99}{2}  
\dhline{116}{99}{}{129}{92}{2}  
\dhline{129}{92}{}{141}{82}{2}  
\dhline{141}{82}{}{150}{70}{2}  
\end{picture}}

\def\dSS{  
\unitlength=0.04ex  
\special{em:linewidth 0.3pt}  
\begin{picture}(180,40)  
\emline{10}{70}{}{170}{70}{}  
\emline{23}{63}{}{23}{77}{}  
\emline{160}{63}{}{160}{77}{}  
\dhline{36}{70}{}{45}{82}{2}  
\dhline{45}{82}{}{57}{92}{2}  
\dhline{57}{92}{}{70}{99}{2}  
\dhline{70}{99}{}{85}{103}{2}  
\dhline{85}{103}{}{101}{103}{2}  
\dhline{101}{103}{}{116}{99}{2}  
\dhline{116}{99}{}{129}{92}{2}  
\dhline{129}{92}{}{141}{82}{2}  
\dhline{141}{82}{}{150}{70}{2}  
\end{picture}}  
  
\def\dOL{  
\unitlength=0.04ex  
\special{em:linewidth 0.3pt}  
\begin{picture}(180,40)  
\emline{-45}{70}{}{220}{70}{}  
\emline{23}{63}{}{23}{77}{}  
\emline{130}{63}{}{130}{77}{}  
\dhline{36}{70}{}{45}{82}{2}  
\dhline{45}{82}{}{57}{92}{2}  
\dhline{57}{92}{}{70}{99}{2}  
\dhline{70}{99}{}{85}{103}{2}  
\dhline{85}{103}{}{101}{103}{2}  
\dhline{101}{103}{}{116}{99}{2}  
\dhline{116}{99}{}{129}{92}{2}  
\dhline{129}{92}{}{141}{82}{2}  
\dhline{141}{82}{}{150}{70}{2}  
\end{picture}}  
  
\def\dOLL{  
\unitlength=0.04ex  
\special{em:linewidth 0.3pt}  
\begin{picture}(180,40)  
\emline{-45}{70}{}{220}{70}{}  
\emline{23}{63}{}{23}{77}{}  
\emline{160}{63}{}{160}{77}{}  
\dhline{36}{70}{}{45}{82}{2}  
\dhline{45}{82}{}{57}{92}{2}  
\dhline{57}{92}{}{70}{99}{2}  
\dhline{70}{99}{}{85}{103}{2}  
\dhline{85}{103}{}{101}{103}{2}  
\dhline{101}{103}{}{116}{99}{2}  
\dhline{116}{99}{}{129}{92}{2}  
\dhline{129}{92}{}{141}{82}{2}  
\dhline{141}{82}{}{150}{70}{2}  
\end{picture}}  
  
\def\palka{  
\unitlength=0.04ex  
\special{em:linewidth 0.3pt}  
\begin{picture}(180,40)  
\emline{5}{70}{}{170}{70}{}  
\end{picture}}

\def\dOLT{  
\unitlength=0.04ex  
\special{em:linewidth 0.3pt}  
\begin{picture}(180,40)  
\emline{-45}{70}{}{220}{70}{}  
\emline{60}{63}{}{60}{77}{}  
\emline{160}{63}{}{160}{77}{}  
\dhline{36}{70}{}{45}{82}{2}  
\dhline{45}{82}{}{57}{92}{2}  
\dhline{57}{92}{}{70}{99}{2}  
\dhline{70}{99}{}{85}{103}{2}  
\dhline{85}{103}{}{101}{103}{2}  
\dhline{101}{103}{}{116}{99}{2}  
\dhline{116}{99}{}{129}{92}{2}  
\dhline{129}{92}{}{141}{82}{2}  
\dhline{141}{82}{}{150}{70}{2}  
\end{picture}}  
  
\begin{document}  
\draft  
  
 \title{Anomalous scaling of a passive scalar in  
 the presence of strong anisotropy}  
  
 \author{L. Ts. Adzhemyan,$^{1}$ N. V. Antonov,$^{1}$  
  M. Hnatich,$^{2}$ and S. V. Novikov$^{1}$}  
 \address{  
 $^{1}$ Department of Theoretical Physics, St~Petersburg University,  
 Uljanovskaja 1, St~Petersburg, Petrodvorez, 198904 Russia,\\  
 $^{2}$ Institute of Experimental Physics, Slovak Academy of Sciences,  
 Watsonova 47, 04011 Kosice, Slovakia}  
  
  
 \maketitle  
  
 \begin{abstract}  
Field theoretic renormalization group and the operator product expansion  
are applied to a model of a passive scalar quantity $\theta(t,{\bf x})$,  
advected by the Gaussian strongly  
anisotropic velocity field with the covariance  
$\propto\delta(t-t')|{\bf x}-{\bf x'}|^{\eps}$. Inertial-range anomalous  
scaling behavior is established, and explicit asymptotic expressions for  
the structure functions $ S_n (\r) \equiv \langle  
[\theta(t,{\bf x}+\r)-\theta(t,{\bf x})]^{n} \rangle $  
are obtained; they are represented by superpositions of power laws with  
nonuniversal (dependent on the anisotropy parameters) anomalous exponents,  
calculated to the first order in $\eps$ in any space dimension $d$.  
In the limit of vanishing anisotropy, the exponents are associated with  
tensor composite operators  
built of the scalar gradients, and exhibit a kind of hierarchy related  
to the degree of anisotropy: the less is the rank, the less is the  
dimension and, consequently, the more important is the contribution  
to the inertial-range behavior. The leading terms of the even (odd)  
structure functions are given by the scalar (vector) operators. For  
the finite anisotropy, the exponents cannot be associated with individual  
operators (which are essentially ``mixed'' in renormalization), but the  
aforementioned hierarchy survives for all the cases studied.  
The second-order structure function $S_{2}$ is studied in more detail  
using the renormalization group and zero-mode techniques;  
the corresponding exponents and amplitudes are calculated within the  
perturbation theories in $\eps$, $1/d$, and in the anisotropy parameters.  
If the anisotropy of the velocity is strong enough, the skewness factor  
$S_{3}/S_{2}^{3/2}$ {\it increases} going down towards to the depth of  
the inertial range; the higher-order odd ratios increase even if the  
anisotropy is weak.  
 \end{abstract}  
  
 \pacs{PACS number(s): 47.27.$-$i, 47.10.+g, 05.10.Cc}  
  
 \section{Introduction} \label{sec:Intro}

Recently, much effort has been invested to understand  
inertial-range anomalous scaling of a passive scalar.  
Both the natural and numerical experiments suggest  
that the violation of the classical Kolmogorov--Obukhov theory  
\cite{Monin,Legacy} is even more strongly pronounced for a  
passively advected scalar field than for the velocity field itself;  
see, e.g., \cite{An,synth} and literature cited therein. At the same time,  
the problem of passive advection appears to be easier tractable  
theoretically. The most remarkable progress has  
been achieved for the so-called Kraichnan's rapid-change model  
\cite{Kraich1}: for the first time, the anomalous exponents  
have been calculated on the basis of a microscopic model and within  
regular expansions in formal small parameters; see, e.g.,  
\cite{Falk1,Falk2,GK,Pumir,Siggia,Eyink,RG} and references therein.  
  
Within the ``zero-mode approach,''  developed in  
\cite{Falk1,Falk2,GK,Pumir,Siggia}, nontrivial anomalous  
exponents are related to the zero modes (unforced solutions)  
of the closed exact equations satisfied by the equal-time  
correlations. Within the approach based on the field theoretic  
renormalization group (RG) and operator product expansion (OPE),  
the anomalous scaling emerges as a consequence of the existence in the  
model of composite operators with {\it negative} critical dimensions,  
which determine the anomalous exponents \cite{Eyink,RG,RG1,RG3,KJW}.  
  
Another important question recently addressed is the effects of  
large-scale anisotropy on inertial-range statistics of passively advected  
fields \cite{Pumir,Siggia,RG3,KJW,CLMV99,Lanotte,Lanotte2,ABP} and the  
velocity field itself \cite{Borue,Arad1}. According to the classical  
Kolmogorov--Obukhov theory, the anisotropy introduced at large scales by  
the forcing (boundary conditions, geometry of an obstacle etc.) dies out  
when the energy is transferred down to the smaller scales owing to the  
cascade  
mechanism \cite{Monin,Legacy}. A number of recent works confirms this  
picture  
for the {\it even} correlation functions, thus giving some quantitative  
support to the aforementioned hypothesis on the restored local isotropy  
of the inertial-range turbulence for the velocity and passive fields  
\cite{RG3,CLMV99,Lanotte,Lanotte2,ABP,Borue,Arad1}. More precisely, the  
exponents describing the inertial-range scaling exhibit universality and  
hierarchy related to the degree of anisotropy, and the leading contribution  
to an even function is given by the exponent from the isotropic shell  
\cite{RG3,KJW,Lanotte,Lanotte2,ABP,Arad1}. Nevertheless, the anisotropy  
survives in the inertial range and reveals itself in {\it odd} correlation  
functions, in disagreement with what was expected on the basis of the  
cascade ideas.  
The so-called skewness factor decreases down the scales much slower than  
expected \cite{An,synth,Pumir,Siggia}, while the higher-order odd  
dimensionless ratios (hyperskewness ets) increase, thus signalling of  
persistent small-scale anisotropy \cite{RG3,CLMV99,Lanotte2}. The effect  
seems rather universal, being observed for the scalar \cite{RG3} and vector  
\cite{Lanotte2} fields, advected by the Gaussian rapid-change  
velocity, and for the scalar advected by the two-dimensional Navier-Stokes  
velocity field \cite{CLMV99}.  
  
In the present paper, we study the anomalous scaling behavior of a  
passive scalar advected by the time-decorrelated strongly anisotropic  
Gaussian velocity field. In contradistinction with the studies of  
\cite{synth,Pumir,Siggia,RG3}, where the velocity was isotropic and  
the large-scale anisotropy was  
introduced by the imposed linear mean gradient, the uniaxial anisotropy in  
our model persists for all scales, leading to nonuniversality of  
the anomalous exponents through their dependence on the anisotropy  
parameters.  
  
The aim of our paper is twofold.  
  
Firstly, we obtain explicit inertial-range expressions for  
the structure functions and correlation functions of the scalar  
gradients and calculate the corresponding anomalous exponents to  
the first order of the $\eps$ expansion. We show that the exponents  
in our model become nonuniversal through the dependence on the  
parameters describing the anisotropy of the velocity field.  
Owing to the anisotropy of the velocity statistics, the composite  
operators of different ranks mix strongly in renormalization, and  
the corresponding anomalous exponents are given by the eigenvalues  
of the matrices which are neither diagonal nor triangular  
(in contrast with the case of large-scale anisotropy). In the language of  
the zero-mode technique this means that the $SO(d)$ decompositions of the  
correlation functions (employed, e.g., in Refs. \cite{ABP})  
do not lead to the diagonalization of the  
differential operators in the corresponding exact equations.  
  
Nevertheless, the hierarchy obeyed by the exponents for the vanishing  
anisotropy survives when the anisotropy becomes strong. This fact  
(along with the nonuniversality) is the main  
qualitative result of the paper.  
  
Another scope of the  
paper is to give the detailed account of the RG approach and to  
compare the results obtained within both the RG and zero-mode  
techniques on the example of the second-order structure function.  
  
The paper is organized as follows.  
  
In Sec. \ref{sec:scenario}, we give the precise formulation of  
the model and outline briefly the general strategy of the RG approach.  
  
In Sec. \ref{sec:QFT}, we give the field theoretic formulation  
of the model and derive the exact equations for the response function  
and pair correlator.  
  
In Sec. \ref{sec:QFT1}, we  
perform the ultraviolet (UV) renormalization of the model and  
derive the corresponding RG equations with exactly known RG  
functions ($\beta$ functions and anomalous dimensions). These equations  
possess an infrared (IR) stable fixed point, which establishes  
the existence of IR scaling with exactly known critical  
dimensions of the basic fields and parameters of the model.  
  
In Sec. \ref{sec:RGE}, we discuss the solution of the RG equations  
for various correlation functions and determine the dependence of  
the latter on the UV scale.  
  
In Sec. \ref{sec:Operators}, we discuss  
the renormalization of various composite operators. In  
particular, we derive the one-loop result for the critical  
dimensions of the tensor operators, constructed of the scalar gradients.  
  
In Sec. \ref{sec:OPE}, we show that these dimensions play the part  
of the anomalous exponents, and present explicit inertial-range  
expressions for the structure functions and correlations of the  
scalar gradients.  
  
In Sec. \ref{sec:Exact}, we give the detailed  
comparison of the RG and zero-mode techniques on the example of the  
second-order structure function, compare the representation obtained  
using RG and OPE with the ordinary perturbation theory, and  
calculate the amplitude factors in the corresponding power laws.  
  
The results obtained are reviewed in Sec. \ref{sec:Con}.

\section{Definition of the model. Anomalous scaling and ``dangerous''  
composite operators.}  \label {sec:scenario}

The advection of a passive scalar field $\theta(x)\equiv\theta(t,{\bf x})$  
in the rapid-change model is described by the stochastic equation  
\begin{equation}  
\nabla _t\theta=\nu _0\Delta \theta+f , \quad  
\nabla _t\equiv \partial _t+ v_{i}\, \partial_{i},  
\label{1}  
\end{equation}  
where $\partial _t \equiv \partial /\partial t$,  
$\partial _i \equiv \partial /\partial x_{i}$, $\nu _0$  
is the molecular diffusivity coefficient, $\Delta$  
is the Laplace operator, ${\bfv}(x)$ is the transverse (owing  
to the incompressibility) velocity field, and $f\equiv f(x)$ is a  
Gaussian scalar noise with zero mean and correlator  
\begin{equation}  
\langle  f(x)  f(x')\rangle = \delta(t-t')\, C(r/\ell), \quad  
r\equiv|{\bf x}-{\bf x'}|.  
\label{2}  
\end{equation}  
Here $\ell$ is an integral scale related to the scalar  
noise and $C(r/\ell)$ is a function finite as $\ell\to\infty$.  
With no loss of generality, we take $C(0)=1$ in what follows.  
The velocity ${\bfv}(x)$ obeys a Gaussian distribution with zero  
mean and correlator  
\begin{equation}  
\langle v_{i}(x) v_{j}(x')\rangle = D_{0}\,  
\frac{\delta(t-t')}{(2\pi)^d}  
\int d{\bf k}\, T_{ij}({\bf k})\, (k^{2}+m^{2})^{-d/2-\eps/2}\,  
\exp [{\rm i}{\bf k}({\bf x}-{\bf x'})] .  
\label{3}  
\end{equation}  
In the isotropic case, the tensor quantity $T_{ij}({\bf k})$ in  
(\ref{3}) is taken to be the ordinary transverse projector,  
$T_{ij}({\bf k})=P_{ij}({\bf k}) \equiv \delta _{ij} - k_i k_j / k^2$,  
$k\equiv |{\bf k}|$, $D_{0}>0$ is an amplitude  
factor, $1/m$ is another integral scale, and $d$  is the  
dimensionality of the ${\bf x}$  space; $0<\eps<2$ is a parameter  
with the real (Kolmogorov) value $\eps=4/3$. The relations  
\begin{equation}  
D_{0}/\nu_0 \equiv g_{0}\equiv  \Lambda^{\eps}  
\label{Lambda}  
\end{equation}  
define the coupling constant $g_{0}$ (i.e., the formal expansion  
parameter in the ordinary perturbation theory)  
and the characteristic UV momentum scale $\Lambda$.  
In what follows, we shall not distinguish the two IR  
scales, setting $m\simeq1/\ell$.

The issue of interest is, in particular, the behavior of the  
equal-time structure functions  
\begin{equation}  
S_{N}(r)\equiv\langle[\theta(t,{\bf x})-\theta(t,{\bf x'})]^{N}\rangle  
\label{struc}  
\end{equation}  
in the inertial range, specified by the inequalities  
$ 1/\Lambda <<r<<1/m \simeq \ell$. In the isotropic case, the  
odd functions $S_{2n+1}$ vanish, while for $S_{2n}$  
dimensionality considerations give  
\begin{equation}  
S_{2n}(r)=  \nu_0^{-n}\, r^{2n}\, R_{2n} (\Lambda r, mr),  
\label{strucdim}  
\end{equation}  
where $R_{2n}$ are some functions of dimensionless parameters. In  
principle, they can be calculated within the ordinary perturbation  
theory (i.e., as series in $g_{0}$), but this is not useful for  
studying inertial-range behavior: the coefficients  
are singular in the limits $\Lambda r \to\infty$  and/or  
$mr\to0$, which compensate the smallness of $g_{0}$  
(assumed in ordinary perturbation theory),  
and in order to find correct IR behavior we have to sum the entire  
series. The desired summation can be accomplished using  
the field theoretic renormalization group (RG) and operator product  
expansion (OPE); see Refs. \cite{RG,RG1,RG3}.

The RG analysis consists of two main stages. On the  
first stage, the multiplicative renormalizability of the model is  
demonstrated and the differential RG equations for its correlation  
functions are obtained. The asymptotic behavior of the functions  
like (\ref{struc}) for $\Lambda r>>1$ and any fixed $mr$ is given  
by IR stable fixed points of the RG equations and has the form  
\begin{equation}  
S_{2n}(r)= \nu_0^{-n}\, r^{2n}\,  
(\Lambda r)^{-\gamma_{n}} R_{2n} (mr),  \quad \Lambda r>>1,  
\label{strucdim2}  
\end{equation}  
with certain, as yet unknown, ``scaling functions'' $R_{2n} (mr)$.  
In the theory of critical phenomena \cite{Zinn,book3}, the quantity  
$\Delta[S_{2n}]\equiv-2n + \gamma_{n}$ is termed the ``critical  
dimension,''  
and the exponent $\gamma_{n}$, the difference between the critical  
dimension $\Delta[S_{2n}]$ and the ``canonical dimension'' $-2n$,  
is called the ``anomalous dimension.'' In the case at hand, the latter  
has an extremely simple form: $\gamma_{n}=n\eps$. Whatever be the functions  
$ R_{n} (mr)$, the representation (\ref{strucdim2})  
implies the existence of a scaling (scale invariance) in  
the IR region ($\Lambda r>>1$, $mr$ fixed)  
with definite critical dimensions of all ``IR relevant''  parameters,  
$\Delta[S_{2n}] =-2n+n \eps$, $\Delta_{r}=-1$, $\Delta_{m}=1$ and  
fixed ``irrelevant'' parameters $\nu_0$ and $\Lambda$.  
This means that the structure functions (\ref{struc}) scale  
as $S_{2n}\to \lambda^{\Delta[S_{2n}]} S_{2n}$ upon the substitution  
$m\to \lambda^{\Delta_{m}} m$,  
$r\to \lambda^{\Delta_{r}} r$. In general, the exponent  
$\Delta[S_{2n}]$ is replaced by the critical dimension of the  
corresponding correlation function. This dimension is calculated  
as a series in $\eps$, so that the exponent $\eps$ plays the part  
analogous to that played by the parameter $\eps=4-d$ in the RG theory  
of critical phenomena, while $\ell$ is an analog of the correlation  
length $\ell\equiv r_{c}$; see \cite{Zinn,book3}.  
  
On the second stage, the small $mr$ behavior of the functions  
$ R_{2n} (mr)$ is studied within the general representation  
(\ref{strucdim2}) using the OPE. It shows that, in the limit $mr\to0$,  
the functions $ R_{2n} (mr)$ have the asymptotic forms  
\begin{equation}  
 R_{2n} (mr) = \sum_{F} C_{F}(mr)\, (mr)^{\Delta[F]},  
\label{ope}  
\end{equation}  
where $C_{F}$ are coefficients regular in $mr$. In general,  
the summation is implied over all possible renormalized composite  
operators $F$ allowed by the symmetry (more precisely, see Sec.  
\ref{sec:OPE}); $\Delta[F]$ being their critical dimensions.  
Without loss of generality, it can be assumed that the expansion is  
made in irreducible tensors; in the isotropic case only scalar  
operators have nonvanishing mean values and contribute to  
the right hand side of Eq. (\ref{ope}).

The peculiarity of the models describing turbulence is the existence  
of the so-called ``dangerous'' composite operators with {\it negative}  
critical dimensions \cite{RG,RG1,RG3,UFN,turbo}. Their contributions  
into the OPE give rise to a singular behavior of the scaling functions for  
$mr\to0$, and the leading term is given by the operator with minimal  
$\Delta[F]$. The leading contributions  
to $S_{2n}$ are determined by scalar gradients  
$F_{n} = (\partial_{i}\theta\partial_{i}\theta)^{n}$ and have the form  
\begin{equation}  
S_{2n}(r) \simeq D_{0}^{-n}\, r^{n(2-\eps)}\, (mr)^{\Delta_{n}},  
\label{HZ1}  
\end{equation}  
where the critical dimensions $\Delta_{n}$ of the operators $F_{n}$  
are given by  
\begin{equation}  
\Delta_{n}=-2n(n-1)\eps/(d+2)+O(\eps^{2})=-2n(n-1)\eps/d+O(1/d^{2}).  
\label{exps}  
\end{equation}  
The expression  
(\ref{HZ1}) agrees with the results obtained earlier in \cite{Falk2,GK}  
using the zero-mode techniques; the $O(\eps^{2})$ contribution to  
$\Delta_{n}$ is obtained in \cite{RG}.  
  
In the theory of turbulence, the singular $m$ dependence of correlation  
functions with the exponents nonlinear in $n$ is referred to as  
anomalous scaling, and $\Delta_{n}$ themselves are termed the anomalous  
exponents; see, e.g., \cite{Monin,Legacy}. The above discussion shows that  
the anomalous exponents (in the sense of the turbulence theory) are not  
simply related to the anomalous or critical dimensions (in the sense of  
the theory of phase transitions) of the structure functions themselves;  
they are determined by the critical dimensions of certain composite  
operators entering into the corresponding OPE.\footnote{  
The OPE and the concept of dangerous operators in the stochastic  
hydrodynamics were introduced and investigated in detail in  
\cite{JETP}; see also \cite{UFN,turbo}.  
For the Kraichnan model, the relationship between the anomalous  
exponents and dimensions of composite operators was anticipated  
in \cite{Falk1,Falk2,GK,Eyink} within certain phenomenological formulation  
of the OPE, the so-called ``additive fusion rules,'' typical  
to the models with multifractal behavior \cite{DL}.  
A similar picture has been discussed in \cite{Burg,Burg1} in  
connection with the Burgers turbulence and growth phenomena.}

In a number of papers, e.g. \cite{synth,Pumir,Siggia,RG3}, the artificial  
stirring force in Eq. (\ref{1}) was replaced by the term $(\h{\bfv})$,  
where $\h$ is a constant vector that determines the distinguished  
direction and therefore introduces large-scale anisotropy.  
The anisotropy gives rise to nonvanishing odd functions $S_{2n+1}$.  
The critical dimensions of all composite operators remain unchanged,  
but the irreducible tensor operators acquire nonzero mean values and  
their contributions appear on the right hand side of Eq. (\ref{ope}); see  
\cite{RG3}. This is easily understood in the language of the zero-mode  
approach: the noise $f$ and the term $(\h{\bfv})$ do not affect  
the differential operators in the equations satisfied by the equal-time  
correlations functions; the zero modes (homogeneous solutions)  
coincide in the two cases, but in the latter case the modes with  
nontrivial angular dependence should be taken into account.

The direct calculation to the order $O(\eps)$ has shown that  
the leading exponent associated with a given rank contribution to  
Eq. (\ref{1}) decreases monotonically with the rank \cite{RG3}. Hence,  
the leading term of the inertial-range behavior of an even structure  
function is determined by the same exponent (\ref{exps}), while  
the exponents related to the tensor operators determine only  
subleading corrections. A similar hierarchy has been established  
recently in Ref. \cite{Lanotte} (see also \cite{Lanotte2})  
for the magnetic field advected passively by the rapid-change velocity  
in the presence of a constant background field, and in \cite{Arad1}  
within the context of the Navier--Stokes turbulence.

Below we shall take the velocity statistics to be anisotropic also  
at small scales. We replace the ordinary transverse projector in  
Eq. (\ref{3}) with the general transverse structure that possesses  
the uniaxial anisotropy:  
\begin{equation}  
T_{ij}({\bf k}) = a(\psi) P_{ij} ({\bf k}) + b (\psi) \tilde n_{i}({\bf k})  
\tilde n_{j}({\bf k}).  
\label{T}  
\end{equation}  
Here the unit vector ${\bf n}$ determines the distinguished direction  
(${\bf n}^{2}=1$),  $\tilde n_{i} ({\bf k})\equiv P_{ij} ({\bf k})n_{j}$,  
and $\psi$ is the angle between the vectors ${\bf k}$ and  
${\bf n}$, so that $({\bf n}{\bf k})=k\cos\psi$ [note that  
$(\tilde{\bf n}{\bf k})=0$]. The scalar functions can be  
decomposed the Gegenbauer polynomials (the $d$-dimensional  
generalization of the Legendre polynomials, see Ref. \cite{Grad}):  
\begin{equation}  
a(\psi) = \sum_{l=0}^{\infty} a_{l} P_{2l}(\cos\psi), \quad  
b(\psi) = \sum_{l=0}^{\infty} b_{l} P_{2l}(\cos\psi)  
\label{Legendre}  
\end{equation}  
(we shall see later that odd polynomials do not affect the scaling  
behavior).  
The positivity of the correlator (\ref{3}) leads to the conditions  
\begin{equation}  
a(\psi) > 0, \quad a(\psi)+ b(\psi)\sin^{2}\psi >0.  
\label{positiv}  
\end{equation}  
In practical calculations, we shall mostly confine ourselves with the  
special case  
\begin{equation}  
T_{ij}({\bf k}) = (1+\rho_{1}\cos^{2}\psi) P_{ij} ({\bf k}) + \rho_{2}  
 \tilde n_{i}({\bf k}) \tilde n_{j}({\bf k}).  
\label{T34}  
\end{equation}  
Then the inequalities (\ref{positiv}) reduce to $\rho_{1,2}>-1$.  
We shall see that this case represents nicely all the main features  
of the general model (\ref{T}).  
  
We note that the quantities (\ref{T}), (\ref{T34}) possess the  
symmetry $\n\to-\n$. The anisotropy makes it possible to introduce  
mixed correlator $\langle{\bfv}f\rangle\propto\n  
\delta(t-t')\, C'(r/\ell)$ with some function $C'(r/\ell)$  
analogous to $C(r/\ell)$ from Eq. (\ref{2}). This violates the  
evenness in $\n$ and gives rise to nonvanishing odd functions  
$S_{2n+1}$. However, this leads to no serious alterations in the  
RG analysis; we shall discuss this case in Sec. \ref{sec:OPE},  
and for now we assume $\langle{\bfv}f\rangle=0$.  
  
In a number of papers, e.g., \cite{Barton,Carati,Denis,Kim,Busa},  
the RG techniques were applied to the anisotropically driven  
Navier--Stokes equation, including passive advection and magnetic  
turbulence, with the expression (\ref{T34}) entering into the  
stirring force correlator. The detailed account can be found in  
Ref. \cite{turbo}, where some errors of the previous treatments  
are also corrected. However, these studies have up to now been  
limited to the first stage, i.e., investigation of the existence  
and stability of the fixed points and calculation of the critical  
dimensions of basic quantities. Calculation of the anomalous  
exponents in those models remains an open problem.

\section{Field theoretic formulation and the Dyson--Wyld equations}  
\label {sec:QFT}

The stochastic problem (\ref{1})--(\ref{3}) is equivalent  
to the field theoretic model of the set of three fields  
$\Phi\equiv\{\theta', \theta, {\bfv}\}$ with action functional  
\begin{equation}  
S(\Phi)=\theta' D_{\theta}\theta' /2+  
\theta' \left[ - \partial_{t} - ({\bfv}\partt)  
+ \nu _0\Delta \right] \theta  
-{\bfv} D_{v}^{-1} {\bfv}/2.  
\label{action}  
\end{equation}  
The first four terms in  Eq. (\ref{action}) represent  
the Martin--Siggia--Rose-type action for the stochastic  
problem (\ref{1}), (\ref{2}) at fixed ${\bfv}$ (see, e.g.,  
\cite{Zinn,book3}), and the last term  
represents the Gaussian averaging over ${\bfv}$. Here $D_{\theta}$  
and $D_{v}$ are the correlators (\ref{2}) and (\ref{3}), respectively,  
the required integrations over $x=(t,{\bf x})$ and summations over  
the vector indices are implied.  
  
The formulation (\ref{action}) means that statistical averages  
of random quantities in stochastic problem (\ref{1})--(\ref{3}) can be  
represented as functional averages with the weight $\exp S(\Phi)$,  
so that the generating functionals of total [$G(A)$] and connected  
[$W(A)$] Green functions of the problem  
are given by the functional integral  
\begin{equation}  
G(A)=\exp  W(A)=\int {\cal D}\Phi \exp [S(\Phi )+A\Phi ]  
\label{field}  
\end{equation}  
with arbitrary sources $A\equiv A^{\theta'},A^{\theta},A^{\bfv}$  
in the linear form  
\[A\Phi \equiv \int dx[A^{\theta'}(x)\theta' (x)+A^{\theta }(x)\theta (x)  
+ A^{\bfv}_{i}(x)v_{i}(x)].\]  
  
The model (\ref{action}) corresponds to a standard Feynman  
diagrammatic technique with the triple vertex  
$-\theta'({\bfv}\partt)\theta\equiv\theta' V_{j}v_{j}\theta$  
with vertex factor (in the momentum-frequency representation)  
\begin{equation}  
V_{j}= - {\rm i} k_{j},  
\label{vertex}  
\end{equation}  
where ${\bf k}$ is the momentum flowing into the vertex via the  
field $\theta$. The bare propagators in the momentum-frequency  
representation have the form  
\begin{eqnarray}  
\langle \theta \theta' \rangle _0=\langle \theta' \theta \rangle _0^*=  
(-i\omega +\nu _0k^2)^{-1} , \quad  
\langle \theta \theta \rangle _0=C(k)\,(\omega ^2+\nu _0^2k^4)^{-1},  
\quad  
\langle \theta '\theta '\rangle _0=0 ,  
\label{lines}  
\end{eqnarray}  
where $C(k)$ is the Fourier transform of the function $C(r/\ell)$ from Eq.  
(\ref{2}) and the bare propagator  
$\langle{\bfv}{\bfv}\rangle _0 \equiv \langle{\bfv}{\bfv}\rangle$  
is given by Eq. (\ref{3}) with the transverse projector from Eqs.  
(\ref{T}) or (\ref{T34}).  
  
The pair correlation functions $\langle\Phi\Phi\rangle$ of the  
multicomponent field $\Phi\equiv\{\theta', \theta, {\bfv}\}$ satisfy  
standard Dyson equation, which in the component notation reduces to  
the system of two equations, cf. \cite{Monin}  
\begin{mathletters}  
\label{Dyson}  
\begin{equation}  
G^{-1}(\omega, \k)= -{\rm i}\omega +\nu_0 k^{2} -  
\Sigma_{\theta'\theta} (\omega, \k),  
\label{Dyson1}  
\end{equation}  
\begin{equation}  
D(\omega, \k)= |G(\omega, \k)|^{2}\,  
[C(k)+\Sigma_{\theta'\theta'} (\omega, \k)],  
\label{Dyson2}  
\end{equation}  
\end{mathletters}  
where $G(\omega, \k)\equiv\langle\theta\theta'\rangle$ and  
$D(\omega, \k)\equiv\langle\theta\theta\rangle$ are the exact response  
function and pair correlator, respectively, and $\Sigma_{\theta'\theta}$,  
$\Sigma_{\theta'\theta'}$ are self-energy operators represented by  
the corresponding 1-irreducible diagrams; all the other functions  
$\Sigma_{\Phi\Phi}$ in the model (\ref{action}) vanish identically.  
  
The feature characteristic of the models like (\ref{action}) is that  
all the skeleton multiloop diagrams entering into the  
self-energy operators  
$\Sigma_{\theta'\theta}$, $\Sigma_{\theta'\theta'}$  
contain effectively closed circuits of retarded  
propagators $\langle\theta\theta'\rangle$  
(it is crucial here that the propagator $\langle  
{\bf v}{\bf v}\rangle_{0}$ in Eq. (\ref{3}) is proportional to the  
$\delta$ function in time) and therefore vanish.  
  
Therefore the self-energy operators  
in (\ref{Dyson}) are given by the one-loop approximation  
exactly and have the form  
\begin{mathletters}  
\label{sigma}  
\begin{equation}  
\Sigma_{\theta'\theta} (\omega, \k)=  
\put(0.00,-56.00){\makebox{\dS}} \hskip1.7cm = - D_{0}\, k_{i}k_{j}  
\int\frac{d\omega'}{2\pi} \int\frac{d{\bf q}}{(2\pi)^{d}}  
\frac{ T_{ij}(\q)} {(q^{2}+m^{2})^{d/2+\eps/2}}\, G(\omega',\q'),  
\label{sigma1}  
\end{equation}  
\begin{equation}  
\Sigma_{\theta'\theta'} (\omega, \k)=  
\put(0.00,-56.00){\makebox{\dSS}} \hskip1.7cm =  D_{0} \, k_{i}k_{j}  
\int\frac{d\omega'}{2\pi} \int\frac{d{\bf q}}{(2\pi)^{d}}  
\frac{ T_{ij}(\q) } {(q^{2}+m^{2})^{d/2+\eps/2}}\, D(\omega',\q'),  
\label{sigma2}  
\end{equation}  
\end{mathletters}  
where $\q'\equiv{\bf k}-{\bf q}$.  
The solid lines in the diagrams denote the exact propagators  
$\langle\theta\theta'\rangle$ and $\langle\theta\theta\rangle$,  
the ends with a slash correspond to the field $\theta'$, and the  
ends without a slash correspond to $\theta$; the dashed lines  
denote the bare propagator (\ref{3}); the vertices correspond  
to the factor (\ref{vertex}).

The integrations over $\omega'$ in the right-hand sides of  
Eqs. (\ref{sigma}) give the equal-time  
response function $G(\q)=(1/{2\pi})\int{d\omega'}\,G(\omega',\q)$  
and the equal-time pair correlator  
$D(\q)=(1/{2\pi})\int{d\omega'}\,D(\omega',\q)$;  
note that both the self-energy operators are in fact independent of  
$\omega$. The only contribution to $G(\q)$ comes from the bare  
propagator $\langle\theta\theta'\rangle_{0}$ from Eq. (\ref{lines}),  
which in the $t$ representation is  
discontinuous at coincident times. Since the correlator (\ref{3}),  
which enters into the one-loop diagram for $\Sigma_{\theta'\theta}$,  
is symmetric in $t$ and $t'$, the response function must be  
defined at $t=t'$ by half the sum of the limits.  
This is equivalent to the convention  
$G(\q)=(1/{2\pi})\int{d\omega'}\,(-i\omega'+\nu _0 q^2)^{-1}=1/2$  
and gives  
\begin{equation}  
\Sigma_{\theta'\theta} (\omega, \k)=- \frac{D_{0}\, k_{i}k_{j}}{2}  
\int\frac{d{\bf q}}{(2\pi)^{d}}  
\frac{ T_{ij}(\q) } {(q^{2}+m^{2})^{d/2+\eps/2}}.  
\label{sigma3}  
\end{equation}  
  
The integration of Eq. (\ref{Dyson2})  over the frequency $\omega$  
gives a closed equation for the equal-time correlator. Using Eq.  
(\ref{sigma3}) it can be written in the form  
\begin{equation}  
2\nu_0 k^{2}D(\k)=C(k) + D_{0}\,k_{i}k_{j}  
\int\frac{d{\bf q}}{(2\pi)^{d}} \,  
\frac{T_{ij}(\q) } {(q^{2}+m^{2})^{d/2+\eps/2}}\,  
\Bigl[D(\q')-D(\k)\Bigr].  
\label{9}  
\end{equation}  
  
Equation (\ref{9}) can also be rewritten as a partial differential  
equation for the pair correlator in the coordinate representation,  
$D({\bf r})\equiv\langle \theta  
(t,{\bf x}) \theta(t,{\bf x}+{\bf r})\rangle$  
[we use the same notation $D$ for the coordinate  
function and its Fourier transform].  
Noting that the integral in Eq. (\ref{9}) involves convolutions of  
the functions $D(\k)$ and $D_{0}\,T_{ij}(\q)/(q^{2}+m^{2})^{d/2+\eps/2}$,  
and replacing the momenta by the corresponding derivatives,  
${\rm i}k_{i} \to \partial_{i}$ and so on, we obtain:  
\begin{eqnarray}  
2\nu_0 \partial^{2}D({\bf r})+C(r/\ell) + D_{0}\, S_{ij}({\bf r}) \,  
\partial_{i}\partial_{j} \, D({\bf r}) =0,  
\label{12}  
\end{eqnarray}  
where the ``effective eddy diffusivity'' is given by  
\begin{equation}  
S_{ij}({\bf r}) \equiv \int\frac{d{\bf q}}{(2\pi)^{d}} \,  
\frac{T_{ij}(\q)} {(q^{2}+m^{2})^{d/2+\eps/2}}  
\Bigl[1-\exp\,({\rm i}{\bf q}\cdot{\bf r})\Bigr]  
\label{10}  
\end{equation}  
(note that $\partial_{i}S_{ij}=0$).  
For $0<\eps <2$, equations (\ref{9})--(\ref{10}) allow for the limit  
$m\to0$: the possible IR divergence of the integrals at ${\bf q}=0$ is  
suppressed by the vanishing of the expressions in the square brackets.  
For the isotropic case (i.e., after the substitution $T_{ij}\to P_{ij}$)  
Eq. (\ref{12}) coincides (up to the notation) with the well-known  
equation for the equal-time pair correlator in the model \cite{Kraich1}.  
  
Equation (\ref{sigma3}) will be used in the next section for the exact  
calculation of the RG functions; solution of Eq. (\ref{12}) will be  
discussed in Sec. \ref{sec:Exact} in detail.  
  
\section{Renormalization, RG functions, and RG equations}  
\label {sec:QFT1}  
  
The analysis of the UV divergences is based on the analysis of  
canonical dimensions. Dynamical models of the type (\ref{action}),  
in contrast to static models, have two scales, so that the canonical  
dimension of some quantity $F$ (a field or a parameter in the action  
functional)  
is described by two numbers, the momentum dimension $d_{F}^{k}$ and  
the frequency dimension $d_{F}^{\omega}$. They are determined so that  
$[F] \sim [L]^{-d_{F}^{k}} [T]^{-d_{F}^{\omega}}$, where $L$ is the  
length scale and $T$ is the time scale. The dimensions are found  
from the obvious  
normalization conditions $d_k^k=-d_{\bf x}^k=1$, $d_k^{\omega }  
=d_{\bf x}^{\omega }=0$, $d_{\omega }^k=d_t^k=0$,  
$d_{\omega }^{\omega }=-d_t^{\omega }=1$, and from the requirement  
that each term of the action functional be dimensionless (with  
respect to the momentum and frequency dimensions separately).  
Then, based on $d_{F}^{k}$ and $d_{F}^{\omega}$,  
one can introduce the total canonical dimension  
$d_{F}=d_{F}^{k}+2d_{F}^{\omega}$ (in the free theory,  
$\partial_{t}\propto\Delta$), which plays in the theory of  
renormalization of dynamical models the same role as  
the conventional (momentum) dimension does in static problems.  
  
The dimensions for the model (\ref{action}) are given in  
Table \ref{table1}, including the parameters  
which will be introduced later on.  
>From Table \ref{table1} it follows that the model is  
logarithmic (the coupling constant $g_{0}$ is dimensionless)  
at $\eps=0$, so that the UV divergences have the form of  
the poles in $\eps$ in the Green functions.  
  
The total canonical dimension of an arbitrary  
1-irreducible Green function $\Gamma = \langle\Phi \cdots \Phi  
\rangle _{\rm 1-ir}$ is given by the relation  
\begin{equation}  
d_{\Gamma }=d_{\Gamma }^k+2d_{\Gamma }^{\omega }=  
d+2-N_{\Phi }d_{\Phi},  
\label{deltac}  
\end{equation}  
where $N_{\Phi}=\{N_{\theta'},N_{\theta},N_{\bfv}\}$ are the  
numbers of corresponding fields entering into the function  
$\Gamma$, and the summation over all types of the fields is  
implied.  
The total dimension $d_{\Gamma}$ is the formal index of the  
UV divergence. Superficial UV divergences, whose removal requires  
counterterms, can be present only in those functions $\Gamma$ for  
which $d_{\Gamma}$ is a non-negative integer.  
  
Analysis of the divergences should be based on the following auxiliary  
considerations:  
  
(i) From the explicit form of the vertex and bare propagators in  
the model (\ref{action}) it follows that $N_{\theta'}- N_{\theta}=2N_{0}$  
for any 1-irreducible Green function, where $N_{0}\ge0$  
is the total number of bare propagators $\langle \theta \theta  
\rangle _0$ entering into the function (obviously, no diagrams  
with $N_{0}<0$ can be constructed). Therefore, the difference  
$N_{\theta'}- N_{\theta}$ is an even non-negative integer for  
any nonvanishing function.  
  
(ii) If for some reason a number of external momenta occurs as an  
overall factor in all the diagrams of a given Green function, the  
real index of divergence $d_{\Gamma}'$ is smaller than $d_{\Gamma}$  
by the corresponding number (the Green function requires  
counterterms only if $d_{\Gamma}'$  is a non-negative integer).  
In the model (\ref{action}), the derivative $\partt$ at the  
vertex $\theta'({\bfv}\partt)\theta$ can be moved onto the  
field $\theta'$ by virtue of the transversality of the field  
${\bfv}$. Therefore, in any 1-irreducible diagram it is always  
possible to move the derivative onto any of the external  
``tails'' $\theta$ or $\theta'$, which decreases the real index  
of divergence: $d_{\Gamma}' = d_{\Gamma}- N_{\theta}-N_{\theta'}$.  
The fields $\theta$, $\theta'$ enter into the counterterms only  
in the form of derivatives $\partial\theta$, $\partial\theta'$.

From the dimensions in Table \ref{table1} we find  
$d_{\Gamma} = d+2 - N_{\bfv} + N_{\theta}- (d+1)N_{\theta'}$  
and $d_{\Gamma}'=(d+2)(1-N_{\theta'}) - N_{\bfv}$.  
It then follows that for any $d$, superficial  
divergences can exist only in the 1-irreducible functions  
$\langle\theta'\theta\dots\theta\rangle_{\rm 1-ir}$ with $N_{\theta'}=1$  
and arbitrary value of $N_{\theta}$, for which $d_{\Gamma}=2$,  
$d_{\Gamma}'=0$. However, all the functions with $N_{\theta}>  
N_{\theta'}$ vanish (see above) and obviously do not require  
counterterms. We are left with the only superficially  
divergent function $\langle\theta'\theta\rangle_{\rm 1-ir}$;  
the corresponding counterterms must contain two symbols  
$\partial$  and in the isotropic case reduce to the only  
structure $\theta'\Delta\theta$. In the presence of anisotropy,  
it is necessary to also introduce new counterterm of the form  
$\theta'(\n\partt)^{2} \theta$, which is absent in the unrenormalized  
action functional (\ref{action}). Therefore, the model  
(\ref{action}) in its original formulation is not multiplicatively  
renormalizable, and in order to use the standard RG techniques it is  
necessary to extend the model by adding the new contribution to the  
unrenormalized action:  
\begin{equation}  
S(\Phi)=\theta' D_{\theta}\theta' /2+ \theta' \left[-\partial_{t}  
-({\bfv}\partt) + \nu _0\Delta + \alpha_0\nu_0 (\n\partt)^{2} \right]  
\theta -{\bfv} D_{v}^{-1} {\bfv}/2.  
\label{extended}  
\end{equation}  
Here $\alpha_{0}$ is a new dimensionless unrenormalized parameter.  
The stability of the system implies the positivity of the total viscous  
contribution  $\nu _0 k^{2} + \alpha_0\nu_0 (\n\k)^{2}$, which  
leads to the inequality $\alpha_{0}>-1$.  
Its real (``physical'') value is zero, but this fact does  
not hinder the use of the RG techniques, in which it is first assumed  
to be arbitrary, and the equality $\alpha_{0}=0$ is imposed as the  
initial condition in solving the equations for invariant variables  
(see Sec. \ref{sec:RGE}). Below we shall see that the zero  
value of $\alpha_{0}$ corresponds to certain nonzero value of  
its renormalized analog, which can be found explicitly.  
  
For the action (\ref{extended}), the nontrivial bare propagators  
in (\ref{lines}) are replaced with  
\begin{eqnarray}  
\langle \theta \theta' \rangle _0=\langle \theta' \theta \rangle _0^*=  
\left(-{\rm i}\omega +\nu_0 k^2+ \alpha_{0}\nu_0 (\n \k)^{2} \right  
)^{-1} ,  \qquad  
\langle \theta \theta \rangle _0= \frac{C(k)}{ \vert -  
{\rm i}\omega + \nu_0k^2+ \alpha_{0}\nu_0 (\n \k)^{2} \vert ^{2}}\, .  
\label{lines2}  
\end{eqnarray}  
  
After the extension, the model has become multiplicatively  
renormalizable: inclusion of the counterterms is reproduced by the  
inclusion of two independent renormalization constants $Z_{1,2}$  
as coefficients in front of the counterterms.  
This leads to the renormalized action of the form  
\begin{equation}  
S_{R}(\Phi)=\theta' D_{\theta}\theta' /2+ \theta' \left[-\partial_{t}  
-({\bfv}\partt) + \nu Z_{1}\Delta + \alpha\nu Z_{2}  
(\n\partt)^{2} \right]  
\theta -{\bfv} D_{v}^{-1} {\bfv}/2,  
\label{renormalized}  
\end{equation}  
or, equivalently, to the multiplicative renormalization  
of the parameters $\nu_0$, $g_{0}$ and $\alpha_{0}$  
in the action functional (\ref{extended}):  
\begin{equation}  
\nu_0=\nu Z_{\nu},\quad g_{0}=g\mu^{\eps}Z_{g}, \quad \alpha_{0}=  
\alpha Z_{\alpha}.  
\label{18}  
\end{equation}  
Here $\mu$ is the reference mass in the minimal subtraction (MS)  
scheme, which we always use in what follows, $\nu$, $g$ and $\alpha$  
are renormalized counterparts of the corresponding bare parameters,  
and the renormalization constants $Z=Z(g,\alpha;\,\rho_{1,2},\,\eps,d)$  
depend only on the dimensionless parameters. The correlator (\ref{3}) in  
(\ref{renormalized}) is expressed in renormalized variables using Eqs.  
(\ref{18}). The comparison of Eqs. (\ref{extended}),  
(\ref{renormalized}), and (\ref{18}) leads to the relations  
\begin{equation}  
Z_{1}=Z_{\nu}, \quad Z_{2}=Z_{\alpha}Z_{\nu}, \quad Z_{g}=Z_{\nu}^{-1}.  
\label{18a}  
\end{equation}  
The last relation in (\ref{18a}) results from the absence of  
renormalization of the contribution with $D_{0}$, so that  
$D_{0}\equiv g_{0}\nu_0 = g\mu^{\eps} \nu$; see (\ref{Lambda}).  
No renormalization of the fields, anisotropy parameters  and the  
``mass'' $m$ is required, i.e., $Z_{\Phi}=1$ for all $\Phi$, and so on.  
  
The relation $ S(\Phi,e_{0})=S_{R}(\Phi,e,\mu)$  
(where $e_{0}\equiv\{\nu_0, g_{0}, \alpha_{0}\} $  
is the complete set of bare parameters, and  
$e \equiv\{\nu, g, \alpha\}$ is the set of their renormalized  
counterparts) for the generating functional $W(A)$ in Eq. (\ref{field})  
yields $ W(A,e_{0})=W_{R}(A,e,\mu)$. We use $\widetilde{\cal D}_{\mu}$  
to denote the differential operation $\mu\partial_{\mu}$ at fixed  
$e_{0}$ and operate on both sides of this equation with it.  
This gives the basic RG differential equation:  
\begin{equation}  
{\cal D}_{RG}\,W_{R}(A,e,\mu)  = 0.  
\label{RG}  
\end{equation}  
Here ${\cal D}_{RG}$ is the operation $\widetilde{\cal D}_{\mu}$  
expressed in the renormalized variables:  
\begin{equation}  
{\cal D}_{RG}\equiv {\cal D}_{\mu} + \beta_{g}\partial_{g} +  
\beta_{\alpha}\partial_{\alpha} - \gamma_{\nu}{\cal D}_{\nu},  
\label{RG2}  
\end{equation}  
where we have written ${\cal D}_{x}\equiv x\partial_{x}$ for  
any variable $x$, and the RG functions (the $\beta$ functions and  
the anomalous dimension $\gamma$) are defined as  
\begin{mathletters}  
\label{RGF}  
\begin{equation}  
\gamma_{F}(g,\alpha)\equiv\widetilde{\cal D}_\mu \ln Z_{F}  
\label{RGF1}  
\end{equation}  
for any renormalization constant $Z_{F}$ and  
\begin{equation}  
\beta_{g}(g,\alpha)\equiv \widetilde{\cal D}_\mu g =  
g\,(-\eps - \gamma_{g}) = g\,(-\eps + \gamma_{\nu})  
= g\,(-\eps + \gamma_{1}),  
\label{RGF2}  
\end{equation}  
\begin{equation}  
\beta_{\alpha}(g,\alpha)\equiv \widetilde{\cal D}_\mu \alpha= -\alpha  
\gamma_{\alpha}= \alpha (\gamma_{1}-\gamma_{2}).  
\label{RGF3}  
\end{equation}  
\end{mathletters}  
The relation between $\beta_{g}$ and $\gamma_{\nu}$ in Eq. (\ref{RGF2})  
results from the definitions and the last relation in (\ref{18a}).

Now let us turn to the calculation of the renormalization constants  
$Z_{1,2}$ in the MS scheme. They are determined by the requirement that  
the Green function $G=\langle\theta'\theta\rangle$, when expressed in  
renormalized variables, be UV finite, i.e., have no  
poles in $\eps$. It satisfies the Dyson equation of the form  
\begin{equation}  
G^{-1} (\omega, \k) = -{\rm i} \omega + \nu_0 k^{2}  
+ \alpha_{0} \nu_0 (\n\k)^{2} - \Sigma_{\theta'\theta} (\omega, \k),  
\label{Dyson3}  
\end{equation}  
which is obtained from Eq. (\ref{Dyson1}) with the obvious  
substitution $\nu_0 k^{2}\to \nu_0 k^{2} + \alpha_{0} \nu_0 (\n\k)^{2}$;  
the self-energy operator $\Sigma_{\theta'\theta}$ is given by the  
same expression (\ref{sigma3}). With $T_{ij}$ from (\ref{T34}) the  
integration in Eq. (\ref{sigma3}) yields  
\begin{eqnarray}  
\Sigma_{\theta'\theta} (k)= - \frac{D_{0} \, J(m)} {2d(d+2)}  
\Bigl[(d-1)(d+2) k^{2} + \rho_{1} (d+1) k^{2} +\rho_{2} k^{2} -  
2 \rho_{1} (\n\k)^{2} + (d^{2}-2 ) \rho_{2} (\n\k)^{2} \Bigr],  
\label{sigma5}  
\end{eqnarray}  
where we have written  
\begin{equation}  
J(m)\equiv\int\frac{d{\bf q}}{(2\pi)^{d}}\,  
\frac{1}{(q^{2}+m^{2})^{d/2+\eps/2}}  
= \frac{\Gamma(\eps/2)\,m^{-\eps}}{(4\pi)^{d/2}\Gamma(d/2+\eps/2)}.  
\label{otvet2}  
\end{equation}  
In deriving Eqs. (\ref{sigma5}), (\ref{otvet2}), we used the relations  
\begin{eqnarray}  
\int d{\bf q}\, f(q)\frac{q_{i}q_{j}}{q^{2}}  =  
\frac{\delta_{ij}}{d} \int d{\bf q}\, f(q), \quad  
\int d{\bf q}\, f(q)\frac{q_{i}q_{j}q_{l}q_{p}}{q^{4}}  =  
\frac {\delta_{ij}\delta_{lp}+\delta_{ip}\delta_{lj}+  
\delta_{il}\delta_{pj}}{d(d+2)} \int d{\bf q}\, f(q) .  
\label{isotropy}  
\end{eqnarray}  
  
The renormalization constants $Z_{1,2}$ are found from the requirement  
that the function $\langle\theta'\theta\rangle$ expressed  
in renormalized variables be finite for $\eps\to0$. This  
requirement determines $Z_{1,2}$ up to an UV finite contribution; the  
latter is fixed by the choice of a renormalization scheme. In the MS  
scheme all renormalization constants have the form ``1 + only poles  
in  $\eps$.''  We substitute Eqs. (\ref{18}), (\ref{18a})  
into Eqs. (\ref{Dyson3}), (\ref{sigma5})  
and choose $Z_{1,2}$ to cancel the pole in $\eps$ in the integral  
$J(m)$. This gives:  
\begin{equation}  
Z_{1}=1- \frac{gC_{d}}{2d(d+2)\eps}  
\Bigl[(d-1)(d+2)+\rho_{1}(d+1)+\rho_{2} \Bigr], \qquad  
Z_{2}=1- \frac{gC_{d}}{2d(d+2)\alpha\eps}  
\Bigl[-2\rho_{1}+\rho_{2}(d^{2}-2) \Bigr],  
\label{Z}  
\end{equation}  
where we have written $C_{d}\equiv S_{d}/(2\pi^d)$ and  
$S_{d} = 2\pi^{d/2}/\Gamma(d/2)$ is the surface area of the unit  
sphere in $d$-dimensional space.  
  
For the anomalous dimension $\gamma_{1}(g)\equiv  
\widetilde{\cal D}_\mu \ln Z_{1}  
=\beta_{g}\partial_{g}\ln Z_{1}$  from the relations (\ref{RGF2})  
and (\ref{Z}) one obtains (note that $Z_{1}$ is independent of  
$\alpha$)  
\begin{eqnarray}  
\gamma_{1}(g)=\frac{-\eps {\cal D}_g \ln Z_{\nu}}{1-{\cal D}_g \ln Z_{\nu}}  
=\frac{gC_{d}}{2d(d+2)} \Bigl[(d-1)(d+2)+\rho_{1}(d+1)+\rho_{2} \Bigr],  
\label{gamma1}  
\end{eqnarray}  
and for  $\gamma_{2}(g,\alpha)\equiv  \widetilde{\cal D}_\mu \ln Z_{2}  
=(\beta_{g}\partial_{g} + \beta_{\alpha}\partial_{\alpha}) \ln Z_{2}$  
one has  
\begin{eqnarray}  
\gamma_{2}(g,\alpha)= \frac {\bigl[ (-\eps+\gamma_{1}) \D_g + \gamma_{1})  
\D_{\alpha} \Bigr] \ln Z_{2} } {1+\D_{\alpha} \ln Z_{2}}=  
 \frac { -\eps \D_g \ln Z_{2} } {1+\D_{\alpha} \ln Z_{2}}=  
\frac{gC_{d}}{2d(d+2)\alpha}  
\bigl[-2\rho_{1}+\rho_{2}(d^{2}-2) \bigr]  
\label{gamma2}  
\end{eqnarray}  
[note that $(\D_g + \D_{\alpha}) \ln Z_{2}=0$].  
The cancellation of the poles in  $\eps$ in Eqs. (\ref{gamma1}),  
(\ref{gamma2}) is a consequence of the UV finiteness of the anomalous  
dimensions $\gamma_{F}$; their independence of $\eps$ is a property  
of the MS scheme. Note also that the expressions  
(\ref{Z})--(\ref{gamma2}) are exact, i.e., have no corrections of  
order $g^{2}$ and higher; this is a consequence of the fact that the  
one-loop approximation (\ref{sigma3}) for the self-energy operator  
is exact.  
  
The fixed points of the RG equations  
are determined from the requirement that all the beta functions  
of the model vanish. In our model the coordinates $g_{*},\alpha_{*}$ of  
the fixed points are found from the equations  
\begin{equation}  
\beta_{g} (g_{*},\alpha_{*})=\beta_{\alpha} (g_{*},\alpha_{*})=0,  
\label{points}  
\end{equation}  
with the beta functions from Eqs. (\ref{RGF}).  
The type of the fixed point is determined by the eigenvalues of  
the matrix $\Omega=\{\Omega_{ik}=\partial\beta_{i}/\partial g_{k}\}$,  
where $\beta_{i}$ denotes the full set of the beta functions and  
$g_{k}$ is the full set of charges. The IR asymptotic behavior  
is governed by the IR stable fixed points, i.e., those for  
which all the eigenvalues are positive. From the explicit expressions  
(\ref{gamma1}), (\ref{gamma2}) it then follows that the RG equations  
of the model have the only IR stable fixed point with the coordinates  
\begin{equation}  
g_{*}C_{d}= \frac{2d(d+2)\eps} {(d-1)(d+2)+\rho_{1}(d+1)  
+\rho_{2}}, \quad \alpha_{*} = \frac {-2\rho_{1}+\rho_{2}(d^{2}-2)}  
{(d-1)(d+2)+\rho_{1}(d+1)+\rho_{2}}.  
\label{FP}  
\end{equation}  
For this point, both the eigenvalues of the matrix $\Omega$ equal to  
$\eps$; the values $\gamma^{*}_{1}=\gamma^{*}_{2}=\gamma^{*}_{\nu} =\eps$  
are also found exactly from Eqs. (\ref{RGF})  
[here and below, $\gamma^{*}_{F} \equiv \gamma_{F} (g_{*},\alpha_{*})$].  
The fixed point (\ref{FP}) is degenerate in the sense that its  
coordinates depend continuously on the anisotropy parameters  
$\rho_{1,2}$.\footnote{ Formally, $\rho_{1,2}$ can be treated as the  
additional coupling constants. The corresponding beta functions  
$\beta_{1,2}\equiv\widetilde{\cal D}_\mu{\rho_{1,2}}$ vanish  
identically owing to the fact that $\rho_{1,2}$ are not renormalized.  
Therefore the equations $\beta_{1,2}=0$ give no additional constraints  
on the values of the parameters $g,\alpha$ at the fixed point.}

 \section{Solution of the RG equations. Invariant variables}  
 \label {sec:RGE}  
  
The solution of the RG equations is discussed in Refs.  
\cite{Zinn,book3,UFN,turbo,JETP} in detail; below we confine  
ourselves to only the information we need.

Consider the solution of the RG equation on the example of the  
even different-time structure functions  
\begin{equation}  
S_{2n}(\r,\tau)\equiv\langle[\theta(t,{\bf x})  
-\theta(t',{\bf x'})]^{2n}\rangle,  
\quad \r\equiv\x-\x', \quad \tau\equiv t-t' .  
\label{differ}  
\end{equation}  
It satisfies the RG equation $\D_{RG} S_{2n}=0$ with the operator  
$\D_{RG}$ from Eq. (\ref{RG2})  
(this fact does not follow automatically from Eq. (\ref{RG})  
and will be justified below in Sec. \ref{sec:Operators}).  
  
In renormalized variables, dimensionality considerations give  
\begin{equation}  
S_{2n}(\r,\tau)= \nu^{-n} r^{2n}  
R_{2n}(\mu r, \tau\nu/r^{2}, mr, g, \alpha),  
\label{differ2}  
\end{equation}  
where $R_{2n}$ is a function of completely dimensionless arguments (the  
dependence on $d$, $\eps$, $\rho_{1,2}$ and the angle between the  
vectors ${\bf r}$ and ${\bf n}$ is also implied).  
>From the RG equation the identical representation follows,  
\begin{equation}  
S_{2n}(\r,\tau)= (\bar\nu)^{-n} r^{2n} R_{2n}(1, \tau\bar\nu/r^{2}, mr,  
\bar g, \bar \alpha),  
\label{differ3}  
\end{equation}  
where the invariant variables $\bar e = \bar e (\mu r, e)$  
satisfy the equation $\D_{RG} \bar e =0$ and the normalization  
conditions $\bar e = e$ at $\mu r=1$   (we recall that  
$e\equiv \{\nu,g,\alpha,m\}$ denotes the full set of renormalized  
parameters). The identity $\bar m \equiv m$ is a consequence of the  
absence of $\D_m$ in the operator $\D_{RG}$ owing to the fact that  
$m$ is not renormalized. Equation (\ref{differ3}) is valid because  
both sides of it satisfy the RG equation and coincide for $\mu r=1$  
owing to the normalization of the invariant variables.  
The relation between the bare and invariant variables has the form  
\begin{equation}  
\nu_0=\bar \nu Z_{\nu}(\bar g),  
\quad g_{0}=\bar g r^{-\eps}Z_{g}(\bar g), \quad \alpha_{0}=  
\bar\alpha Z_{\alpha}(\bar g, \bar \alpha).  
\label{exo1}  
\end{equation}  
Equation (\ref{exo1}) determines  
implicitly the invariant variables as functions of the bare parameters;  
it is valid because both sides of it satisfy the RG equation, and  
because Eq. (\ref{exo1}) at $\mu r=1$ coincides with  
(\ref{18}) owing to the normalization conditions.  
  
In general, the large $\mu r$ behavior of the invariant variables is  
governed by the IR stable fixed point: $\bar g\to g_{*}$,  
$\bar\alpha\to\alpha_{*}$ for $\mu r\to\infty$. However, in multicharge  
problems one has to take into account that even when the IR point  
exists, not every phase trajectory (i.e., solution of Eq. (\ref{exo1}))  
reaches it in the limit $\mu r\to\infty$. It may first pass outside  
the natural region of stability [in our case, $g>0$, $\alpha>-1$]  
or go to infinity within this region. Fortunately, in our case the  
constants $Z_{F}$ entering into Eq. (\ref{exo1}) are known exactly  
from Eqs. (\ref{Z}), and it is readily checked that the RG flow indeed  
reaches the fixed point (\ref{FP}) for any initial conditions  
$g_{0}>0$, $\alpha_{0}>-1$, including the physical case $\alpha_{0}=0$.  
Furthermore, the large $\mu r$ behavior of the invariant viscosity  
$\bar\nu$ is also found explicitly from Eq. (\ref{exo1}) and the last  
relation in (\ref{18a}): $\bar\nu = D_{0} r^{\eps} /\bar g \to  
D_{0} r^{\eps} / g_{*}$  (we recall that $D_{0}=g_{0}\nu_0$).  
Then for $\mu r\to\infty$  and any fixed $mr$ we obtain  
\begin{equation}  
S_{2n}(\r,\tau)= D_{0}^{-n} r^{n(2-\eps)} g_{*}^{n}\,  
\xi_{2n}( \tau D_{0} r^{\Delta_{t}},mr),  
\label{differ4}  
\end{equation}  
where  
\begin{equation}  
\xi_{2n}( D_{0} \tau r^{\Delta_{t}},mr)\equiv  
R_{2n}(1, D_{0} \tau r^{\Delta_{t}}, mr, g_{*},  \alpha_{*}),  
\label{differ5}  
\end{equation}  
and $\Delta_{t}\equiv -2+\gamma_{\nu}^{*} =-2+\eps$  
is the critical dimension of time. The dependence of the scaling  
function $\xi_{2n}$ on its arguments is {\it not} determined by the  
RG equation (\ref{RG}) itself. For the equal-time structure function  
(\ref{struc}), the first argument of $\xi_{2n}$ in the representation  
(\ref{differ5}) is absent:  
\begin{equation}  
S_{2n}(\r)= D_{0}^{-n} r^{n(2-\eps)} g_{*}^{n}\,  \xi_{2n}(mr),  
\label{100}  
\end{equation}  
where the definition of $\xi_{2n}$ is obvious from (\ref{differ5}).  
It is noteworthy that Eqs. (\ref{differ4})--(\ref{100})  
prove the independence of the structure functions in the  
IR range (large $\mu r$  and any  $mr$) of the viscosity coefficient or,  
equivalently of the UV scale:  
the parameters $g_{0}$ and $\nu_0$ enter into Eq. (\ref{differ4})  
only in the form of the combination $D_{0}=g_{0}\nu_0$. A similar  
property was established in Ref. \cite{FF} for the stirred Navier--Stokes  
equation and is related to the Second Kolmogorov hypothesis; see also  
the discussion in \cite{turbo,JETP}.  
  
Now let us turn to the general case. Let $F(r,\tau)$ be some  
multiplicatively renormalized quantity (say, a correlation function  
involving composite operators), i.e., $F=Z_{F}F_{R}$ with  
certain renormalization constant $Z_{F}$. It satisfies the RG  
equation of the form $[\D_{RG} + \gamma_{F}] F_{R}=0$ with $\gamma_{F}$  
from (\ref{RGF1}). Dimensionality considerations give  
\begin{equation}  
F_{R}(r,\tau)= \nu^{d^{\omega}_{F}} r^{-d_{F}}  
R_{F}(\mu r, \tau\nu/r^{2}, mr, g, \alpha),  
\label{differ6}  
\end{equation}  
where $d^{\omega}_{F}$ and $d_{F}$ are the frequency and total canonical  
dimensions of $F$ (see Sec. \ref{sec:QFT1}) and $R_{F}$ is a function of  
dimensionless arguments. The analog of Eq. (\ref{differ3}) has the form  
\begin{equation}  
F(r,\tau)= Z_{F} (g, \alpha) F_{R} = Z_{F} (\bar g, \bar\alpha) \,  
(\bar\nu)^{d^{\omega}_{F}} r^{-d_{F}}  
R_{F}(1, \tau\bar\nu/r^{2}, mr, \bar g, \bar\alpha).  
\label{differ7}  
\end{equation}  
In the large $\mu r$ limit, one has  $Z_{F} (\bar g, \bar\alpha)  
\simeq \const (\Lambda r)^{-\gamma_{F}^{*}}$; see, e.g., \cite{AV}.  
The UV scale appears in this relation from Eq. (\ref{Lambda}). Then  
in the IR range ($\Lambda r\sim \mu r$ large, $mr$ arbitrary) Eq.  
(\ref{differ7}) takes on the form  
\begin{equation}  
F(r,\tau)\simeq \const \Lambda ^{-\gamma_{F}^{*}}\,  
D_{0}^{d^{\omega}_{F}} \, r^{-\Delta[F]}  
\xi_{F}( D_{0} \tau r^{\Delta_{t}},mr) .  
\label{RGR}  
\end{equation}  
Here  
\begin{equation}  
\Delta[F]\equiv\Delta_{F} = d_{F}^{k}- \Delta_{t}  
d_{F}^{\omega}+\gamma_{F}^{*}, \quad \Delta_{t}=-2+\eps  
\label{32B}  
\end{equation}  
is the critical dimension of the function $F$ and the scaling  
function $\xi_{F}$ is related to $R_{F}$ as in Eq. (\ref{differ4}).  
For nontrivial $\gamma_{F}^{*}$, the function $F$ in the IR range  
retains the dependence on $\Lambda$ or, equivalently, on $\nu_0$.

 \section{Renormalization and critical dimensions of  
 composite operators} \label {sec:Operators}

Any local (unless stated to be otherwise) monomial or polynomial  
constructed of primary fields and their derivatives at a single  
spacetime point $x\equiv (t,{\bf x})$ is termed a composite operator.  
Examples are $\theta^{N}(x)$,  
$[\partial_{i}\theta(x)\partial_{i}\theta(x)]^{N}$,  
$\partial_{i}\theta(x)\partial_{j}\theta(x)$,  
$\theta'(x)\nabla_{t}\theta(x)$ and so on.  
  
Since the arguments of the fields coincide, correlation functions with  
such operators contain additional UV divergences, which are removed  
by additional renormalization procedure; see,  
e.g., \cite{Zinn,book3}. For the renorma\-li\-zed correlation functions  
standard RG equations are obtained, which describe IR scaling of  
certain ``basis'' operators  $F$ with definite critical dimensions  
$\Delta_{F}\equiv\Delta[F]$. Due to the renormalization,  
$\Delta[F]$ does not coincide in general with the naive sum of  
critical dimensions of the fields and derivatives entering into $F$.  
As a rule, composite operators mix in renormalization,  
i.e., an UV finite renormalized operator $F^{R}$ has the form  
$F^{R}=F+$ counterterms, where the contribution of the  
counterterms is a linear combination of $F$ itself and,  
possibly, other unrenormalized operators which ``admix''  
to $F$ in renormalization.  
  
Let $F\equiv\{F_{\alpha}\}$ be a closed set, all of whose  
monomials mix only with each other in renormalization.  
The renormalization matrix $Z_{F}\equiv\{Z_{\alpha\beta}\}$  
and the  matrix of anomalous dimensions  
$\gamma_{F}\equiv\{\gamma_{\alpha\beta}\}$  
for this set are given by  
\begin{equation}  
F_{\alpha }=\sum _{\beta} Z_{\alpha\beta}  
F_{\beta }^{R},\qquad \gamma _F=Z_{F}^{-1}\D_{\mu }Z_{F},  
\label{2.2}  
\end{equation}  
and the corresponding matrix of critical dimensions  
$\Delta_{F}\equiv\{\Delta_{\alpha\beta}\}$ is given by Eq. (\ref{32B}),  
in which $d_{F}^{k}$ and $d_{F}^{\omega}$ are understood as  
the diagonal matrices of canonical dimensions of the operators in  
question (with the diagonal elements equal to sums of corresponding  
dimensions of all fields and derivatives constituting $F$) and  
$\gamma^{*}_{F}\equiv\gamma_{F} (g_{*},\alpha_{*})$ is the  
matrix (\ref{2.2}) at the fixed point (\ref{FP}).  
  
Critical dimensions of the set $F\equiv\{F_{\alpha}\}$ are  
given by the eigenvalues of the matrix $\Delta_{F}$. The ``basis''  
operators that possess definite critical dimensions have the form  
\begin{equation}  
F^{bas}_{\alpha}=\sum_{\beta} U_{\alpha \beta}F^{R}_{\beta}\ ,  
\label{2.5}  
\end{equation}  
where the matrix $ U_{F} =  \{U_{\alpha \beta} \}$ is such that  
$\Delta'_{F}= U_{F} \Delta_{F} U_{F}^{-1}$ is diagonal.  
  
In general, counterterms to a given operator $F$ are  
determined by all possible 1-irreducible Green functions  
with one operator $F$ and arbitrary number of primary fields,  
$\Gamma=\langle F(x) \Phi(x_{1})\dots\Phi(x_{2})\rangle_{\rm 1-ir}$.  
The total canonical dimension (formal index of UV divergence)  
for such function is given by  
\begin{equation}  
d_\Gamma = d_{F} - N_{\Phi}d_{\Phi},  
\label{index}  
\end{equation}  
with the summation over all types of fields entering into  
the function. For superficially divergent diagrams,  
$d_\Gamma$ is a non-negative integer, cf. Sec. \ref{sec:QFT1}.  
  
Let us consider operators of the form $\theta^{N}(x)$  
with the canonical dimension $d_{F}=-N$, entering into the  
structure functions (\ref{struc}). From Table \ref{table1}  
in Sec. \ref{sec:QFT1} and Eq. (\ref{index}) we obtain  
$d_\Gamma = -N+N_{\theta}-N_{\bfv} -(d+1)N_{\theta'}$,  
and from the analysis of the diagrams it follows that the total  
number of the fields $\theta$ entering into the function  
$\Gamma$ can never exceed the number of the fields $\theta$  
in the operator $\theta^{N}$ itself, i.e., $N_{\theta}\le N$  
(cf. item (i) in Sec. \ref{sec:QFT1}).  
Therefore, the divergence can only exist in the functions  
with $N_{\bfv}= N_{\theta'}=0$, and arbitrary value of  
$N=N_{\theta}$, for which the formal index vanishes, $d_\Gamma =0$.  
However, at least one of $N_{\theta}$ external ``tails''  
of the field $\theta$ is attached to a vertex  
$\theta'({\bfv}\partt)\theta$ (it is impossible to construct  
nontrivial, superficially divergent diagram of the desired  
type with all the external tails attached to the vertex $F$),  
at least one derivative $\partial$ appears as an extra  
factor in the diagram, and, consequently, the real index of  
divergence $d_\Gamma'$ is necessarily negative.  
  
This means that the operator $\theta^{N}$ requires no counterterms  
at all, i.e., it  is in fact UV finite, $\theta^{N}=Z\,[\theta^{N}]^{R}$  
with $Z=1$. It then follows that the critical dimension of  
$\theta^{N}(x)$ is simply given by the expression (\ref{32B})  
with no correction from $\gamma_{F}^{*}$ and is therefore reduced  
to the sum of the critical dimensions of the factors:  
\begin{equation}  
\Delta [\theta^{N}] = N\Delta[\theta] =N (-1+\eps/2).  
\label{2.6}  
\end{equation}  
Since the structure functions (\ref{struc}) or (\ref{differ}) are linear  
combinations of pair correlators involving the operators $\theta^{N}$,  
equation (\ref{2.6}) shows that they indeed satisfy the RG equation  
of the form (\ref{RG}), discussed in Sec. \ref{sec:RGE}.  
We stress that the relation (\ref{2.6}) was not clear {\it a priori};  
in particular, it is violated if the velocity field becomes  
nonsolenoidal \cite{RG1}.

In the following, an important role will be also played by the  
tensor composite operators  
$ \partial_{i_{1}}\theta\cdots\partial_{i_{p}}\theta\,  
(\partial_{i}\theta\partial_{i}\theta)^{n}$  
constructed solely of the scalar gradients. It is convenient to deal  
with the scalar operators obtained by contracting the tensors with  
the appropriate number of the vectors $\n$,  
\begin{equation}  
F[N,p]\equiv [(\n\partt)\theta]^{p} (\partial_{i}\theta\partial_{i}  
\theta)^{n}, \quad N\equiv 2n+p.  
\label{Fnp}  
\end{equation}  
Their canonical dimensions depend only on the total number of the fields  
$\theta$ and have the form $d_{F}=0$, $d_{F}^{\omega}=-N$.  
  
In this case, from Table \ref{table1} and Eq. (\ref{index}) we obtain  
$d_\Gamma = N_{\theta}-N_{\bfv} -(d+1)N_{\theta'}$,  
with the necessary condition $N_{\theta}\le N$, which follows  
from the structure of the diagrams. It is also clear from the  
analysis of the diagrams that the counterterms to these operators  
can involve the fields $\theta$, $\theta'$ only in the form of  
derivatives, $\partial\theta$, $\partial\theta'$,  
so that the real index of divergence has the form  
$d_\Gamma' = d_\Gamma -N_{\theta}-N_{\theta'}=  
-N_{\bfv} -(d+2)N_{\theta'}$.   It then follows that  
superficial divergences can exist only in the Green functions with  
$N_{\bfv}=N_{\theta'}=0$  and any $N_{\theta}\le N$,  
and that the corresponding operator counterterms reduce to the  
form $F[N',p']$ with $N'\le N$. Therefore, the operators  
(\ref{Fnp}) can mix only with each other in renormalization, and  
the corresponding infinite renormalization matrix  
\begin{equation}  
F[N,p] = \sum_{N',p'} \, Z_{[N,p]\,[N',p']} \,F^{R}[N',p']  
\label{Matrix}  
\end{equation}  
is in fact block-triangular, i.e., $Z_{[N,p]\,[N',p']} =0$ for $N'> N$.  
It is then obvious that the critical dimensions associated with the  
operators $F[N,p]$ are completely determined by the eigenvalues of the  
finite subblocks with $N'= N$. In the following, we shall not be interested  
in the precise form of the basis operators (\ref{2.5}), we rather shall  
be interested in the anomalous dimensions themselves. Therefore, we  
can neglect all the elements of the matrix (\ref{Matrix}) other than  
$Z_{[N,p]\,[N,p']}$.

In the isotropic case, as well as in the presence of large-scale  
anisotropy, the elements $Z_{[N,p]\,[N,p']}$ vanish for $p<p'$,  
and the block $Z_{[N,p]\,[N,p']}$ is triangular along with the  
corresponding blocks of the matrices $U_{F}$ and $\Delta_{F}$  
from Eqs. (\ref{2.5}), (\ref{32B}). In the isotropic case it can be  
diagonalized by changing to irreducible operators (scalars, vectors,  
and traceless tensors), but even for nonzero imposed gradient its  
eigenvalues are the same as in the isotropic case. Therefore, the  
inclusion of large-scale anisotropy does not affect critical  
dimensions of the operators (\ref{Fnp}); see \cite{RG3}. In the  
case of small-scale anisotropy, the operators with different values  
of $p$ mix heavily in renormalization, and the matrix  
$Z_{[N,p]\,[N,p']}$ is neither diagonal nor triangular here.

Now let us turn to the calculation of the renormalization constants  
$Z_{[N,p]\,[N,p']}$ in the one-loop approximation.  
Let $\Gamma(x;\theta)$ be the generating functional of the  
1-irreducible Green functions with one composite operator $F[N,p]$  
from Eq. (\ref{Fnp}) and any number of fields $\theta$. Here $x\equiv  
(t,{\bf x})$ is the argument of the operator and $\theta$ is  
the functional argument, the ``classical counterpart'' of the random  
field $\theta$. We are interested in the $N$-th term of the  
expansion of $\Gamma(x;\theta)$ in $\theta$, which we denote  
$\Gamma_{N}(x;\theta)$; it has the form  
\begin{equation}  
\Gamma_{N}(x;\theta) = \frac{1}{N!} \int dx_{1} \cdots \int dx_{N}  
\, \theta(x_{1})\cdots\theta(x_{N})\,  
\langle F[N,p](x) \theta(x_{1})\cdots\theta(x_{N})\rangle_{\rm 1-ir}.  
\label{Gamma1}  
\end{equation}  
In the one-loop approximation the function (\ref{Gamma1}) is  
represented diagrammatically in the following manner:  
\begin{equation}  
\Gamma_{N}= F[N,p] + \frac{1}{2} \put(-20.00,-50.00){\makebox{\dA}}  
\hskip1.4cm .  
\label{Gamma2}  
\end{equation}  
Here the solid lines denote the {\it bare} propagator  
$\langle\theta\theta'\rangle_{0}$ from Eq. (\ref{lines2}),  
the ends with a slash correspond to the field $\theta'$, and the  
ends without a slash correspond to $\theta$; the dashed line  
denotes the bare propagator (\ref{3}); the vertices correspond  
to the factor (\ref{vertex}).  
  
The first term is the ``tree'' approximation, and the black circle with  
two attached lines in the diagram denotes the variational derivative  
$ V(x;\, x_{1}, x_{2}) \equiv \delta^{2} F[N,p] /  
{\delta\theta(x_{1})\delta\theta(x_{2})}$. It is convenient to  
represent it in the form  
\begin{equation}  
V(x;\, x_{1}, x_{2})=  
\partial_{i} \delta(x-x_{1})\, \partial_{j} \delta(x-x_{2})\,  
\frac{\partial^{2}}{\partial a_{i} \partial a_{j}}\,  
\Bigl[ (\n\a)^{p} (a^2)^{n} \Bigr],  
\label{Vertex}  
\end{equation}  
where $a_{i}$ is a constant vector, which {\it after the differentiation}  
is substituted with $\partial_{i} \theta(x)$.  
The diagram in Eq. (\ref{Gamma2}) is written analytically in the form  
\begin{equation}  
\int dx_{1} \cdots \int dx_{4} V(x;\, x_{1}, x_{2})  
\langle \theta(x_{1}) \theta'(x_{3}) \rangle_{0}  
\langle \theta(x_{2}) \theta'(x_{4}) \rangle_{0}  
\langle v_{k}(x_{3}) v_{l} (x_{4}) \rangle_{0}  
\partial_{k} \theta(x_{3})  \partial_{l} \theta(x_{4}),  
\label{Coor}  
\end{equation}  
with the bare propagators from Eqs. (\ref{3}), (\ref{lines2});  
the derivatives appear from the ordinary vertex factors  
(\ref{vertex}).

In order to find the renormalization constants, we need not the entire  
exact expression (\ref{Coor}), rather we need its UV divergent part.  
The latter is proportional to a polynomial built of $N$ factors  
$\partial\theta$ at a single spacetime point $x$. The needed $N$  
gradients has already been factored out from the expression (\ref{Coor}):  
$(N-2)$ factors from the vertex (\ref{Vertex}) and 2 factors  
from the ordinary vertices (\ref{vertex}). Therefore, we can neglect the  
spacetime inhomogeneity of the gradients and replace them with the  
constant vectors $a_{i}$. Expression (\ref{Coor}) then can be written,  
up to an UV finite part, in the form  
\begin{equation}  
a_{k}a_{l} \, \frac{\partial^{2}}{\partial a_{i} \partial a_{j}}\,\Bigl[  
(\n\a)^{p} (a^{2}) ^{n}\Bigr]\,  X_{ij,\,kl},  
\label{Coor2}  
\end{equation}  
where we have denoted  
\begin{equation}  
X_{ij,\,kl} \equiv  
\int dx_{3} \, \int dx_{4} \,\,  
\partial_{i} \langle \theta(x) \theta'(x_{3}) \rangle_{0}\,  
\partial_{j}\langle \theta(x) \theta'(x_{4}) \rangle_{0} \,  
\langle v_{k}(x_{3}) v_{l} (x_{4}) \rangle_{0},  
\label{X}  
\end{equation}  
or, in the momentum-frequency representation, after the integration over  
the frequency,  
\begin{equation}  
X_{ij,\,kl} =  \frac{D_{0}}{2\nu_0} \, \int \frac{d\q}{(2\pi)^d}\,  
\frac{q_{i}q_{j}}{(q^{2}+\alpha (\q\n)^{2})}\,  
\frac {T_{kl}(\q)} {(k^{2}+m^{2})^{-d/2-\eps/2}},  
\label{X2}  
\end{equation}  
with $D_{0}$ from Eq. (\ref{3}) and $T_{kl}$ from Eq. (\ref{T34}).  
The tensor $X_{ij,\,kl}$ can be decomposed in certain basic structures:  
\begin{equation}  
X_{ij,\,kl} =  \sum_{s=1}^{6}  B_{s}\,  S^{(s)}_{ij,\,kl} \,  ,  
\label{X3}  
\end{equation}  
where $B_{s}$ are scalar coefficients and $S^{(s)}$ are all possible  
fourth rank tensors constructed of the vector $\n$ and Kronecker  
delta symbols, and symmetric with respect to the permutations within  
the pairs $\{ij\}$ and $\{kl\}$:  
\begin{eqnarray}  
S^{(1)}_{ij,\,kl} &=& \delta_{ij}\delta_{kl} , \quad  
S^{(2)}_{ij,\,kl} =(\delta_{ik}\delta_{jl} + \delta_{il}\delta_{jk})/2,  
                                            \quad  
S^{(3)}_{ij,\,kl} = \delta_{ij} n_{k}n_{l}, \quad  
S^{(4)}_{ij,\,kl} = \delta_{kl} n_{i}n_{j}, \nonumber \\  
S^{(5)}_{ij,\,kl} &=& (n_{i}n_{k}\delta_{jl} + n_{i}n_{l}\delta_{jk} +  
           n_{j}n_{l}\delta_{ik} + n_{j}n_{k}\delta_{il} )/4, \quad  
S^{(6)}_{ij,\,kl} = n_{i}n_{j}n_{k}n_{l}.  
\label{X4}  
\end{eqnarray}  
It is now convenient to change from the tensors (\ref{X2}), (\ref{X3})  
to the scalar integrals $A_{s}$ defined by the relations  
\begin{equation}  
A_{s}= \sum_{ijkl}\, X_{ij,\,kl} \, S^{(s)}_{ij,\,kl}, \quad  
s=1,\cdots,6  
\label{X5}  
\end{equation}  
[it is obvious from Eqs. (\ref{X3}), (\ref{X4}) that the coefficients  
$A_{2}$ and $A_{5}$ vanish identically by virtue of the relation  
$q_{l} T_{lk}(\q) =0$].  
The UV divergence and the angular integration in Eqs. (\ref{X2}),  
(\ref{X3}) can be disentangled using the relations  
\begin{eqnarray}  
A_{s} = \frac{D_{0}}{2\nu_0} \,  J(m)  \, H_{s},  
\label{X8}  
\end{eqnarray}  
where the integral $J(m)$ from (\ref{otvet2}) contains a pole in $\eps$,  
and $H_{s}$ are UV finite and $\eps$ independent quantities given by  
only the angular integrations:  
\begin{eqnarray}  
H_{1} &=& \biggl\langle \frac{T_{ii}} {1+\alpha\cos^{2}\psi}  
\biggr\rangle_{S}\, , \quad  
H_{3} = \biggl\langle \frac{n_{i}n_{j}T_{ij}} {1+\alpha\cos^{2}\psi}  
\biggr\rangle_{S} \, , \nonumber \\ \  
\nonumber \\  
H_{4} &=& \biggl\langle \frac{T_{ii}\,\cos^{2}\psi}  
{1+\alpha\cos^{2}\psi} \biggr\rangle_{S} \, , \quad  
H_{6} = \biggl\langle \frac{n_{i}n_{j}T_{ij}\,\cos^{2}\psi}  
{1+\alpha\cos^{2}\psi} \biggr\rangle_{S}\, ,  
\label{X9}  
\end{eqnarray}  
where $\psi$ is the angle between the vectors ${\bf q}$ and ${\bf n}$,  
so that $({\bf n}{\bf q})=q\cos\psi$, and the quantities $T_{ii}$  
and $n_{i}n_{j}T_{ij}$ with $T_{ij}$ from (\ref{T34})  
depend only on $\psi$ (the dependence on $d$ and the anisotropy  
parameters $\rho_{1,2}$ is of course also implied).  The brackets  
$\langle\cdots \rangle_{S}$ denote averaging over the unit  
$d$-dimensional sphere, which for a function dependent only on  
$\psi$ takes on the form  
\begin{equation}  
\langle f(\psi) \rangle_{S} =\frac{S_{d-1}}{S_{d}} \int^{\pi}_{0}  
d\psi\, \sin^{d-2}\psi\, f(\psi),  
\label{sphere}  
\end{equation}  
with $S_{d} = 2\pi^{d/2}/\Gamma(d/2)$. Finally, the change of  
variables $y=\cos^{2} \psi$ in the integrals (\ref{X9}) gives:  
\begin{equation}  
\int^{1}_{0} dy\, \frac{y^{a-1} (1-y)^{b-1}}{(1+\alpha y)^{c}} =  
B(a,b) \, F(c,a; a+b; -\alpha),  
\label{hyperpuper}  
\end{equation}  
where $B$ is the Euler beta function and $F$ is  
the hypergeometric function, see, e.g., \cite{Grad}. This gives:  
\begin{eqnarray}  
H_{1} &=& (d-1+\rho_{2})\, F_{0} + \Bigl[(d-1)\rho_{1}-\rho_{2}\Bigr] \,  
\frac{F_{1}}{d},  
\nonumber \\  
H_{3} &=& (1+\rho_{2})\, F_{0} + (-1+\rho_{1}-2\rho_{2})  
\, \frac{F_{1}}{d} + (\rho_{2}-\rho_{1}) \frac{3F_{2}} {d(d+2)},  
\nonumber \\  
H_{4} &=& (d-1+\rho_{2})\, \frac{F_{1}}{d} +  
\Bigl[(d-1)\rho_{1}-\rho_{2}\Bigr] \,  
\frac{3F_{2}} {d(d+2)},  
\nonumber \\  
H_{6} &=&(1+\rho_{2})\, \frac{F_{1}}{d} + (-1+\rho_{1}-2\rho_{2})\,  
\frac{3F_{2}} {d(d+2)}+ (\rho_{2}-\rho_{1})\,  
\frac{15F_{3}}{d(d+2)(d+4)},  
\label{hyperpuper2}  
\end{eqnarray}  
where we have denoted  
$F_{n} \equiv F(1,1/2+n; d/2+n; -\alpha)$.  
  
In principle, all the hypergeometric functions entering into Eqs.  
(\ref{hyperpuper2}) can be reduced to the only function, for example  
$F_{3}$, but the coefficients then become too complex.  
  
After quite lengthy but straightforward calculations, the quantity  
in Eq. (\ref{Coor2}) can be expressed through the coefficients $B_{s}$,  
then through $A_{s}$, and finally through $H_{s}$:  
\begin{equation}  
\frac{D_{0}}{2\nu_0} \, J(m) \, \Bigl\{ Q_{1}\, F[N,p-2] +  
Q_{2}\, F[N,p] + Q_{3}\, F[N,p+2] + Q_{4}\, F[N,p+4] \Bigr\} ,  
\label{X6}  
\end{equation}  
where we have substituted $a_{i}$ back with the gradients  
$\partial_{i}\theta(x)$, used the notation (\ref{Fnp}), and denoted  
\begin{eqnarray}  
Q_{1} &\equiv& \frac{p(p-1)}{(d-1)} \Bigl(H_{4}-H_{6}\Bigr),  
\nonumber \\  
\nonumber \\  
Q_{2} &\equiv& \frac{p(p-1)}{(d-1)} \Bigl(dH_{6}-H_{4}\Bigr) +  
\frac{(N-p)} {(d-1)(d+1)}\, \Bigl\{ (N-p+d-1) (H_{1}-H_{3}) +  
\Bigr.  \nonumber \\  
\Bigl. &+& H_{4}  
\bigl[ -(N-2)+p(2d+3) \bigr] +3 H_{6} \bigl[ (N-2)-p(d+2) \bigr]  
\Bigr\} ,  
\nonumber \\  
\nonumber \\  
Q_{3} &\equiv& \frac{(N-p)}{(d-1)(d+1)}  \Bigl\{  -(2N-2p+d-3) H_{1} +  
H_{3} \bigl[ (N-p)(d+3) +(d+2)(d-3) \bigr]  +  
\Bigr.  \nonumber \\   \Bigl.  
&+&  H_{4} \bigl[ (N-2)(d+3)-p(3d+8)  \bigr]  + 2 H_{6}  (d+2)  
\bigl[ -3(N-2)+p(d+4)  \bigr] \Bigr\},  
\nonumber \\  
\nonumber \\  
Q_{4} &\equiv& \frac{(N-p)(N-p-2)}{(d-1)(d+1)} \Bigl[ H_{1}  
- (d+2) (H_{3}+ H_{4}) +(d+2)(d+4)H_{6} \Bigr].  
\label{X7}  
\end{eqnarray}

The renormalization constants $Z_{[N,p]\,[N,p']}$  
can be found from the requirement that the function (\ref{Gamma2})  
after the replacement $F[N,p]\to F^{R}[N,p]$ with $F^{R}[N,p]$  
from (\ref{Matrix}) be UV finite, i.e., have no poles in  $\eps$,  
when expressed in renormalized variables using the formulas  
(\ref{18}). In the MS scheme, the diagonal element has the form  
$Z_{[N,p]\,[N,p']}=1+$  only poles in $\eps$, while the nondiagonal  
ones contain only poles. Then from Eqs. (\ref{Gamma2}), (\ref{X8})  
and (\ref{X6}) one obtains  
\begin{eqnarray}  
Z_{[N,p][N,p]} = 1+ \frac{g\,C_{d}} {4\eps}\, Q_{2}, \quad  
Z_{[N,p][N,p-2]} &=& \frac{g\,C_{d}} {4\eps}\, Q_{1},  \nonumber \\  
Z_{[N,p][N,p+2]} = \frac{g\,C_{d}} {4\eps}\, Q_{3},  \quad  
Z_{[N,p][N,p+4]} &=& \frac{g\,C_{d}} {4\eps}\, Q_{4},  
\label{Znp}  
\end{eqnarray}  
with coefficients $Q_{i}$ from Eq. (\ref{X7}) and $C_{d}$  
from (\ref{Z}). The corresponding anomalous dimensions  
$\ \gamma_{[N,p]\,[N',p']} = Z_{[N,p]\,[N'',p'']}^{-1}  \Dm  
Z_{[N'',p'']\,[N',p']}\ $ have the form  
\begin{eqnarray}  
\gamma_{[N,p][N,p]} = - g\,C_{d}\, Q_{2}/ 4, \quad  
\gamma_{[N,p][N,p-2]} &=&- g\,C_{d}\, Q_{1}/ 4, \nonumber \\  
\gamma_{[N,p][N,p+2]} = - g\,C_{d}\, Q_{3}/ 4, \quad  
\gamma_{[N,p][N,p+4]} &=&- g\,C_{d}\, Q_{4}/ 4.  
\label{Gnp}  
\end{eqnarray}  
In contrast to exact expressions (\ref{Z}), the quantities in  
Eqs. (\ref{Znp}), (\ref{Gnp}) have nontrivial corrections of  
order $g^{2}$ and higher.  
The matrix of critical dimensions (\ref{32B}) is given by  
\begin{equation}  
\Delta_{[N,p][N,p']} = N\eps/2 + \gamma^{*}_{[N,p][N,p']},  
\label{Dnp}  
\end{equation}  
where the asterisk implies the substitution (\ref{FP}).  
The set of equations (\ref{FP}), (\ref{hyperpuper2}), (\ref{X7}),  
(\ref{Gnp}), (\ref{Dnp}) gives the desired  expression for  
the matrix of critical dimensions of the composite operators  
(\ref{Fnp}) in the one-loop approximation, i.e., in the first order  
in $\eps$ (we recall that $g_{*}=O(\eps)$ and $\alpha_{*}=O(1)$).  
In contrast to simpler expressions like (\ref{2.6}),  
the dimensions in Eq. (\ref{Dnp}) depend on the anisotropy  
parameters $\rho_{1,2}$ and have nontrivial corrections of order  
$\eps^{2}$ and higher.  
  
In the one-loop approximation, the expression (\ref{X6}) contains  
only four terms, and therefore all the matrix elements  
$\gamma_{[N,p][N,p']}$ other than (\ref{Gnp}) vanish. However, they  
become nontrivial in higher orders, so that the basis operators  
(\ref{2.5}), associated with the eigenvalues of the matrix (\ref{Dnp}),  
are given by mixtures of the monomials (\ref{Fnp}) with all possible  
values of the index $p$, or, in other words, mixtures of the tensors  
belonging to different representations of the rotation group.

As already said above, the critical dimensions themselves are given  
by the eigenvalues of the matrix (\ref{Dnp}). One can check that for  
the isotropic case ($\rho_{1,2}=0$), its elements with $p'>p$ vanish,  
the matrix becomes triangular, and its eigenvalues are simply given by  
the diagonal elements $\Delta[N,p]\equiv\Delta_{[N,p][N,p]}$. They  
are found explicitly and have the form  
\begin{equation}  
\Delta[N,p] = N\eps/2+ \frac{2p(p-1)-(d-1)(N-p)(d+N+p)}{2(d-1)(d+2)}\,  
\eps +O(\eps^{2}).  
\label{Qnp}  
\end{equation}  
It is easily seen from Eq. (\ref{Qnp}) that for fixed $N$ and any  
$d\ge2$, the dimension $\Delta[N,p]$ decreases monotonically with $p$  
and reaches its minimum for the minimal possible value of $p=p_{N}$,  
i.e., $p_{N}=0$ if $N$ is even and $p_{N}=1$ if $N$ is odd:  
\begin{mathletters}  
\label{hier}  
\begin{equation}  
\Delta[N,p] > \Delta[N,p']  \quad {\rm if} \quad  p>p' \, .  
\label{hier1}  
\end{equation}  
Furthermore, this minimal value $\Delta[N,p_{N}]$ decreases monotonically  
as $N$ increases for odd and even values of $N$ separately, i.e.,  
\begin{equation}  
0\ge\Delta[2n,0]>\Delta[2n+2,0] , \quad \Delta[2n+1,1]> \Delta[2n+3,1].  
\label{hier2}  
\end{equation}  
A similar hierarchy is demonstrated by the critical dimensions of  
certain tensor operators in the stirred Navier--Stokes turbulence;  
see Ref. \cite{Triple} and Sec. 2.3 of \cite{turbo}. However, no  
clear hierarchy is demonstrated by neighboring even  
and odd dimensions:  
from the relations  
\begin{equation}  
\Delta[2n+1,1]-\Delta[2n,0]=\frac{\eps(d+2-4n)}{2(d+2)}, \quad  
\Delta[2n+2,0]- \Delta[2n+1,1] =\frac{\eps(2-d)}{2(d+2)}  
\label{hier3}  
\end{equation}  
\end{mathletters}  
it follows that the inequality $\Delta[2n+1,1]>\Delta[2n+2,0]$  
holds for any $d>2$, while the relation $\Delta[2n,0]>\Delta[2n+1,1]$  
holds only if $n$ is sufficiently large, $n>(d+2)/4$.  
\footnote{The situation is different in the presence of the linear  
mean gradient: the first term $N\eps/2$ in Eq. (\ref{Qnp}) is then  
absent owing to the difference in canonical dimensions, and the  
complete hierarchy relations hold,  
$\Delta[2n,0]>\Delta[2n+1,1]>\Delta[2n+2,0]$; see \cite{RG3}. }

In what follows, we shall use the notation $\Delta[N,p]$ for the  
eigenvalue of the matrix (\ref{Dnp}) which coincide with (\ref{Qnp})  
for $\rho_{1,2}=0$. Since the eigenvalues depend continuously on  
$\rho_{1,2}$, this notation is unambiguous at least for small  
values of $\rho_{1,2}$.

The dimension $\Delta[2,0]$ vanishes identically for any $\rho_{1,2}$  
and to all orders in $\eps$. Like in the isotropic model, this can be  
demonstrated using the Schwinger equation of the form  
\begin{equation}  
\int{\cal D}\Phi {\delta} \left[ \theta(x) \exp S_{R}( \Phi)  
+ A \Phi\right]/{\delta\theta'(x)}  =0,  
\label{Schwi}  
\end{equation}  
(in the general sense of the word, Schwinger equations are any relations  
stating that any functional integral of a total variational derivative  
is equal to zero; see, e.g., \cite{Zinn,book3}). In (\ref{Schwi}),  
$S_{R}$ is the renormalized action (\ref{renormalized}), and the notation  
introduced in (\ref{field}) is used. Equation (\ref{Schwi}) can be  
rewritten in the form  
\begin{eqnarray}  
\Big\langle \theta' D_{\theta} \theta - \nabla_{t}[\theta^{2}/2] +  
\nu Z_{1}\Delta[\theta^{2}/2] + \alpha \nu Z_{2} (\n\partt)^{2}  
[\theta^{2}/2] -\nu Z_{1} F[2,0] - \alpha \nu Z_{2} F[2,2]  
\Big\rangle_{A}  
= -A_{\theta'} \delta W_{R}(A)/\delta A_{\theta}.  
\label{Schwi2}  
\end{eqnarray}  
Here $D_{\theta}$ is the correlator (\ref{2}), $\langle \cdots \rangle  
_{A}$  
denotes the averaging with the weight $ \exp [S_{R}( \Phi) +  
A \Phi]$, $W_{R}$ is determined by Eq. (\ref{field}) with  
the replacement $S\to S_{R}$, and the argument $x$ common to all  
the quantities in (\ref{Schwi2}) is omitted.  
  
The quantity $\langle F \rangle _{A}$ is the generating functional  
of the correlation functions with one operator $F$ and any number of  
the primary fields $\Phi$, therefore the UV finiteness of the  
operator $F$ is equivalent to the finiteness of the functional  
$\langle F\rangle _{A}$. The quantity in the right hand side of Eq.  
(\ref{Schwi2}) is UV finite (a derivative of the renormalized functional  
with respect to finite argument), and so is the operator in the  
left hand side. Our operators $F[2,0]$, $F[2,2]$ do not admix in  
renormalization to $\theta' D_{\theta}\theta$  
(no needed diagrams can be constructed), and to the  
operators $\nabla_{t}[\theta^{2}/2]$ and $\triangle[\theta^{2}/2]$  
(they have the form of total derivatives, and $F[N,p]$ do not  
reduce to this form). On the other hand, all the operators in  
(\ref{Schwi2}) other than $F[N,p]$ do not admix to $F[N,p]$,  
because the counterterms of the operators (\ref{Fnp}) can involve  
only operators of the same type; see above. Therefore,  
the operators $F[N,p]$ entering into Eq. (\ref{Schwi2}) are independent  
of the others, and so they must be UV finite separately:  
$\nu Z_{1} F[2,0] + \alpha \nu Z_{2} F[2,2] = $ UV  finite.  
Since the operator in (\ref{Schwi2}) is UV finite, it coincides  
with its finite part,  
$$  
\nu Z_{1} F[2,0] + \alpha \nu Z_{2} F[2,2] =  
\nu F^{R}[2,0] + \alpha \nu  F^{R}[2,2],  
$$  
which along with the relation (\ref{Matrix}) gives  
$$  
Z_{1} Z_{[2,0][2,0]} + \alpha Z_{2} Z_{[2,2][2,0]} =1, \quad  
Z_{1} Z_{[2,0][2,2]} + \alpha Z_{2} Z_{[2,2][2,2]} =\alpha,  
$$  
and therefore for the anomalous dimensions in the MS scheme one obtains  
$$  
\gamma_{1} + \gamma_{[2,0][2,0]} +\alpha \gamma_{[2,2][2,0]} =0, \quad  
\gamma_{[2,0][2,2]} + \alpha \gamma_{2} + \alpha \gamma_{[2,2][2,2]} =0.  
$$  
These relations can indeed be checked from the explicit first-order  
expressions (\ref{Znp}), (\ref{Gnp}). Bearing in mind that  
$\gamma_{1}^{*}=\gamma_{2}^{*}=\eps$ (see Sec.  
\ref{sec:QFT1}), we conclude that among the four elements of the  
matrix $\gamma_{F}^{*}$ only two, which we take to be  
$\gamma^{*}_{[2,2][2,0]}$ and $\gamma^{*}_{[2,2][2,2]}$,  
are independent. Then the matrix of critical dimensions (\ref{Qnp})  
takes on the form  
\begin{eqnarray}  
\Delta_{[2,p][2,p']} = \eps + \pmatrix{ -\eps-\alpha_{*}  
\gamma^{*}_{[2,2][2,0]}  
& -\alpha_{*}\eps-\alpha_{*} \gamma^{*}_{[2,2][2,2]} \cr  
\gamma^{*}_{[2,2][2,0]} & \gamma^{*}_{[2,2][2,2]} \cr} .  
\label{Schwi3}  
\end{eqnarray}  
It is then easily checked that the eigenvalue of the matrix  
(\ref{Schwi3}), which is identified with $\Delta[2,0]$, does not  
involve unknown anomalous dimensions and vanishes identically,  
$\Delta[2,0]\equiv0$, while the second one is represented as  
$$  
\Delta[2,2]=\eps-\alpha_{*}\gamma^{*}_{[2,2][2,0]}+\gamma^{*}_{[2,2][2,2]}.  
$$  
Using the explicit $O(\eps)$ expressions (\ref{X7}), (\ref{Gnp}),  
one obtains to the order $O(\eps)$:  
\begin{eqnarray}  
\Delta[2,2]/\eps = 2+ \Bigl\{  
-(d-2)d(d+2)(d+4)F^{*}_{0} - (d+2)(d+4) (2+(d-2)\rho_{1}+d\rho_{2})  
F^{*}_{1}+  
\nonumber \\  
+3(d+4)(d-2\rho_{1}+2d\rho_{2}) F^{*}_{2}+  
15d (\rho_{1}-\rho_{2})  F^{*}_{3} \Bigr\} \Bigl/  
\Bigl\{(d-1)(d+4)[(d-1)(d+2)+(d+1)\rho_{1}+\rho_{2}]\Bigr\} \, ,  
\label{Delta22}  
\end{eqnarray}  
where $F^{*}_{n} \equiv F(1,1/2+n; d/2+n; -\alpha_{*})$ with  
$\alpha_{*}$ from Eq. (\ref{FP}) and the hypergeometric function from  
Eq. (\ref{hyperpuper}).  
  
In Fig. 1, we present the levels of the dimension (\ref{Delta22})  
on the $\rho_{1}$--$\rho_{2}$ plane for $d=3$.  
We note that the dependence on $\rho_{1,2}$  
is quite smooth, and that $\Delta[2,2]$ remains positive  
on the whole of the $\rho_{1,2}$ plane, i.e., the first of the  
hierarchy relations (\ref{hier}) remains valid also in the presence  
of anisotropy. A similar behavior takes place also for $d=2$.  
  
For $N>2$, the eigenvalues can be found analytically only within the  
expansion in $\rho_{1,2}$. Below we give such expansions up to  
the terms $O(\rho_{1,2}^{2})$ for all the eigenvalues with $N\le4$  
(with the notation $\rho_{3}\equiv 2\rho_{1}+d \rho_{2}$):  
\begin{mathletters}  
\label{expansions}  
\begin{eqnarray}  
\gamma^{*}[2,2]/\eps = \frac{2}{(d-1)(d+2)} +  
\frac{4(d-2)(d+1)\rho_{3}} {(d-1)^{2}(d+2)^{2}(d+4)} +  
\frac{2(d+1)}{(d-1)^{3}(d+2)^{3}(d+4)(d+6)} \times  
\nonumber \\ \times  
\biggl[  
-6d(d^{2}+6d-15)\rho_{1}^{2}+  
(d^{5}+6d^{4}-41d^{3}+ 42d^{2}+16d+24)\rho_{1}\rho_{2} +  
(-11d^{4}+23d^{3}+2d^{2}-26d+24)\rho_{2}^{2} \biggr],  
\label{expansion22}  
\end{eqnarray}  
\begin{eqnarray}  
\gamma^{*}[3,1]/\eps=- \frac{(d+4)}{(d+2)} -  
\frac{2(d-2)(d+1)\rho_{3}} {(d-1)(d+2)^{3}} +  
\frac{2(d-2)(d+1)}{(d-1)^{2}(d+2)^{6}(d+4)^{2}} \times  
\nonumber \\ \times  
\biggl[  2(d^{5}+15d^{4}+88d^{3}+204d^{2}+16d-96) \rho_{1}^{2}+  
\nonumber \\  
+d(-d^{6}+d^{5}+80d^{4}+296d^{3}+ 64d^{2}-48d+64)  
\rho_{1}\rho_{2} +  
 (3d^{6}+29d^{5}+74d^{4}-32d^{3}-72d^{2}+48d+64)  
\rho_{2}^{2} \biggr],  
\label{expansion31}  
\end{eqnarray}  
\begin{eqnarray}  
\gamma^{*}[3,3]/\eps= \frac{6}{(d-1)(d+2)} +  
\frac{12(d-2)(d+1)^{2}\rho_{3}} {(d-1)^{2}(d+2)^{3}(d+4)} -  
\frac{6(d-2)(d+1)}{(d-1)^{3}(d+2)^{6}(d+4)^{2}(d+6)} \times  
\nonumber \\  
\times\biggl[  
 2(3d^{6}+54d^{5}+321d^{4}+722d^{3}+420d^{2}+184d+96)  
\rho_{1}^{2}+  
\nonumber \\  
+d(-d^{7}-16d^{6}-49d^{5}+158d^{4}+844d^{3}+ 904d^{2}+1120d+640)  
\rho_{1}\rho_{2} +  
\nonumber \\  
+(13d^{7}+111d^{6}+258d^{5}+26d^{4}-124d^{3}+88d^{2}+336d+192)  
\rho_{2}^{2} \biggr],  
\label{expansion33}  
\end{eqnarray}  
\begin{eqnarray}  
\gamma^{*}[4,0]/\eps= \frac{-2(d+4)}{(d+2)} -  
\frac{4(d-2)^{2}(d+1)(d+6)\rho_{3}^{2}} {(d-1)^{2}d(d+2)^{4}(d+4)} ,  
\label{expansion40}  
\end{eqnarray}  
\begin{eqnarray}  
\gamma^{*}[4,2]/\eps= \frac{-d^{2}-5d+8}{(d-1)(d+2)} -  
\frac{4(d-2)(d+1)(d^{2}+3d-16)\rho_{3}} {(d-1)^{2}(d+2)^{2}(d+4)^{2}} -  
\frac{2(d+1)}{(d-1)^{3}d(d+2)^{3}(d+4)^{5}(d+6)} \times  
\nonumber \\ \times  
\biggl[-2 (2d^{9}+47d^{8}+  
422d^{7}+1253d^{6}-  
3684d^{5}-23344d^{4}-4032d^{3}+57856d^{2}  
+86016d-73728) \rho_{1}^{2}+  
\nonumber \\  
+d(d^{9}-28d^{8}-759d^{7}-4654d^{6}-2712d^{5}+ 40232  
d^{4}+42592d^{3}- 76928d^{2}- 202240d+122880) \rho_{1}\rho_{2}-  
\nonumber \\  
 - d (8d^{9}+149d^{8}+699d^{7}-1074d^{6}-  
10994 d^{5}-2240d^{4}+ 29248 d^{3}  
+29696 d^{2}-49664 d+24576) \rho_{2}^{2} \biggr],  
\label{expansion42}  
\end{eqnarray}  
\begin{eqnarray}  
\gamma^{*}[4,4]/\eps= \frac{12}{(d-1)(d+2)} +  
\frac{24(d-2)(d+1)\rho_{3}} {(d-1)^{2}(d+2)(d+4)^{2}} -  
\frac{12(d-2)}{(d-1)^{3}(d+2)^{4}(d+4)^{5}(d+6)} \times  
\nonumber \\  
\times\biggl[  
2d(3d^{7}+81d^{6}+ 823d^{5}+4023d^{4}+10230d^{3}+14536d^{2}+11040d+3456)  
\rho_{1}^{2}+  
\nonumber \\  
+(-d^{10}-23d^{9}-163d^{8}-225d^{7}+2040d^{6}+9948d^{5}+  
21240d^{4}+25424d^{3}+ 18112d^{2}+8960d+3072) \rho_{1}\rho_{2} +  
\nonumber \\  
+(15d^{9}+232d^{8}+1309d^{7}+3306d^{6}+  
3782d^{5}+780d^{4}-768d^{3}+3712d^{2}+6656d+3072) \rho_{2}^{2} \biggr].  
\label{expansion44}  
\end{eqnarray}  
\end{mathletters}  
  
These expressions illustrate two facts which seem to hold for all $N$:  
  
(i) The leading anisotropy correction is of order $O(\rho_{1,2})$  
for $p\ne0$ and $O(\rho_{1,2}^{2})$ for $p=0$, so that the dimensions  
$\gamma^{*}[N,0]$ are anisotropy independent in the {\it linear}  
approximation, and  
  
(ii) This leading contribution depends on $\rho_{1,2}$ only through the  
combination $\rho_{3}\equiv 2\rho_{1}+d \rho_{2}$.  
  
This conjecture is confirmed by the following expressions for $N=6, 8$  
and $p=0$:  
\begin{mathletters}  
\label{expansions-2}  
\begin{eqnarray}  
\gamma^{*}[6,0]/\eps= \frac{-2(d+6)}{(d+2)} - \frac{12(d-2)^{2}(d+1)  
(d^{2}+14d+48)\rho_{3}^{2}} {(d-1)^{2}d(d+2)^{4}(d+4)^{2}} ,  
\label{expansion60}  
\end{eqnarray}  
\begin{eqnarray}  
\gamma^{*}[8,0]/\eps= \frac{-4(d+8)}{(d+2)} -  
\frac{24(d-2)^{2}(d+1)  
(d^{2}+18d+80)\rho_{3}^{2}} {(d-1)^{2}d(d+2)^{4}(d+4)^{2}} ,  
\label{expansion80}  
\end{eqnarray}  
\end{mathletters}  
  
The eigenvalues beyond the small $\rho_{1,2}$ expansion have been obtained  
numerically. Some of them are presented in Figs. 2--5, namely,  
the dimensions $\Delta[n,p]$ for $n=3,4,5,6$ {\it vs} $\rho_{1}$  
for $\rho_{2}=0$, {\it vs} $\rho_{1}=\rho_{2}$, and {\it vs} $\rho_{2}$  
for $\rho_{1}=0$. The main conclusion that  
can be drawn from these diagrams is that the hierarchy (\ref{hier})  
demonstrated by the dimensions for the isotropic case ($\rho_{1,2}=0$)  
holds valid for all the values of the anisotropy parameters.

  \section{Operator product expansion and anomalous scaling}  
  \label {sec:OPE}

Representations (\ref{100})  for any scaling  
functions $\xi(mr)$ describes the behavior of the correlation functions  
for $\Lambda r>>1$ and any fixed value of $mr$; see Sec. \ref{sec:RGE}.  
The inertial range corresponds to the additional condition $mr<<1$.  
The form of the functions $\xi(mr)$ is not determined by the RG equations  
themselves; in analogy with the theory of critical phenomena, its  
behavior for $mr\to0$ is studied using the well-known Wilson operator  
product expansion (OPE); see, e.g., \cite{Zinn,book3,UFN,turbo,JETP}.  
  
According to the OPE, the equal-time product $F_{1}(x)F_{2}(x')$  
of two renormalized operators for  
${\bf x}\equiv ({\bf x} + {\bf x'} )/2 = {\const}$ and  
${\bf r}\equiv {\bf x} - {\bf x'}\to 0$ has the representation  
\begin{equation}  
F_{1}(x)F_{2}(x')=\sum_{F} C_{F} ({\bf r}) F(t,{\bf x}) ,  
\label{OPE}  
\end{equation}  
where the functions $C_{F}$ are coefficients regular in $m^{2}$  
and $F$ are all possible renormalized local composite operators  
allowed by symmetry (more precisely, see below).  
Without loss of generality, it can be assumed that the expansion  
is made in basis operators of the type (\ref{2.5}), i.e., those  
having definite critical dimensions $\Delta_{F}$.  
The renormalized correlator $\langle F_{1}(x)F_{2}(x') \rangle$  
is obtained by averaging Eq. (\ref{OPE}) with the weight  
$\exp S_{R}$ with the renormalized action (\ref{renormalized});  
the quantities $\langle F\rangle \propto  m^{\Delta_{F}}$  
appear on the right hand side.  
  
From the operator product expansion (\ref{OPE}) we therefore  
find the following expression  for the scaling function  
$\xi(mr)$ in the representation (\ref{100}) for the correlator  
$\langle F_{1}(x)F_{2}(x') \rangle$:  
\begin{equation}  
\xi(mr)=\sum_{F} A_{F}\,\, (mr)^{\Delta_{F}}, \quad mr <<1,  
\label{OR}  
\end{equation}  
with the coefficients $A_{F}$ regular in $(mr)^2$.

Now let us turn to the equal-time structure functions $S_{N}$ from  
(\ref{struc}). From now on, we assume that the mixed correlator  
$ \langle {\bfv} f \rangle $ differs from zero  
(see Sec. \ref{sec:scenario}); this does not affect the  
critical dimensions, but gives rise to nonvanishing odd structure  
functions. In general, the operators entering into the OPE are  
those which appear in the corresponding Taylor expansions, and also  
all possible operators that admix to them in renormalization  
\cite{Zinn,book3}.  
The leading term of the Taylor expansion for the function $S_{N}$  
is obviously given by the operator $F[N,N]$ from Eq. (\ref{Fnp});  
the renormalization gives rise to all the operators $F[N',p]$ with  
$N'\le N$ and all possible values of $p$. The operators with  
$N'> N$  (whose contributions would be more important) do not  
appear in Eq. (\ref{OR}), because they do not enter into the  
Taylor expansion for $S_{N}$ and do not admix in renormalization  
to the terms of the Taylor expansion; see Sec. \ref{sec:Operators}.  
Therefore, combining the RG representation (\ref{differ4}) with  
the OPE representation (\ref{OR}) gives the desired asymptotic  
expression for the structure function in the inertial range:  
\begin{equation}  
S_{N}(\r)= D_{0}^{-N/2} r^{N(1-\eps/2)}\, \sum_{N'\le N} \sum_{p}  
\Bigl\{ C_{N',p}\, (mr)^{\Delta[N',p]}+\cdots \Bigr\}  \,.  
\label{struc2}  
\end{equation}  
The second summation runs over all values of $p$, allowed for a given  
$N'$; $C_{N',p}$ are numerical coefficients dependent on $\eps$, $d$,  
$\rho_{1,2}$ and the angle $\vartheta$ between $\r$ and $\n$. The dots  
stand for the contributions of the operators  
other than $F[N,p]$, for example, $\partial^{2}\theta\partial^{2}\theta$;  
they give rise to the terms of order $(mr)^{2+O(\eps)}$ and higher  
and will be neglected in what follows.  
  
Some remarks are now in order.  
  
(i) If the mixed correlator $\langle{\bfv}f\rangle$ is absent,  
the odd structure functions vanish, while the contributions to  
even functions are given only by the operators with even values  
of $N'$. In the isotropic case ($\rho_{1,2}=0$) only  
the contributions with $p=0$ survive; see \cite{RG}.  
In the presence of the anisotropy, $\rho_{1,2}\ne0$, the  
operators with $p\ne0$ acquire nonzero mean values, and their  
dimensions $\Delta[N',p]$ also appear on the right hand side of  
Eq. (\ref{struc2}).  
  
(ii) The leading term of the small $mr$ behavior is obviously  
given by the contribution with the minimal possible value of  
$\Delta[N',p]$. Now we recall the hierarchy relations  
(\ref{hier1}), (\ref{hier2}), which hold for $\rho_{1,2}=0$  
and therefore remain valid at least for $\rho_{1,2}<<1$. This means that,  
if the anisotropy is weak enough, the leading term in Eq.  
(\ref{struc2}) is given by the dimension $\Delta[N,0]$ for any $S_{N}$.  
For all the special cases studied in Sec. \ref{sec:Operators},  
this hierarchy persists also for finite values of the anisotropy  
parameters, and the contribution with $\Delta[N,0]$ remains the  
leading one for such $N$ and $\rho_{1,2}$.  
  
(iii) Of course, it is not impossible that the inequalities  
(\ref{hier1}), (\ref{hier2}) break down for some values of  
$n$, $d$ and $\rho_{1,2}$, and the leading contribution to Eq.  
(\ref{struc2}) is determined by a dimension with $N'\ne N$ and/or  
$p>0$.  
  
Furthermore, it is not impossible that the matrix (\ref{Dnp})  
for some $\rho_{1,2}$ had a pair of complex conjugate eigenvalues,  
$\Delta$ and $\Delta^{*}$. Then the small $mr$ behavior of the  
scaling function $\xi(mr)$ entering into Eq. (\ref{struc2})  
would involve oscillating terms of the form  
$$(mr)^{{\rm Re}\, \Delta}  
\Bigl\{ C_{1} \cos \bigl[{\rm Im}\, \Delta\, (mr)\bigr] +  
C_{2} \sin \bigl[{\rm Im}\, \Delta\, (mr)\bigr] \Bigr\}, $$  
with some constants $C_{i}$.  
  
Another exotic situation emerges if the matrix (\ref{Dnp})  
cannot be diagonalized and is only reduced to the Jordan  
form. In this case, the corresponding contribution to the scaling  
function would involve a logarithmic correction to  the powerlike  
behavior, $\, (mr)^{\Delta}\, \bigl[C_{1}\ln (mr)+C_{2}\bigr]$,  
where $\Delta$ is the eigenvalue related to the Jordan cell.  
However, these interesting hypothetical possiblities are not  
actually realized for the special cases studied above in  
Sec. \ref{sec:Operators}.  
  
(iv) The inclusion of the mixed correlator  
$\langle{\bfv}f\rangle\propto\n \delta(t-t')\, C'(r/\ell)$ violates  
the evenness in $\n$ and gives rise to nonvanishing odd functions  
$S_{2n+1}$ and to the contributions with odd $N'$ to the expansion  
(\ref{struc2}) for even functions. If the hierarchy relations  
(\ref{hier1}),  
(\ref{hier2})  hold, the leading term for the even functions will still  
be given by the contribution with $\Delta[N,0]$. If the relations  
(\ref{hier3}) hold, the leading term for the odd function $S_{2n+1}$  
will be given by the dimension $\Delta[2n,0]$ for $n<(d+2)/4$ and by  
$\Delta[2n+1,1]$ for $n>(d+2)/4$. Note that for the model with an imposed  
gradient, the leading terms for $S_{2n+1}$ are given by the dimensions  
$\Delta[2n+1,1]$ for all $n$; see \cite{RG3}. This can be related to  
the observation of Ref. \cite{SA} that the odd structure functions  
of the velocity field appear more sensitive to the anisotropy  
than the even functions.  
  
Representations similar to Eqs. (\ref{100}), (\ref{struc2}) can easily  
be written down for any equal-time pair correlator, provided its canonical  
and critical dimensions are known. In particular, for the operators  
$F[N,p]$ in the IR region ($\Lambda r \to\infty$, $mr$ fixed)  
one obtains  
\begin{equation}  
\langle F[N_{1},p_{1}] F[N_{2},p_{2}] \rangle  = \nu_0 ^{-(N_{1}+N_{2})/2}  
\sum_{N,p}  \sum_{N',p'} (\Lambda r)^{-\Delta_{[N,p]}-\Delta_{[N',p']}}  
  \xi_{N,p;N',p'}(mr),  
\label{102}  
\end{equation}  
where the summation indices $N$, $N'$ satisfy the inequalities  
$N\le N_{1}$, $N'\le N_{2}$, and the indices $p$, $p'$ take on  
all possible values allowed for given $N$, $N'$. The small $mr$  
behavior of the scaling functions $\xi_{N,p;N',p'}(mr)$ has the  
form  
\begin{equation}  
\xi_{N,p; N',p'}(mr) = \sum_{N'',p''}\, C_{N'',p''}\, (mr)^{  
\Delta[N'',p'']},  
\label{103}  
\end{equation}  
with the restriction $N''\le N+N'$ and corresponding values of  
$p''$; $C_{N'',p''}$ are some numerical coefficients.  
  
So far, we have discussed the special case of the velocity correlator  
given by Eqs. (\ref{3}), (\ref{T34}). Let us conclude this Section  
with a brief discussion of the general case (\ref{T}).  
The RG analysis given above in Secs. \ref{sec:QFT}--\ref{sec:OPE}  
can be extended directly to this case; no serious alterations are  
required. From the expressions (\ref{sigma3}), (\ref{X9})  
it immediately follows that only even polynomials in the  
expansion (\ref{Legendre}) can give contributions to the  
renormalization constants, and consequently, to the coordinates of the  
fixed point and the anomalous dimensions. For this reason,  
the odd polynomials were omitted in Eq. (\ref{Legendre})  
from the very beginning. Moreover, it is clear from Eq.  
(\ref{sigma3}) that only the coefficients $a_{l}$ with $l=0,1$ and  
$b_{l}$ with $l=0,1,2$ contribute to the constants $Z_{1,2}$ in Eq.  
(\ref{renormalized}) and therefore to the basic RG functions  
(\ref{RGF}) and to the coordinates of the fixed point in Eq.  
(\ref{FP}). Therefore, the fixed point in the general model (\ref{T})  
is parametrized completely by these five coefficients; the higher  
coefficients enter only via the positivity conditions (\ref{positiv}).  
  
Furthermore, it is clear from Eqs. (\ref{X9}) that for $\alpha=0$,  
only coefficients $a_{l}$ with $l\le2$ and $b_{l}$ with $l\le3$  
can contribute to the integrals $H_{n}$ and, consequently, to the  
one-loop critical dimensions (\ref{Dnp}). Therefore, the calculation  
of the latter essentially simplifies for the special case  
$a_{0}=1$, $a_{1}=0$ and $b_{l}=0$ for $l\le2$ in Eq. (\ref{T}).  
Then the coordinates of the fixed point (\ref{FP}) are the same  
as in the isotropic model, in particular, $\alpha_{*}=0$, and the  
anomalous exponents will depend on the only two parameters  
$a_{2}$ and $b_{3}$. We have performed a few sample calculations  
for this situation; the results are presented in Figs. 6--9 for  
$\Delta[n,p]$ with $n=3,4,5,6$ {\it vs} $a_{2}$ for $b_{3}=0$,  
{\it vs} $a_{2}=b_{3}$, and {\it vs} $b_{3}$ for $a_{2}=0$.  
In all cases studied, the general picture has  
appeared similar to that outlined above for the case (\ref{T34}).  
In particular, the hierarchy of the critical dimensions, expressed  
by the inequalities (\ref{hier}), persists also for this case.  
We may conclude that the special case (\ref{T34}) case represents  
nicely all the main features of the general model (\ref{T}).

  \section{Exact solution for the second-order structure function  
  and calculation of the amplitudes} \label{sec:Exact}  
  
In renormalized variables, dimensionality considerations give,  
see Eq. (\ref{differ2})  
\begin{equation}  
S_{2}(\r)= (r^{2}/\nu) R(\mu r, mr, g, \alpha),  
\label{two-differ2}  
\end{equation}  
where $R\equiv R_{2}$ is some function of dimensionless variables; the  
dependence on $d$, $\eps$, $\rho_{1,2}$ and the angle $\vartheta$  
between the vectors ${\bf r}$ and ${\bf n}$ is also implied.  
The asymptotic behavior of the function (\ref{two-differ2}) in  
the IR region ($\Lambda r>>1$, $mr$ fixed) is found from the  
solution of the RG equation, see Eqs. (\ref{differ4}) and (\ref{differ5})  
in Sec. \ref{sec:RGE}:  
\begin{equation}  
S_{2}(\r)= D_{0}^{-1} r^{2-\eps} g_{*}\, \xi(mr),  
\label{two-differ4}  
\end{equation}  
where the scaling function $\xi(u)$ is related to the function  
$R$ from Eq. (\ref{two-differ2}) as follows:  
\begin{equation}  
\xi(u)\equiv  R(1, u, g_{*},  \alpha_{*}),  
\label{two-differ5}  
\end{equation}  
with $g_{*}$ and $\alpha_{*}$ from Eq. (\ref{FP}). The function $R$  
in Eq. (\ref{two-differ2}) can be calculated within the  
renormalized perturbation theory as a series in $g$,  
\begin{equation}  
R(g,\cdots) = \sum_{n=0}^{\infty} g^{n} R_{n}(\cdots),  
\label{1.63}  
\end{equation}  
where the dots stand for the arguments of $R$ other than $g$. Making  
the substitutions $\mu r\to1$, $g\to g_{*}$ and $\alpha\to\alpha_{*}$,  
expanding $R_{n}$ in $\eps$, and grouping  
together contributions of the same order, from Eq. (\ref{1.63}) we  
obtain the $\eps$ expansion for the scaling function:  
\begin{equation}  
\xi(u) = \sum_{n=0}^{\infty} \eps ^{n} \xi_{n}(u).  
\label{1.64}  
\end{equation}  
[It is important here that the calculation of any finite order in  
$\eps$ requires only a finite number of terms in Eq. (\ref{1.63}),  
because $g_{*}=O(\eps)$, and the coefficients $R_{n}$ for the  
renormalized quantity do not contain poles in $\eps$.] However, the  
expansion (\ref{1.64}) is not suitable for the analysis of the small  
$u$ behavior of $\xi(u)$: we shall  
see below that the coefficients $\xi_{n}$ contain IR  
singularities of the form $u^{p} \ln^{q}u$, these ``large IR  
logarithms'' compensate for the smallness of $\eps$, and the actual  
expansion parameter appears to be $\eps\ln u$ rather than $\eps$ itself.  
  
The formal statement of the problem is to sum up the expansion  
(\ref{1.64}) assuming that $\eps$ is small with the additional  
condition that $\eps\ln u=O(1)$. The desired solution is given by  
the OPE; it shows that the small $u$ behavior of $\xi(u)$ for the  
second-order structure function has the form,  
see Eqs. (\ref{OR}) and (\ref{struc2}) in Sec. \ref{sec:OPE}:  
\begin{equation}  
\xi(u) = C_{1}  + C_{2} u^{\Delta[2,2]} + \cdots .  
\label{two-struc2}  
\end{equation}  
The first term is the contribution of the operators 1 and $F[2,0]$  
with the dimension $\Delta[2,0]=0$, see the text below Eq.  
(\ref{Schwi3}), and $\Delta[2,2]$ is given to the order $O(\eps)$  
in Eq. (\ref{Delta22}); the dots in Eq. (\ref{two-struc2}) stand for  
the contributions of order $u^{2+O(\eps)}$ and higher. The  
coefficients $C_{i}$ depend on $\eps$, $d$, $\rho_{1,2}$,  
and the angle $\vartheta$. They are sought in the form of  
series in $\eps$:  
\begin{equation}  
C_{n} = \sum_{k=0}^{\infty} C_{n}^{(k)} \, \eps^{k}.  
\label{sought}  
\end{equation}  
Equation (\ref{two-struc2}) can be expanded in $\eps$ and compared  
with (\ref{1.64}). Such a comparison with known $\eps$ expansions  
for the coefficients $\xi_n$ in (\ref{1.64}) and exponents in  
(\ref{two-struc2}) allows one to verify the representation  
(\ref{two-struc2}) directly from the perturbation theory, and to  
determine the coefficients in Eq. (\ref{sought}).

In order to find the coefficients $C_{1,2}$, related to the exponents of  
order $O(\eps)$ in Eq. (\ref{two-struc2}), we do not need the entire  
functions $\xi_n(u)$, we only need the terms of the form  
$\eps^{p} \ln^{q} u$ in their small $u$ asymptotic expansions.  
Knowledge of the terms 1 and $\eps \ln u$ is sufficient for the  
calculation of the lowest (zero) order coefficients $C^{(0)}_{1,2}$.  
Calculation of the terms $(\eps \ln u)^{p}$ with $p>1$ gives  
additional equations for the same quantities $C^{(0)}_{1,2}$, which  
must be satisfied automatically and can therefore be used to verify  
the representation (\ref{two-struc2}).  
Calculation of the terms $\eps^{p} \ln^{q} u$ with $p>q$ gives  
the equations for the higher terms $C_{1,2}^{(k)}$ with $k\ge1$.  
Finally, the terms of the form $\eps^{p} u^{k} \ln^{q} u$  
with $k>0$ are related to the contributions of the form $u^{k+O(\eps)}$  
in Eq. (\ref{two-struc2}).  
  
In what follows, we confine ourselves with the calculation of the  
coefficients $C^{(0)}_{1,2}$. To this end, we have to find the  
function $\xi(u)$ with the accuracy of  
\begin{equation}  
\xi (u) = A_{0} + A_{1}\, \eps \ln u ,  
\label{accur}  
\end{equation}  
where $A_{i} = A_{i} (d,\rho_{1,2},\vartheta)$.  
>From the definition of the structure functions (\ref{struc})  
and the representations (\ref{two-differ4}), (\ref{two-differ5})  
it follows that $\xi(u)= \langle D(\omega,k) \rangle_{RG}$,  
where by the brackets we have denoted the following operation  
\begin{equation}  
\langle D(\omega,k) \rangle_{RG}\equiv \frac{2\nu}{r^{2}}  
\int \frac{d\omega}{(2\pi)}\int \frac{d\k}{(2\pi)^{d}}\,  
D(\omega,k) \, \Bigl[ 1- \exp ({\rm i}\k\r) \Bigr]  
\Bigl\vert _{\mu=1/r,\, g=g_{*},\, \alpha=\alpha_{*}}  
\label{opera}  
\end{equation}  
and $D(\omega,k) = \langle \theta\theta \rangle $ is the exact pair  
correlator of the field $\theta$, which we need to know here up to the  
one-loop approximation:  
\begin{equation}  
D(\omega,k) =  
\put(0.00,-56.00){\makebox{\palka}}  
\put(245.00,00.00){\makebox{+}}  
\put(330.00,-56.00){\makebox{\dOL}}  
\put(625.00,00.00){\makebox{+}}  
\put(710.00,-56.00){\makebox{\dOLT}}  
\put(1005.00,00.00){\makebox{+}}  
\put(1090.00,-56.00){\makebox{\dOLL}}  
\put(1385.00,00.00){\makebox{+ $\,\cdots$}}  
\hskip10.5cm  
\label{oneloop}  
\end{equation}  
  
In what follows, we set $\ell=\infty$ in Eq. (\ref{2}); the IR  
regularization is provided by the ``mass'' $m$ in the correlator (\ref{3}).  
In the momentum-frequency representation, the correlator (\ref{2})  
takes on the form $C(k)=(2\pi)^{d} \delta(\k)$, and the bare propagator  
in the renormalized perturbation theory is obtained from  
$\langle \theta \theta \rangle _0$ in  
Eq. (\ref{lines}) with the substitution $\nu_0 k^{2} \to  
\epsilon(k)\equiv \nu k^{2} + \nu \alpha (\k\n)^{2}$.  
  
Therefore, the contribution of the first (loopless) diagram in Eq.  
(\ref{oneloop}) to the function (\ref{opera}) has the form  
\begin{equation}  
\biggl\langle \frac{(2\pi)^{d} \delta(\k)}{\omega^{2}+\epsilon^{2}(k)}  
\biggr \rangle_{RG} = \frac{1}{2d} \left[1-\frac  
{\alpha_{*}(1+2\cos^{2}\vartheta)}{(d+2)}  
+ O(\rho_{1,2}^{2}) \right] .  
\label{opera1}  
\end{equation}  
For the sake of simplicity, here and below we confine ourselves  
with the first order in the anisotropy parameters $\rho_{1,2}$  
(we recall that $\alpha_{*}=O(\rho_{1,2})$); only the final results  
will be given for general $\rho$. Note also that all the integrals  
entering into the left hand side of Eq. (\ref{opera1}) are  
well-defined for $\ell=\infty$, for example:  
\begin{equation}  
\int d\k \frac{\delta(\k)}{k^{2}} \, \Bigl[ 1- \exp ({\rm i}\k\r) \Bigr]  
= \frac{r^{2}}{2d},\  
\int d\k \frac{\delta(\k)}{k^{2}} \frac{k_{i}k_{j}}{k^{2}}  
 \, \Bigl[ 1- \exp ({\rm i}\k\r) \Bigr] =  
\frac {r^{2}}{2d(d+2)} \left[ \delta_{ij} +2 \frac{r_{i}r_{j}}{r^{2}}  
\right].  
\label{opera2}  
\end{equation}  
  
It is also not too difficult to take into account the contributions  
from the second and third (asymmetric) diagrams in Eq. (\ref{oneloop}).  
It follows from Eq. (\ref{Dyson3}) and (\ref{sigma5}) that one simply  
has to replace $\epsilon(k)$ in the left hand side of Eq. (\ref{opera1})  
with the expression  
$\nu_0 k^{2} + \alpha_{0} \nu_0 (\n\k)^{2} - \Sigma_{\theta'\theta}$  
and change to the renormalized variables using the relations  
(\ref{18}) and (\ref{Z}). The pole part of the integral $J(m)$  
in Eq. (\ref{sigma5}) is then subtracted; the $\eps$ expansion of the  
difference gives rise to the logarithm of $m$. The substitution  
$g\to g_{*}$, $\alpha\to \alpha_{*}$ leads to drastic simplification  
(this is due to the fact that the expressions (\ref{FP}) for the fixed  
point originated from the same Eq. (\ref{sigma5})), and the total  
contribution of the first three diagrams in Eq. (\ref{oneloop})  
reduces to the loopless contribution (\ref{opera1}) multiplied with  
the factor $(1-\eps \ln u)$.  
  
Now let us turn to the last diagram in Eq. (\ref{oneloop}).  
The quantity to be ``averaged'' in Eq. (\ref{opera}) has the form  
\[\frac{1}{\omega^{2}+ \epsilon^{2}(k)} \,  
\int \frac{d\omega'}{(2\pi)} \int \frac{d\q}{(2\pi)^{d}}  
\frac{g\nu \mu^\eps T_{ij}(\q')} {((q')^{2}+m^{2})^{d/2+\eps/2}} \,  
\frac{C(q)q_{i}q_{j}} {(\omega')^{2}+ \epsilon^{2}(q)},\]  
where $\q'=\k-\q$. We perform the integrations over $\omega'$  
and $\omega$ (the latter is implicit in Eq. (\ref{opera}))  
and replace $\q'$ with $\k$ in the integral over $\q$  
(we recall that $C(k)\propto\delta(\k)$); this gives  
\begin{eqnarray}  
\frac{g\nu\mu^{\eps}\,T_{ij}(\k)} {4\epsilon(k)  
(k^{2}+m^{2})^{d/2+\eps/2}}\,  
\int d\q\, \delta(\q)\, \frac {q_{i}q_{j}} {\epsilon(q)}.  
\label{opera4}  
\end{eqnarray}  
Using Eq. (\ref{T34}) and the relation  
\[\int d\q\, \delta(\q)\, \frac {q_{i}q_{j}} {\epsilon(q)}=  
\frac{\delta_{ij}}{\nu d} - \frac{\alpha}{\nu d(d+2)} \Bigl[ \delta_{ij} +  
2 n_{i}n_{j} \Bigr]+O(\alpha^{2}) \]  
we obtain for the quantity (\ref{opera4})  
\begin{eqnarray}  
\frac{g\mu^\eps} {4d\nu\, k^{2}\,(k^{2}+m^{2})^{d/2+\eps/2} }  
\Bigl[ c_{0} + c_{1} \cos^{2} \psi \Bigr]  
\label{opera5}  
\end{eqnarray}  
with the coefficients  
\[c_{0}= d-1+\rho_{2}-\alpha (d+1) /(d+2), \quad  
c_{1}= (d-1)\rho_{1} - \rho_{2} -\alpha (d^{2}+d-4) /(d+2) \]  
(we also recall that $\psi$ is the angle between $\k$ and $\n$).  
Substituting Eq. (\ref{opera4}) into (\ref{opera})  gives  
\begin{eqnarray}  
\frac{g_{*}}{2d} \bigl[c_{0} I_{0}(u)+c_{1} I_{1}(u) \bigr]  
\label{opera6}  
\end{eqnarray}  
with the dimensionless convergent integrals  
\begin{eqnarray}  
I_{n} (u) = \frac{1}{r^{2}} \int \frac{d\k}{(2\pi)^{d}}\frac  
{\bigl(\cos^{2} \psi\bigr)^{n}}{k^{2} (k^{2}+m^{2})^{d/2}}  
\Bigl[1-\exp ({\rm i}\k\r) \Bigr].  
\label{opera7}  
\end{eqnarray}  
[within our accuracy, we have set $\eps=0$ in all exponents in  
Eqs. (\ref{opera6}), (\ref{opera7})]. The leading terms of their  
small $u$ behavior are found as follows: one differentiates  
$I_{n} (u)$ with respect to $u$, performs the change of variable  
$\k\to u \k$, expands in $u$ the quantity in the square brackets,  
and selects the leading terms of order $1/u$. This gives  
\begin{eqnarray}  
I_{0} (u) \simeq -\frac{C_{d}} {2d}\, \ln u,\quad  
I_{1} (u) \simeq -\frac{C_{d}\,\ln u} {2d(d+2)}  \Bigl[1+  
2\cos^{2} \vartheta \Bigr],  
\label{opera8}  
\end{eqnarray}  
with $C_{d}$ from Eq. (\ref{Z}).

Collecting all the contributions of the form (\ref{accur}) in Eq.  
(\ref{opera}) within our accuracy gives  
\begin{eqnarray}  
A_{0} = \frac{1}{2d} \left[1- \frac{\alpha_{*}(1+2\cos^{2}\vartheta)}  
{d+2} \right], \quad  
A_{1} =- \frac{(d+1)[(d^{2}+d-4)\rho_{1}-d\rho_{2}]}{(d-1)^{2}  
(d+2)^{3}} \,P_{2}(\cos\vartheta),  
\label{accur2}  
\end{eqnarray}  
where $P_{2}(z) = z^{2}-1/d$ is the second  
Gegenbauer polynomial.  
  
It is sufficient to take $\Delta\equiv\Delta[2,2]$ in the isotropic  
approximation, $\Delta=\eps d(d+1)/(d-1)(d+2)$, see Eq. (\ref{Qnp})  
[it is important here that $A_{1} =O(\rho_{1,2})$].  
  
Comparing Eq. (\ref{two-struc2}) with Eq. (\ref{accur}) gives  
$C_{2} = A_{1}\eps/\Delta$  
and $C_{1}=A_{0}-C_{2}$. Substituting these expressions into  
Eq. (\ref{two-struc2}) and then into Eq. (\ref{two-differ4})  
and using Eq. (\ref{FP}) we obtain the final expression for $S_{2}$:  
\begin{eqnarray}  
S_{2} (\r) &=&  \frac{\eps\, D_{0}^{-1} r^{2-\eps}} {(d-1)C_{d}}  
\Biggl\{ 1- (\rho_{1}+\rho_{2})/d +  
2\frac{(\rho_{1}-\rho_{2})} {(d+2)}\,  P_{2} (\cos\vartheta) -  
(mr)^{\Delta}\, \frac{2[(d^{2}+d-4)\rho_{1}-d\rho_{2}]}  
{(d-1)(d+2)^{2}}\, P_{2} (\cos\vartheta) \Biggr\} \, .  
\label{Lo1}  
\end{eqnarray}  
  
Similar (but much more cumbersome) calculations can also be performed if  
the anisotropy parameters $\rho_{1,2}$  
are not supposed to be small. The result has the form:  
\begin{eqnarray}  
S_{2} (\r) =  \frac{d\eps\, D_{0}^{-1} r^{2-\eps}} {(d-1)  
(d+\rho_{1}+\rho_{2})C_{d}} \Biggl\{ 1+  
\frac{g_{*}(d+\alpha_{*})(d+1)(\rho_{2}-\rho_{1})}  
{2(d-1)(d+2)\Delta} \, \widetilde P_{2} (\cos\vartheta) \,  
\left[ 1-\frac{3(d+\alpha_{*})}{(d+2)} F^{*}_{2} \right] -  
\nonumber \\  -  (mr)^{\Delta}\,  
\frac{(d+\alpha_{*})} {(d-1)d(d+2)} \, \widetilde P_{2} (\cos\vartheta) \,  
\left[ \alpha_{*} (3d F^{*}_{2}-(d+2)F^{*}_{1}) +  
\frac{g_{*}(d+1)(\rho_{2}-\rho_{1})} {2\Delta} \left(  
1-\frac{3(d+\alpha_{*})}{(d+2)} F^{*}_{2}\right)\right] \Biggr\}\, ,  
\label{Lo2}  
\end{eqnarray}  
where the quantities $g_{*}$, $\alpha_{*}$ and $\Delta\equiv\Delta[2,2]$  
are given in Eqs. (\ref{FP}) and  (\ref{Delta22}),  
$\widetilde P_{2} (z) \equiv z^{2} - (1+\alpha_{*})/(d+\alpha_{*})$  
and $F^{*}_{n} = F(1,1/2+n;d/2+n;-\alpha_{*})$ (as in Eq. (\ref{Delta22})).

Now let us turn to the exact equation  satisfied by the second-order  
structure function $S_{2} (\r)$.  
>From the definition of the latter and Eq. (\ref{12}) we obtain  
\begin{eqnarray}  
2\nu_0 \Delta S_{2} (\r)  
+ D_{0} S_{ij} (\r) \partial_{i}\partial_{j} S_{2} (\r) =2C(r/\ell).  
\label{Lo4}  
\end{eqnarray}  
In the inertial range we neglect the viscosity and set  
$C(r/\ell) \simeq C(0)=1$; this gives:  
\begin{eqnarray}  
 S_{ij} (\r) \partial_{i}\partial_{j} S_{2} (\r) = 2 D_{0}^{-1}.  
\label{Lo5}  
\end{eqnarray}  
  
We should also set $mr<<1$. The integral (\ref{10}) has a finite limit for  
$m=0$; the explicit calculation with $T_{ij}$  from Eq. (\ref{T34}) gives:  
\begin{eqnarray}  
S_{ij} (\r) &=&  \frac{2r^{\eps}} {C(\eps)} \biggl\{  
(d+\eps-1) \delta_{ij} -\eps r_{i}r_{j}/r^{2}+\rho_{1} (1+\eps z^{2})+  
\rho_{2} \Bigl[(d+\eps-2) n_{i}n_{j}-\eps z (n_{i}r_{j}+n_{j}r_{i})/r  
\Bigr] +  
\nonumber \\   &+&  
\frac{(\rho_{2}-\rho_{1})}{(d+\eps+2)}\Bigl[\delta_{ij} +2n_{i}n_{j} +  
\eps \bigl( \delta_{ij} z^{2} + r_{i}r_{j}/r^{2} +2z  
(n_{i}r_{j}+n_{j}r_{i})/r \bigr) +\eps(\eps-2) z^{2}  
r_{i}r_{j}/r^{2} \Bigr] \biggr\} \, ,  
\label{Lo6}  
\end{eqnarray}  
where $z\equiv (\n\r)/r$ and  
\begin{equation}  
C(\eps) \equiv -2^{2+\eps}\, (4\pi)^{d/2}\,  \Gamma(d/2+\eps/2+1) /  
\Gamma(-\eps/2) = 2\eps d /C_{d}\, +O(\eps^{2}).  
\label{Lo7}  
\end{equation}  
  
For what follows, we need to know the action of the differential  
operator in the left hand side of Eq. (\ref{Lo5}) onto a function of the  
form $r^{2-\eps+\gamma} \psi(z)$, where $\gamma$ is an arbitrary exponent  
(below we use $\gamma$ to denote different exponents; the precise meaning  
in each case will be clear from the context):  
\begin{equation}  
S_{ij} (\r) \partial_{i}\partial_{j} \Bigl[ r^{2-\eps+\gamma} \psi(z)  
\Bigr] = \frac{r^{\gamma}}{C(\eps)} \, \L(\gamma) \,\psi(z) .  
\label{Lo8}  
\end{equation}  
Here  $\L(\gamma)$ is a second-order differential operator with respect to  
$z$, whose explicit form is  
\begin{eqnarray}  
\L(\gamma)&=& (2+\gamma-\eps) \Bigl\{ (d-1) (d+\gamma)(d+2+\eps)+(d-1)(d+2)  
(\rho_{1}+\rho_{2}) +2\eps (d\rho_{2}-\rho_{1})+ z^{2}\eps d(d+1)  
(\rho_{1}-\rho_{2}) +  
\nonumber \\   &+&  
\gamma \bigl [ (d+1) \rho_{1} +(\eps+1) \rho_{2} +z^{2} \bigl(  
\rho_{1}(\eps d-\eps-2)\bigr)+\rho_{2}(d^{2}-2-\eps) \bigr]\Bigr\} +  
\Bigl\{ (1-d)(d-1+\eps)(d+2+\eps)-  
\nonumber \\   &-&  
(d^{2}+1)\rho_{1}+(d^{2}-d-1)\rho_{2} -\eps(d-1)\rho_{1}-\eps(d^{2}-2)  
\rho_{2} +4\eps^{2}\rho_{1} - (2d+1)\eps^{2}\rho_{2}+  
z^{2} \bigl[ 2\rho_{1}- (d^{2}-2) \rho_{2} +  
\nonumber \\   &+&  
\eps (d+1) \bigl( (1-d)\rho_{1} +(2d-3)\rho_{2}  \bigr)+\eps^{2}  
\bigl( -(d+3) \rho_{1} + (2d+1)\rho_{2} \bigr)\bigr] -  
\gamma (1-z^{2})  \bigl[ 4 \rho_{1}-2(d^{2}-2) \rho_{2} +  
\nonumber \\   &+&  
\eps (4 \rho_{1}-2d \rho_{2}) \bigr] \Bigr\} \, z \partial_{z} +  
\Bigl\{ (d+\eps-1) \bigl[ d+2+\eps+\rho_{1}+(d+1+\eps) \rho_{2}  
\bigr] + z^{2} \bigl[ \bigl(2+\eps(d+1+\eps)  \bigr) \rho_{1} -  
\nonumber \\   &-&  
\bigl( (d+\eps)^{2} -2-\eps) \rho_{2} \bigr]\Bigr\}  
\, (1-z^{2}) \partial^{2}_{z}\, .  
\label{Lo9}  
\end{eqnarray}  
  
We seek the solution to Eq. (\ref{Lo5}) in the form  
\begin{equation}  
S_{2}(\r) = r^{2-\eps} \, 2D_{0}^{-1} C(\eps)\, \psi(z),  
\label{Lo10}  
\end{equation}  
which gives the following differential equation for $\psi(z)$:  
\begin{equation}  
\L \psi(z)=1,\qquad \L\equiv \L(\gamma=0).  
\label{Lo11}  
\end{equation}  
Here the operator $\L$ has the form  
\begin{equation}  
\L = \ell_{2} (z) \partial^{2}_{z} + \ell_{1} (z) \partial_{z} +  
\ell_{0} (z),  
\label{Lo12}  
\end{equation}  
with some functions $\ell_{i} (z)$ whose explicit form is readily obtained  
from  
Eq. (\ref{Lo9}) and will not be given here for the sake of brevity.  
The operator $\L$ is self-adjoint on the interval $-1 \le z \le 1$ with  
respect to the scalar product  
\begin{equation}  
(f,g) \equiv \int^{1}_{-1} dz \rho(z) f(z) g(z)  
\label{Lo13}  
\end{equation}  
with the weight function  
\begin{equation}  
\rho(z) = \ell_{2}^{-1} (z)\, \exp \int^{z} dz' \ell_{1}(z')  
/ \ell_{2}(z') .  
\label{Lo14}  
\end{equation}  
The integration gives  
\begin{mathletters}  
\label{Lo15}  
\begin{equation}  
\rho(z) = (1-z^{2})^{(d-3)/2} \biggl\{ 1+ \frac  
{[2+\eps(d+1+\eps)]\rho_{1} -[(d+\eps)^{2}-2-\eps]\rho_{2}}  
{(d+\eps-1) [2+d+\eps +\rho_{1}+ (d+1+\eps)\rho_{2}]} \, z^{2} \biggr\}  
^{\wp} ,  
\label{Lo15A}  
\end{equation}  
where  
\begin{equation}  
\wp \equiv \frac  
{(2+d-\eps) [-(1+\eps)\rho_{1}+ (d(d+\eps)/2)-1) \rho_{2}] }  
{[2+\eps(d+1+\eps)]\rho_{1} + [2+\eps-(d+\eps)^{2}]\rho_{2}}  
\label{Lo15B}  
\end{equation}  
\end{mathletters}  
(the lower limit in Eq. (\ref{Lo14}) is chosen such that $\rho(z=0)=1$).  
  
The self-adjoint operator $\L$ possesses a complete set of eigenfunctions,  
but it is not possible to find them explicitly in the general case.  
However, Eq. (\ref{Lo10}) can be treated within regular perturbation  
expansions in $\eps$, $\rho_{1,2}$ and $1/d$. In what follows, we shall  
discuss them separately.  
  
{\it Perturbation theory in} $\eps$.  
We write the operator $\L=\L(\gamma=0)$ in Eq. (\ref{Lo9}) and the  
desired solution $\psi(z)$ as series in $\eps$:  
\[ \L=\L_{0}+\eps\, \L_{\eps} +\cdots , \qquad  
\psi(z) = \psi^{(0)}(z) +\eps \psi^{(1)}(z) +\cdots . \]  
The leading operator $\L_{0}$ has the form (\ref{Lo12}) with the functions  
$$ \ell_{i} (z)\equiv \widetilde \ell_{i} (z) \Bigr[ (d-1)(d+2)+  
\rho_{1} (d+1)+\rho_{2} \Bigl]\, , $$  
where  
$$ \widetilde \ell_{0} = 2(d+\alpha_{*}), \quad  
\widetilde \ell_{1} = -z [d-1-\alpha_{*}(1-z^{2})], \quad  
\widetilde \ell_{2} =(1-z^{2}) [1+\alpha_{*}(1-z^{2})] \, . $$  
  
It is easy to see that the constant is one of the eigenfunctions of the  
operator $\L_0$ (a consequence of the fact that $\ell_{0}$ is independent  
of $z$). The other eigenfunction (with the eigenvalue of zero) is  
$\widetilde P_{2} (z)$ from (\ref{Lo2}):  
\[\L_0 \, \widetilde P_{2} (z) =0.\]  
Therefore, a solution to the equation  $\L_0  \psi^{(0)}(z) =1$  
can be written in the form  
\begin{equation}  
\psi^{(0)} (z) = \ell^{-1}_{0}+ C \widetilde P_{2} (z) .  
\label{Lo18}  
\end{equation}  
In order to find the constant $C$, consider the equation for the correction  
$\psi^{(1)} (z)$:  
\begin{equation}  
\L_{\eps} \psi^{(0)} + \L_0 \psi^{(1)}  =0.  
\label{Lo19}  
\end{equation}  
  
Using the scalar product (\ref{Lo13}) with the weight  
(\ref{Lo15}) at $\eps=0$,  
\begin{equation}  
\rho^{(0)}(z) = (1-z^{2})^{(d-3)/2} \left( 1-\frac {\alpha_{*}}  
{1+\alpha_{*}} \, z^{2} \right) ^{-1-d/2} ,  
\label{Lo20}  
\end{equation}  
(the operator $\L_0$ is self-adjoint with respect  
to it) we obtain $(\widetilde P_{2}, \L_0 \psi^{(1)})  =  
(\L_0 \widetilde P_{2}, \psi^{(1)}) =0$.  
Then from (\ref{Lo19}) it follows  
that $(\widetilde P_{2}, \L_{\eps} \psi^{(0)})=0$. From the latter  
relation and Eq. (\ref{Lo18}) we find  
$$ C=- \frac {(\widetilde P_{2}, \L_{\eps} \ell^{-1}_{0})}  
{(\widetilde P_{2}, \L_{\eps}\widetilde P_{2})}\, , $$  
which gives  
\begin{equation}  
\psi^{(0)}(z) = \ell_{0}^{-1} \left[ 1-  
\frac {(\widetilde P_{2}, \L_{\eps}\cdot 1 )}  
{(\widetilde P_{2}, \L_{\eps}\widetilde P_{2})}\,  
\widetilde P_{2} (z) \right].  
\label{Lo21}  
\end{equation}  
  
Below we shall show that the solution (\ref{Lo21}) coincides with the RG  
result (\ref{Lo2}) for $m=0$. To this end, we shall show first that the  
last term in Eq. (\ref{Lo2}), proportional to $(mr)^{\Delta}$, is an  
eigenfunction of the operator in the left hand side of Eq. (\ref{Lo4})  
with the eigenvalue of zero. We seek the zero mode of that operator in the  
form $r^{2-\eps+\gamma}\,\Phi(z)$ with some exponent $\gamma$  
and function $\Phi(z)$. According to Eq. (\ref{Lo8}), we arrive at the  
equation $\L(\gamma)\Phi(z)=0$ with $\L(\gamma)$ from (\ref{Lo9}).  
For $\eps=0$, the solution of the latter equation is given by  
$\gamma=0$ and $\Phi(z)=\widetilde P_{2}(z)$ from Eq. (\ref{Lo2}).  
The $O(\eps)$ correction to the solution $\gamma=0$ is sought from the  
equation  
$$ (\L_0 +\eps \L_{\eps} +\gamma \L_{\gamma}) \Phi(z)=0, $$  
where $\gamma \L_{\gamma}$ is the part of operator (\ref{Lo9})  
linear in $\gamma$. We require that the correction operator  
$\eps \L_{\eps} +\gamma \L_{\gamma}$ do not shift the original (zero)  
eigenvalue. This leads to the equation  
$(\widetilde P_{2}, (\eps \L_{\eps} +\gamma \L_{\gamma}) \widetilde P_{2})  
=0$, from which it follows  
\begin{equation}  
\gamma =- \eps \frac {(\widetilde P_{2}, \L_{\eps} \widetilde P_{2})}  
{(\widetilde P_{2}, \L_{\gamma} \widetilde P_{2})} \, .  
\label{Lo22}  
\end{equation}  
Using Eqs. (\ref{Lo9}) and (\ref{Lo20}) it is not difficult to check that  
the expression (\ref{Lo22}) coincides with the dimension  
(\ref{Delta22}), e.g., $\gamma=\Delta[2,2]$. Therefore, we have shown that  
the last term in (\ref{Lo2}) is indeed an eigenfunction of the operator  
from (\ref{Lo5}), and the exponent and function $\Phi(z)$ are found in the  
first and zeroth order in $\eps$, respectively.  
  
Using Eq. (\ref{Lo22}), expression (\ref{Lo21}) can be rewritten in the  
form  
\begin{equation}  
\psi^{(0)} (z) = \ell^{-1}_{0}\, \left[ 1+ \frac{\eps\,  
(\widetilde P_{2}, \L_{\eps}\cdot 1)} {\Delta[2,2]\,  
(\widetilde P_{2}, \L_{\gamma} \widetilde P_{2}) } \,  
\widetilde P_{2} (z) \right]\, .  
\label{Lo23}  
\end{equation}  
Calculation of the scalar products using Eqs. (\ref{Lo9}) and (\ref{Lo20})  
shows that Eq. (\ref{Lo23}) indeed coincides with the solution (\ref{Lo2})  
for $m=0$.  
  
{\it Perturbation theory in the anisotropy parameters} $\rho_{1,2}$.  
In this case it follows from Eq. (\ref{Lo9}) with $\rho_{1,2}=0$,  
that the leading operator $\L_0 (\gamma)$ has the form (\ref{Lo12})  
with  
\begin{eqnarray}  
\ell_{0} &=& (2+\gamma-\eps)(d-1)(d+\gamma)(d+2+\eps),\nonumber\\  
\ell_{1} &=& -(d-1)(d-1+\eps)(d+2+\eps)z,\nonumber\\  
\ell_{2} &=& (d-1+\eps)(d+2+\eps) (1-z^{2}).  
\label{Lo24}  
\end{eqnarray}  
It can be represented in the form  
\begin{equation}  
\L_0 (\gamma)= - (d-1+\eps)(d+2+\eps) \widetilde \L + \ell_{0}.  
\label{Lo25}  
\end{equation}  
The eigenfunctions of the operator $\widetilde \L = (z^{2}-1)  
\partial^{2}_{z}+(d-1)z\partial_{z}$ are nothing other than the well known  
Gegenbauer polynomials $P_{n}(z)$, the corresponding eigenvalues being  
$n(n+d-2)$; see \cite{Grad}. They form an orthogonal system on the interval  
$[-1,1]$ with the weight $(1-z^{2}) ^{(d-3)/2}$ [see Eq. (\ref{Lo15}) with  
$\rho_{1,2}=0$]. Since the term $\ell_{0}$ in Eq. (\ref{Lo25}) is  
independent of $z$, the eigenfunctions of the operator $\L_0 (\gamma)$ are  
the same polynomials $P_{n}(z)$, and the eigenvalues have the form:  
\begin{equation}  
-(d-1+\eps)(d+2+\eps)n(n+d-2)+\ell_{0},  
\label{Lo26}  
\end{equation}  
with $\ell_{0}$ from Eq. (\ref{Lo24}).  
  
The knowledge of the eigenfunctions and eigenvalues of the operator $\L_0  
= \L_0 (\gamma=0)$ in Eq. (\ref{Lo11}) allows one to easily construct the  
iterative solution of that equation. In particular, in the leading  
order one has  
\[ \psi^{(0)} (z) = 1/ \ell_{0} = 1/(2-\eps)d(d-1)(d+\eps).\]  
  
For $m\ne0$, the corrections to the solution of Eq. (\ref{Lo5}) are  
determined by the zero modes of the operator $\L(\gamma)$. In the leading  
(zeroth) order in $\rho_{1,2}$ they are simply found by equating the  
eigenvalues (\ref{Lo26}) to zero. The ``admissible''  
(in the sense of \cite{Falk1,Falk2,GK}) solution is:  
\begin{equation}  
2\gamma^{(0)}_{n}=(\eps-d-2) + \sqrt { (\eps+d-2)^{2} +4n(n+d-2)(d+\eps-1)  
/(d-1) } .  
\label{Lo27}  
\end{equation}  
Therefore, in such approximation the zero modes of the operator from Eq.  
(\ref{Lo5}) have the form $r^{2-\eps+\gamma^{(0)}_{n}} P_{n}(z)$,  
in agreement with Ref. \cite{Falk1} (owing to the $z\to-z$ symmetry of our  
model, only even $n$ can contribute). From the RG viewpoints, the exponents  
in Eq. (\ref{Lo27}) are related to the $n$-th  rank composite operators  
built  
of two fields $\theta$ and $n$ derivatives; in particular,  
$\gamma_{2}$ (which gives the leading correction for $mr<<1$)  
coincides with the dimension $\Delta[2,2]$.  
  
In order to find the $O(\rho_{1,2})$ correction to $\gamma^{(0)}_{2}$,  
which we denote $\gamma^{(1)}_{2}$, we expand  
the operator $\L(\gamma)$ from Eq. (\ref{Lo9}) around the point  
$\rho_{1,2}=0$ and $\gamma=\gamma^{(0)}_{2}$ up to the terms linear in  
$\rho_{1,2}$ and $\gamma^{(1)}_{2}$:  
$$ \L(\gamma) \simeq  \L _0 +\sum_{i=1}^{2} \rho_{i} \L^i  
 + \gamma^{(1)}_{2} \L_1, $$  
which defines the operators $\L^i$ and $\L_1$.  
Proceeding as for the previous case, we obtain  
$$ \gamma^{(1)}_{2} = - \sum_{i=1}^{2} \rho_{i}  \frac  
{(P_{2}, \L^i P_{2})} {(P_{2}, \L_{1} P_{2})} , $$  
where the scalar product is defined with the weight  
$(1-z^{2})^{(d-3)/2}$.  
The explicit calculation gives the desired result:  
\begin{equation}  
\gamma^{(1)}_{2} = \frac {4(d-2)(d+1)\eps \, [2\rho_{1}+(d+\eps)\rho_{2}]}  
{(d-1)(d+4)(d+2+\eps) \, [4d(d^{2}-1)\eps+(d-1)^{2}(d+2-\eps)^{2}]^{1/2}} .  
\label{Lo29}  
\end{equation}  
  
{\it Perturbation theory in} $1/d$. For $d\to\infty$, the weight  
(\ref{Lo15}) takes on the form  
\begin{equation}  
\rho(z) \simeq \left(\frac {1-z^{2}} {1- \rho_{2}z^{2}/(1+\rho_{2})}  
\right) ^{d/2}.  
\label{Lo30}  
\end{equation}  
>From Eq. (\ref{Lo30}) it follows that, for $d>>1$, the function  
$\rho(z)$ differs remarkably from zero only in a small vicinity of the  
point  
$z\simeq0$, more precisely, $z^{2} \le 1/d$. This becomes especially clear  
if  
one changes to the new variable $x^{2}\equiv dz^{2}/2(1+\rho_{2})$. In  
terms  
of $x$, the function (\ref{Lo30}) has a finite limit for $d\to\infty$:  
\begin{equation}  
\rho(x) \simeq e^{-x^{2}} .  
\label{Lo31}  
\end{equation}  
The integration domain in the scalar product (\ref{Lo13}) becomes  
$-\sqrt{d}\le x \le \sqrt{d} $, which for $d\to\infty$ gives  
$-\infty\le x \le \infty $.  
  
According to (\ref{Lo9}),  equation (\ref{Lo11}) in terms of the variable  
$x$ and in the leading approximation in $1/d$ can be written in the form  
\begin{equation}  
\L_0 \psi(x) =2/d^{3},  
\label{Lo32}  
\end{equation}  
where  
\begin{equation}  
\L_0 = \L_0(\gamma=0),\qquad \L_0(\gamma)=\partial^{2}_{x}-2x  
\partial_{x}+2(2+\gamma-\eps).  
\label{Lo33}  
\end{equation}  
The eigenfunctions of the operator $\partial^{2}_{x}-2x\partial_{x}$  
are the well-known Hermit polynomials $H_{n}(x)$ (see, e.g., \cite{Grad}):  
$$ (\partial^{2}_{x}-2x\partial_{x}) H_{n}(x) =2n\,H_{n}(x),  
\quad n=0,1,2,\cdots, $$  
which form an orthogonal system on the axis $-\infty\le x \le \infty $  
with the weight (\ref{Lo31}). It is clear from Eq. (\ref{Lo33}) that the  
eigenfunctions of the operator $\L_0(\gamma)$ are also  
the polynomials $H_{n}(x)$, the corresponding eigenvalues being  
$\lambda_{n}=2(n+\eps-2-\gamma)$.  
  
The zero modes ($\lambda_{n}=0$) correspond to the values  
\begin{equation}  
\gamma=n-2+\eps,  
\label{Lo36}  
\end{equation}  
where $n=2$, 4, 6 and so on, owing to the symmetry $z\to-z$ of our model;  
the minimal eigenvalue is $\gamma=\eps$ with $n=2$.  
Therefore, the operator  $\L_0 = \L_0(\gamma=0)$ has no zero modes,  
and the solution to Eq. (\ref{Lo32}) is simply given by  
$\psi(x) =1/d^{3}(2-\eps)$.  
  
The corrections to the solution $\gamma=\eps$ in $1/d$ are found as above  
for the perturbation theories in $\eps$ and $\rho_{1,2}$. This gives:  
$$ \gamma=\eps+2\eps(1+2\rho_{2}+\rho_{1}\rho_{2})/d^{2}+O(1/d^{3}) $$  
[the $O(1/d)$ term is absent], and the corresponding zero mode is  
$$ (1-x^{2}) - \rho_{2} (2+\rho_{1}+\rho_{2})/d (1+\rho_{2}) +  
O(1/d^{2}). $$  
  
  \section{Conclusion} \label{sec:Con}  
  
Let us recall briefly the main points of the paper.

We have studied the anomalous scaling behavior of a  
passive scalar advected by the time-decorrelated strongly anisotropic  
Gaussian velocity field. The statistics of the latter is given by  
Eqs. (\ref{3}) and (\ref{T}). The general case (\ref{T})  
involves infinitely many parameters; most practical calculations have  
been performed for the truncated two-parametric model (\ref{T34}),  
which seems to represent nicely the main features of the general  
case (\ref{T34}).  
  
The original stochastic problem (\ref{1})--(\ref{3}) can be cast as a  
renormalizable field theoretic model [Eqs. (\ref{action})--(\ref{lines})],  
which allows one to apply the RG and OPE techniques to it.  
The corresponding RG equations have an IR stable fixed point,  
Eq. (\ref{FP}), which leads to the asymptotic expressions of the type  
(\ref{100}), (\ref{RGR}) for various correlation functions in the  
region $\Lambda r>>1$.  
  
Those expressions involve certain scaling functions of the variable $mr$,  
whose behavior at $mr<<1$ is determined by the OPE. It establishes the  
existence of the inertial-range anomalous scaling behavior. The structure  
functions are given by superpositions of power laws with nonuniversal  
(dependent on the anisotropy parameters $\rho_{1,2}$) exponents.  
  
The exponents are determined by the critical dimensions of composite  
operators (\ref{Fnp}) built of the scalar gradients.  
In contrast with the isotropic velocity field, these operators in our model  
mix in renormalization such that the matrices of their critical dimensions  
are neither diagonal nor triangular. These matrices are calculated  
explicitly  
to the order $O(\eps)$ [Eqs. (\ref{hyperpuper2}), (\ref{X7}),  
(\ref{Gnp}), (\ref{Dnp})], but their eigenvalues (anomalous exponents)  
can be found explicitly only as series in $\rho_{1,2}$  
[Eqs. (\ref{expansions}), (\ref{expansions-2})] or  
numerically [Figs. 1--9].

In the limit of vanishing anisotropy, the exponents can be associated  
with definite tensor composite operators  
built of the scalar gradients, and exhibit a kind of hierarchy related  
to the degree of anisotropy: the less is the rank, the less is the  
dimension and, consequently, the more important is the contribution  
to the inertial-range behavior [see Eqs. (\ref{hier})].  
  
The leading terms of the even (odd)  
structure functions are given by the scalar (vector) operators. For  
the finite anisotropy, the exponents cannot be associated with individual  
operators (which are essentially ``mixed'' in renormalization), but,  
surprising enough, the aforementioned hierarchy survives for all the  
cases studied, as is shown in Figs. 2--9.

The second-order structure function $S_{2}(\r)$ is studied in more detail  
using the RG and zero-mode techniques; like in the isotropic case  
\cite{Kraich1}, its leading term has the form $S_{2} \propto r^{2-\eps}$,  
but the amplitude now depends on $\rho_{1,2}$ and the angle between  
the vectors $\r$ and $\n$ from Eq. (\ref{T34}). The first anisotropic  
correction has the form $(mr)^{\Delta[2,2]}$ with the exponent  
$\Delta[2,2]=O(\eps)$ from Eq. (\ref{Delta22}). The function $S_{2}$  
satisfies the exact equation (\ref{Lo4}); this allows for an alternative  
derivation of the perturbation theory in $\eps$, and also gives a  
basis for perturbation theories in $1/d$ or the anisotropy  
parameters $\rho_{1,2}$; see the discussion in Sec.~\ref{sec:Exact}.  
  
It is well known that, for the isotropic velocity field,  
the anisotropy introduced at large  
scales by the external forcing or imposed mean gradient, persists  
in the inertial range and reveals itself in {\it odd} correlation  
functions: the skewness factor $S_{3}/S_{2}^{3/2}$ decreases for  
$mr\to0$ but slowly (see Refs. \cite{An,synth,Pumir,Siggia}),  
while the higher-order ratios $S_{2n+1}/S_{2}^{n+1/2}$ increase  
(see, e.g., \cite{RG3,CLMV99,Lanotte2}).  
  
In the case at hand, the inertial-range behavior of the skewness  
is given by $S_{3}/S_{2}^{3/2}\propto (mr)^{\Delta[3,1]}$. For  
$\rho_{1,2}\to0$, the exponent $\Delta[3,1]$ is given by Eq. (\ref{Qnp})  
with $n=3$ and $p=1$; it is positive and coincides with the result of  
Ref. \cite{Pumir}. The levels of the dimension $\Delta[3,1]$ on the  
$\rho_{1}$--$\rho_{2}$ plane are shown in Fig. 10. One can see that,  
if the anisotropy becomes strong enough, $\Delta[3,1]$ becomes negative  
and the skewness factor {\it increases} going down towards to the depth of  
the inertial range; the higher-order odd ratios increase already when the  
anisotropy is weak.

\acknowledgments  
N.\,V.\,A. is thankful to Juha Honkonen, Antti Kupiainen,  
Andrea Mazzino and Paolo Muratore-Ginanneschi for discussions.  
The work was supported by the Russian Foundation for Fundamental  
Research (Grant No. 99-02-16783), the Slovak Academy of Science  
(Grants No. 2/7232/2000) and the Grant Center  
for Natural Sciences of the Russian State Committee for Higher  
Education (Grant No. 97-0-14.1-30).  
N.V.A. is also thankful to the Department of Mathematics of the  
University of Helsinki for hospitality during his visit,  
financed by the project ``Extended Dynamical Systems.''

\newpage
\begin{table}  
\caption{Canonical dimensions of the fields and parameters in the  
model (\protect\ref{extended}).}  
\label{table1}  
\begin{tabular}{cccccccc}  
$F$ & $\theta$ & $\theta'$ & $ {\bfv} $ & $\nu$, $\nu _{0}$  
& $m$, $\mu$, $\Lambda$ & $g_{0}$ & $g$, $\alpha$, $\alpha_{0}$ \\  
\tableline  
$d_{F}^{k}$ & 0 & $d$ & $-1$ & $-2$ & 1& $\eps $ & 0 \\  
$d_{F}^{\omega}$ & $-1/2$ & $1/2$ & 1 & 1 & 0 & 0 & 0 \\  
$d_{F}$ & $-1$ & $d+1$ & 1 & 0 & 1 & $\eps $ & 0 \\  
\end{tabular}  
\end{table}  
  

\newpage  
\begin{figure}  
 \centerline{\epsfig{file=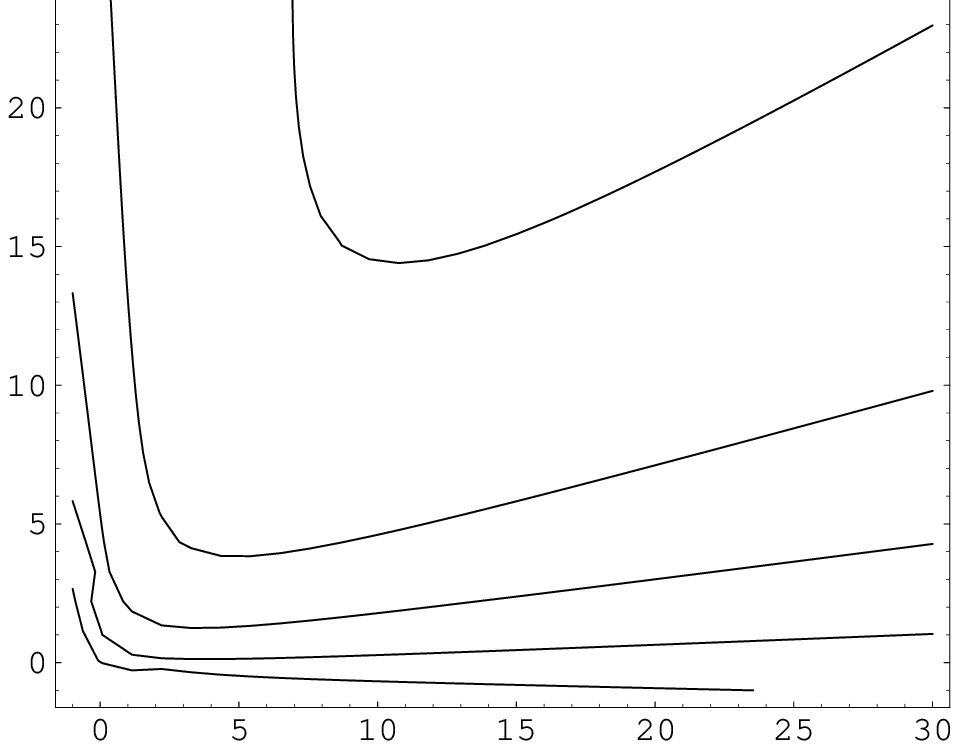,width=8cm}}  
 \vspace{1mm}  
\caption{Levels of the dimension $\Delta[2,2]$ for $d=3$ on the plane  
$\rho_1$--$\rho_2$. Value changes from~$1.15$ (left-bottom)  
to~$1.4$ (right-top) with step~$0.05$.}  
\end{figure}  
  
\begin{figure}  
\centerline{  
\hbox{  
\epsfig{file=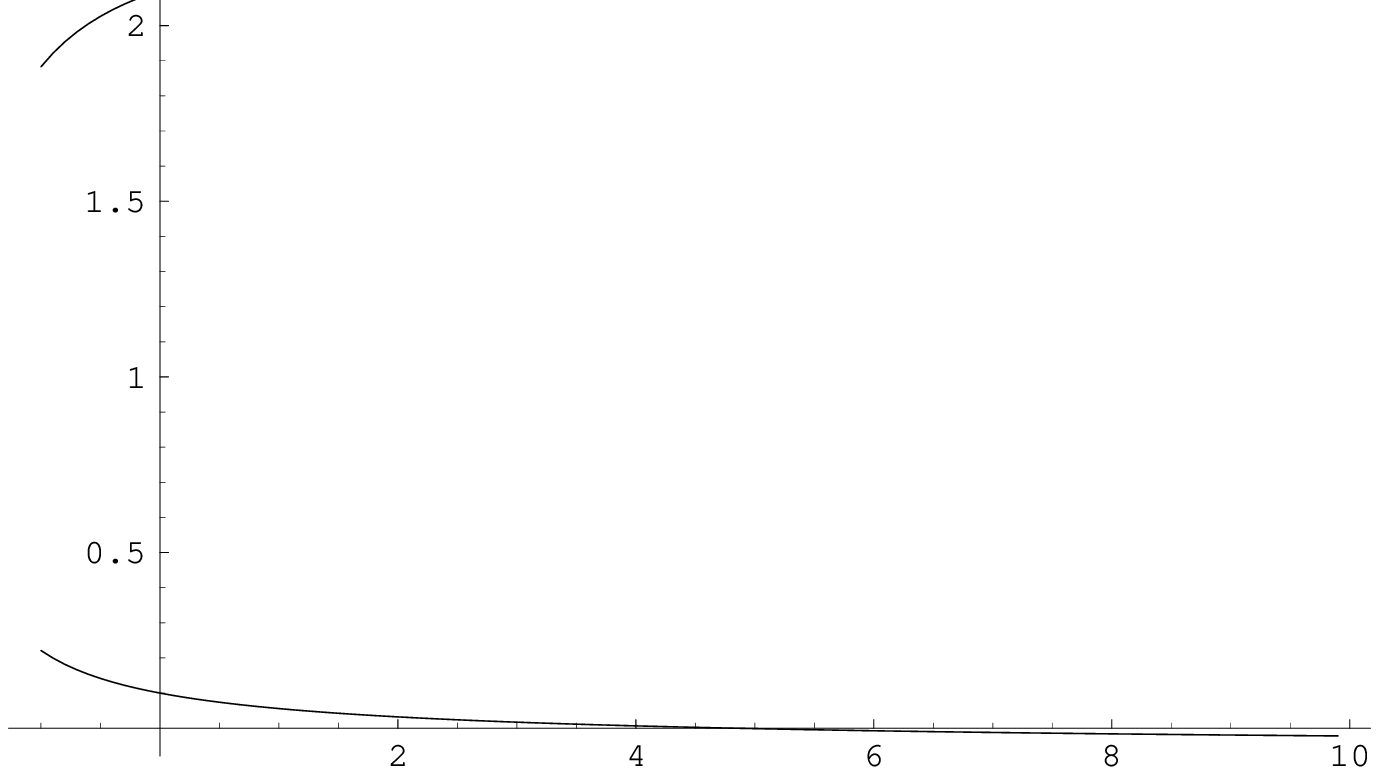,width=5cm}\hskip .5 cm  
\epsfig{file=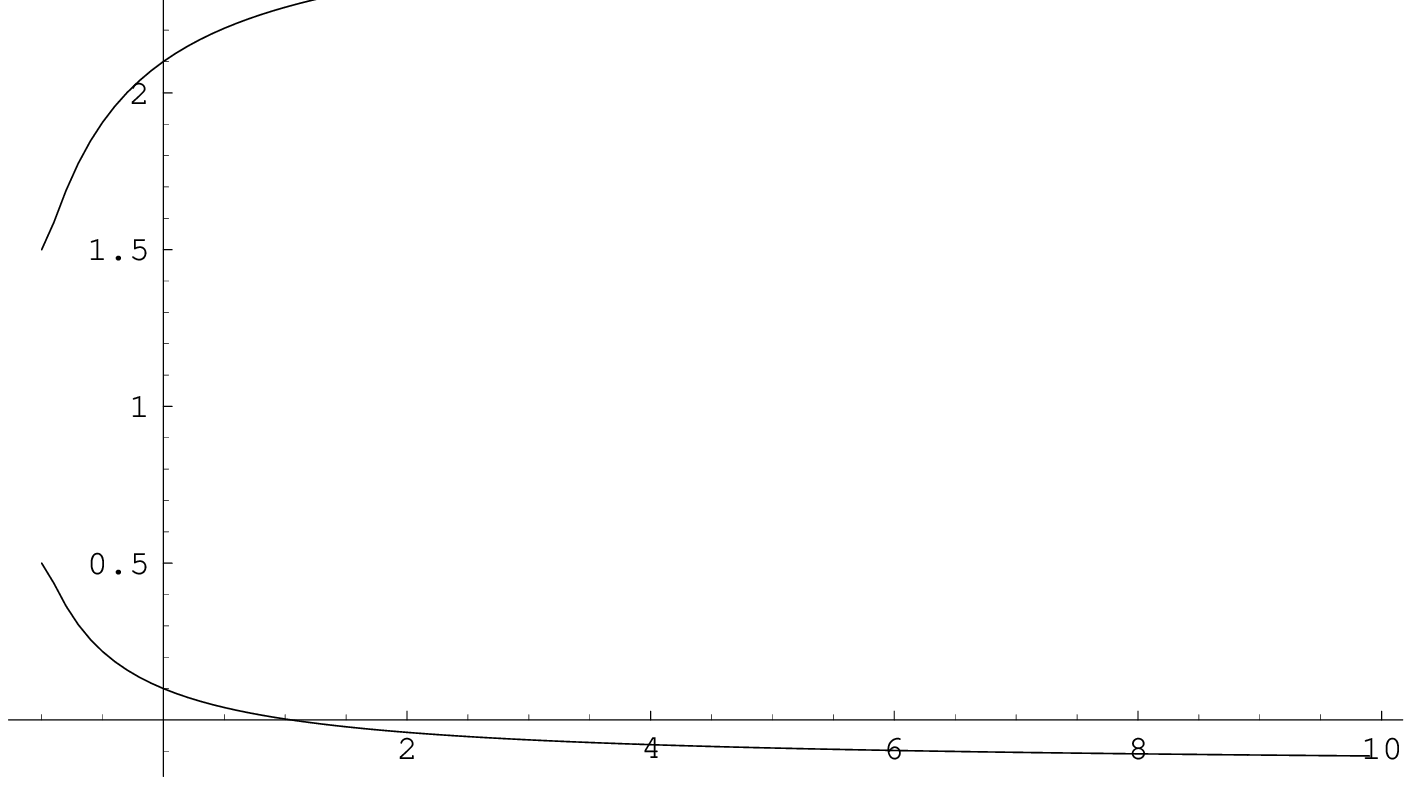,width=5cm}\hskip .5 cm  
\epsfig{file=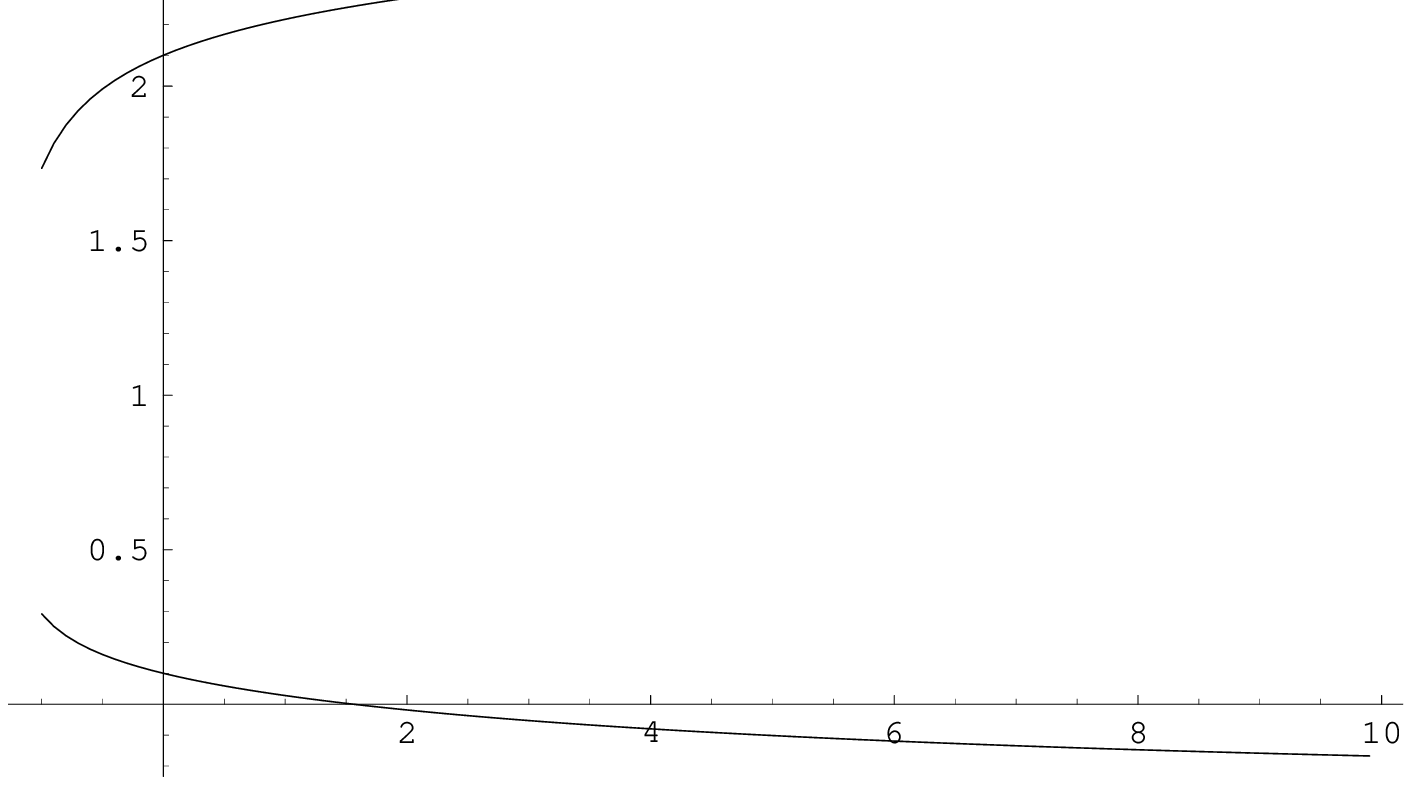,width=5cm}  
} }  
\vspace{1mm}  
\caption{Behavior of the critical dimension  
$\Delta[3,p]$ for $d=3$ with $p=1,3$ (from below to above) {\it  
vs} $\rho_1$ for $\rho_2=0$---{\it left}, {\it vs}  
$\rho\equiv\rho_1=\rho_2$---{\it center},{\it vs} $\rho_2$ for  
$\rho_1=0$---{\it right}.}  
\end{figure}

\begin{figure}  
\centerline{  
\hbox{  
\epsfig{file=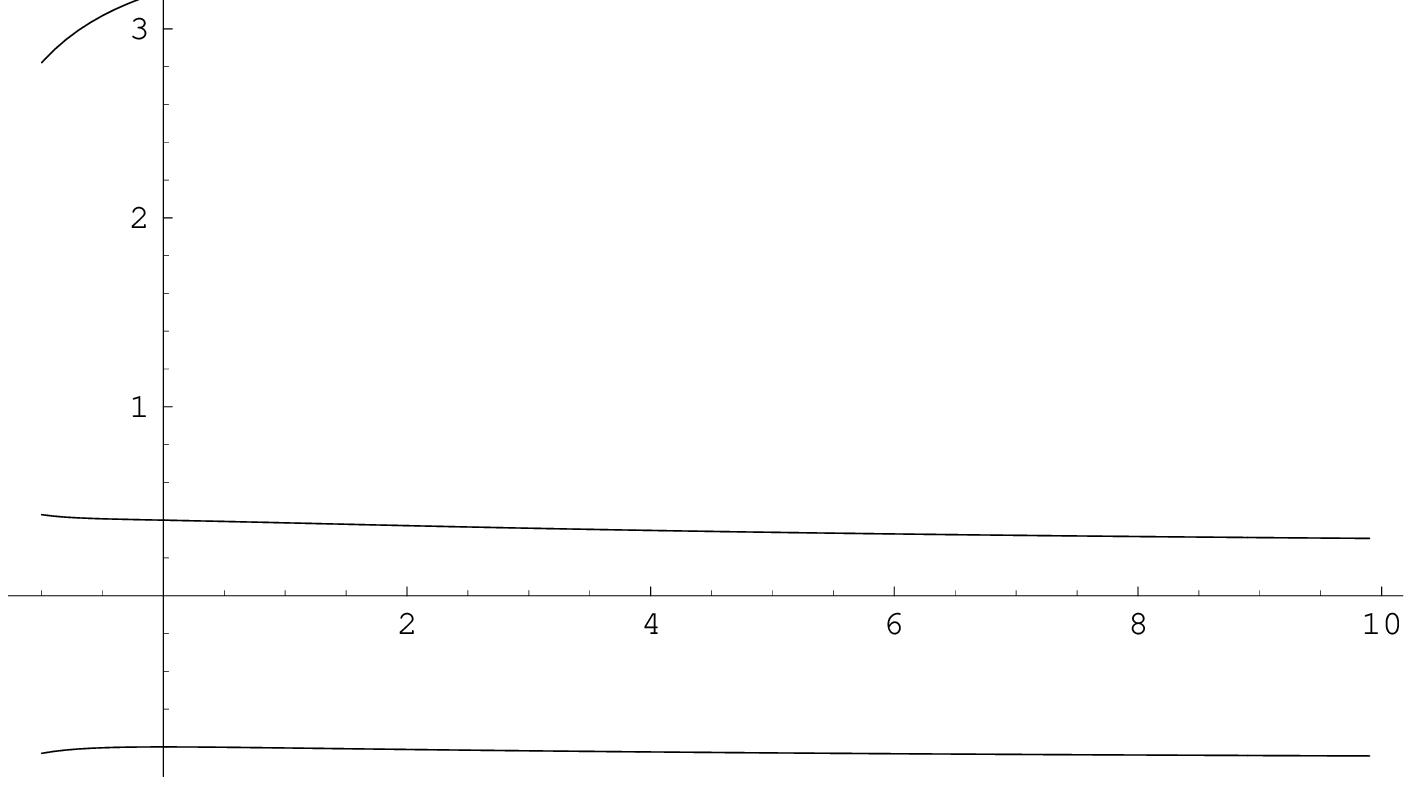,width=5cm}\hskip .5 cm  
\epsfig{file=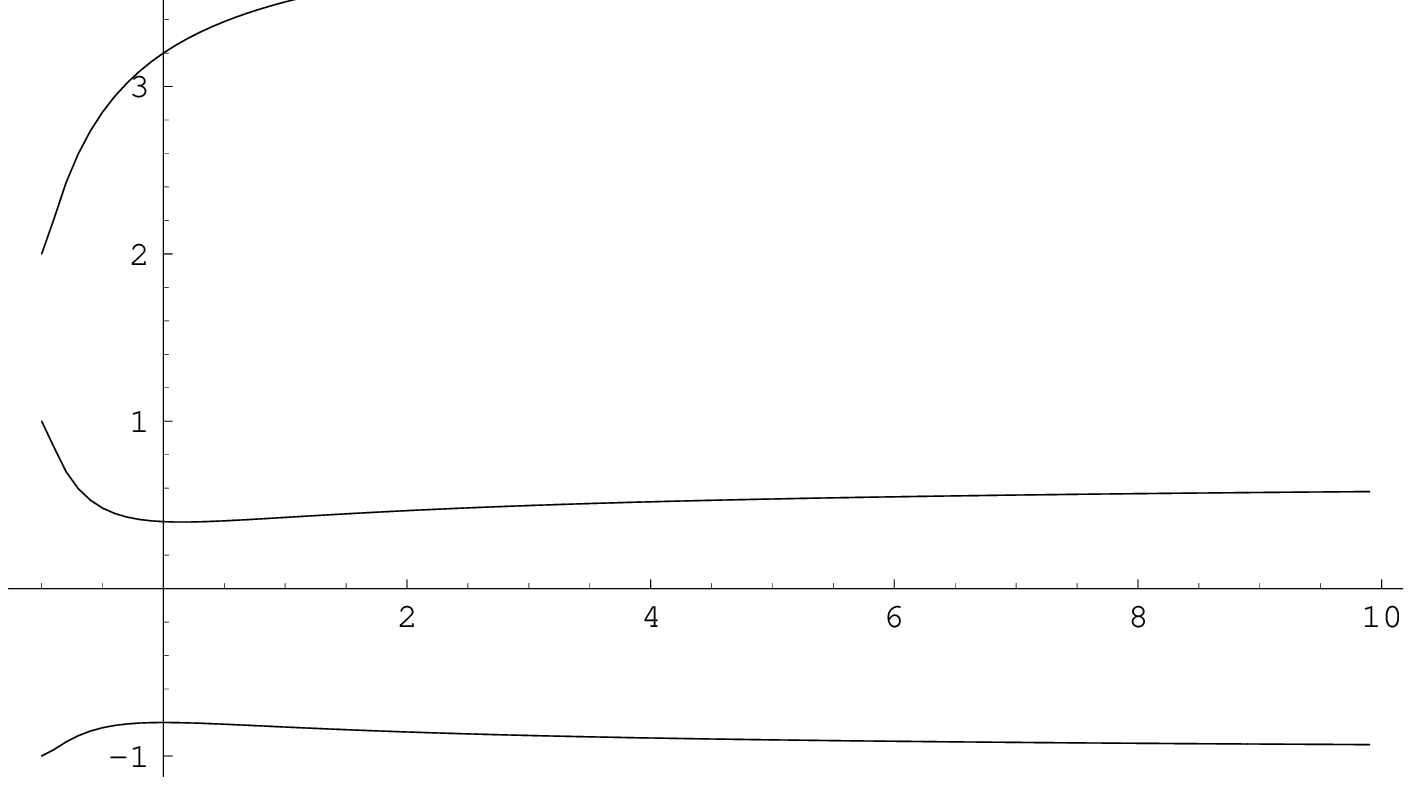,width=5cm}\hskip .5 cm  
\epsfig{file=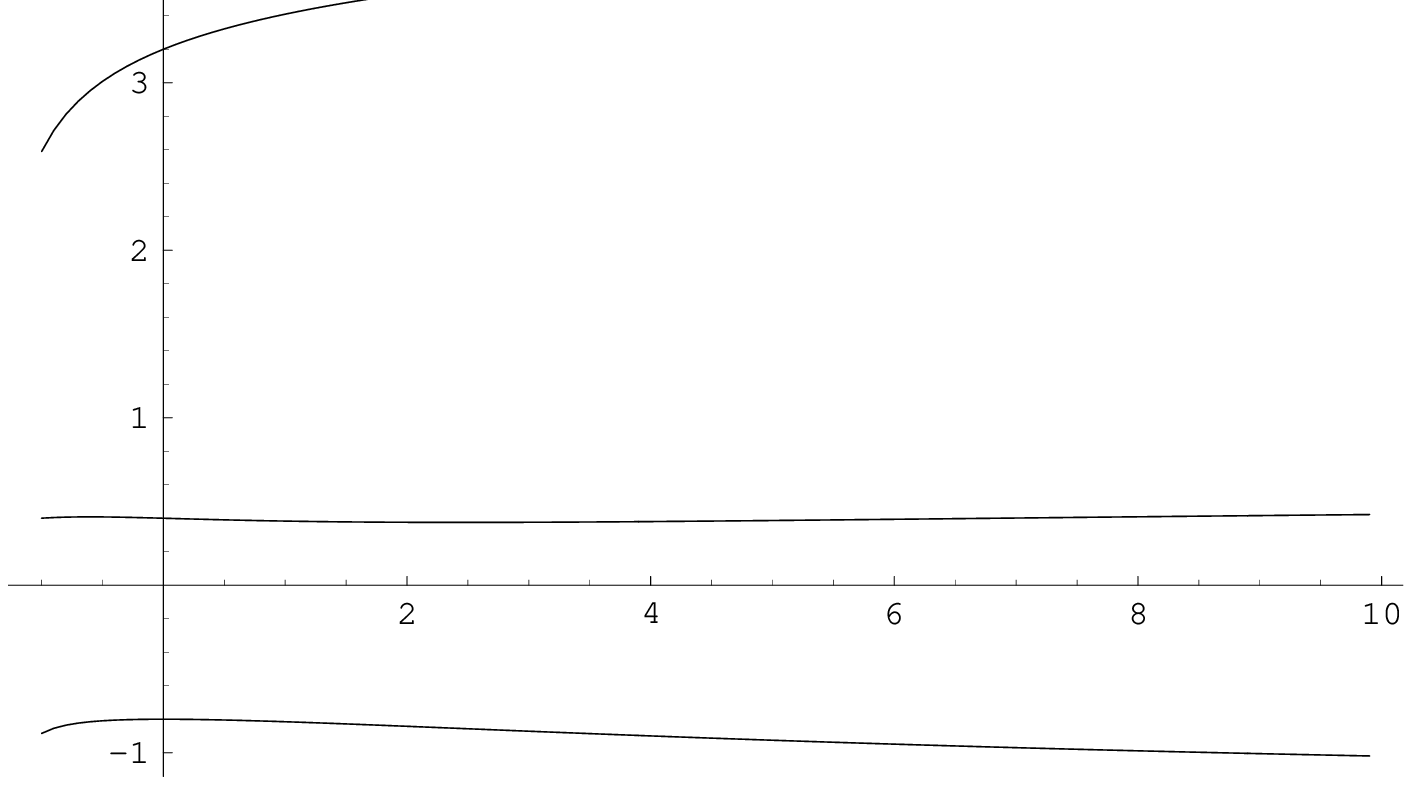,width=5cm}  
} }  
\vspace{1mm}  
\caption{Behavior of the critical dimension  
$\Delta[4,p]$ for $d=3$ with $p=0,2,4$ (from below to above) {\it  
vs} $\rho_1$ for $\rho_2=0$---{\it left}, {\it vs}  
$\rho\equiv\rho_1=\rho_2$---{\it center},{\it vs} $\rho_2$ for  
$\rho_1=0$---{\it right}.}  
\end{figure}

\begin{figure}  
\centerline{  
\hbox{  
\epsfig{file=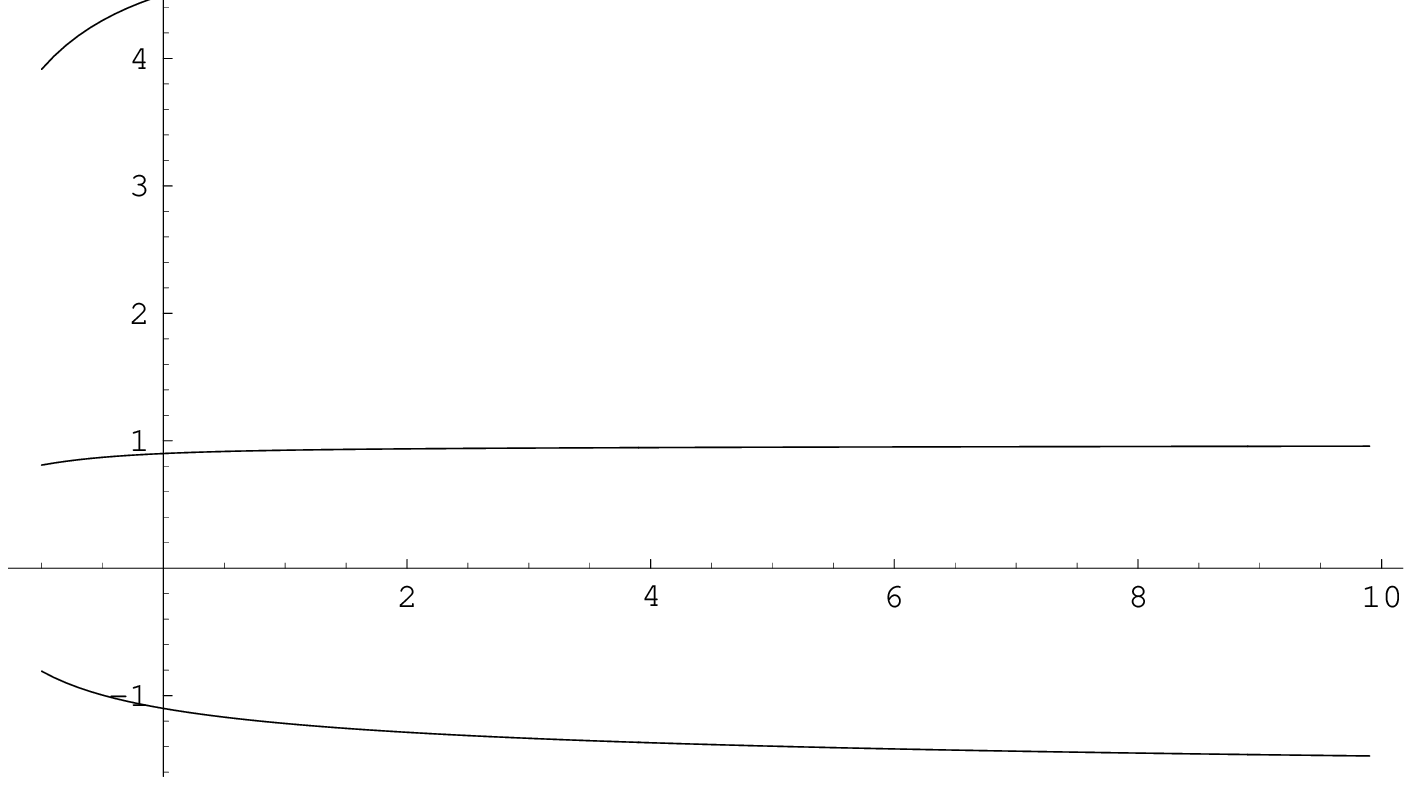,width=5cm}\hskip .5 cm  
\epsfig{file=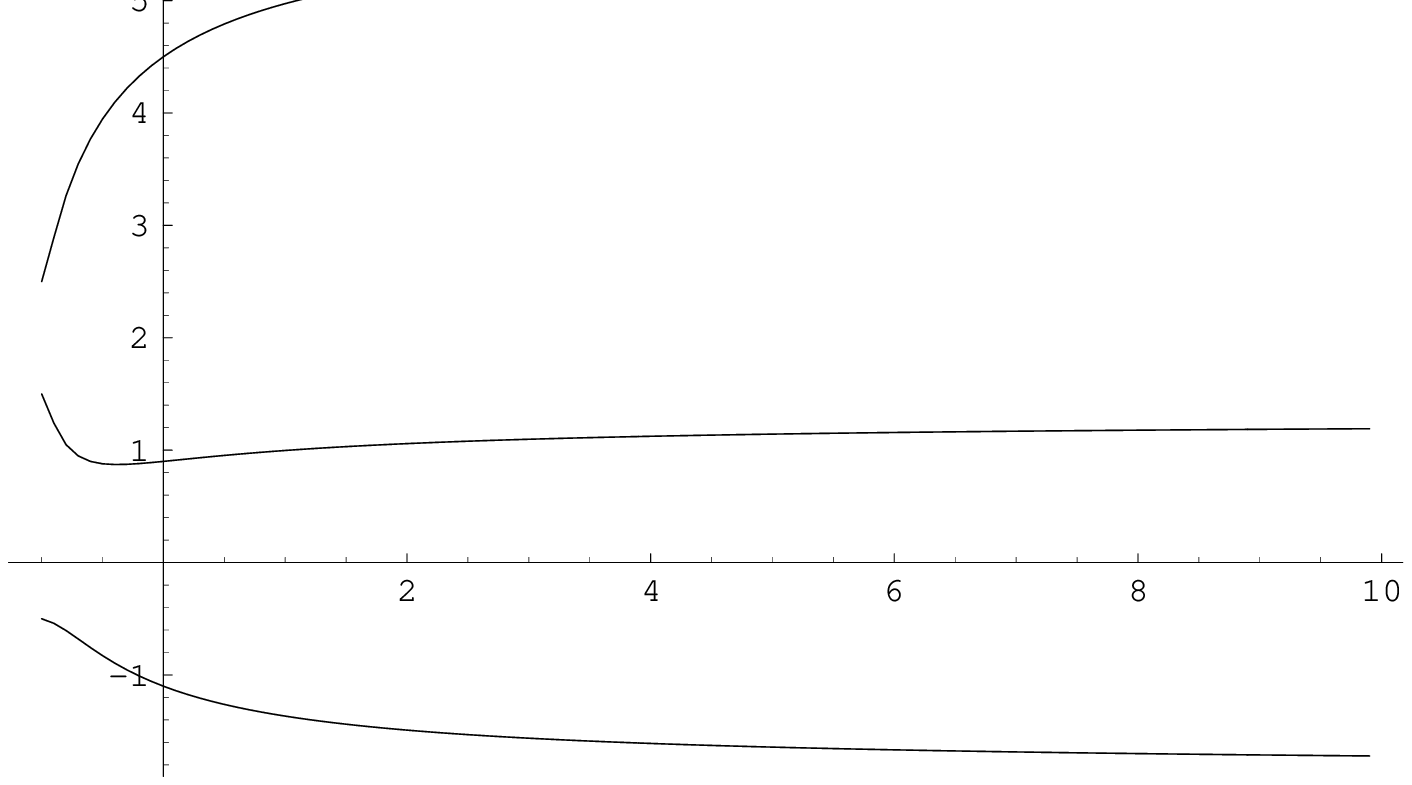,width=5cm}\hskip .5 cm  
\epsfig{file=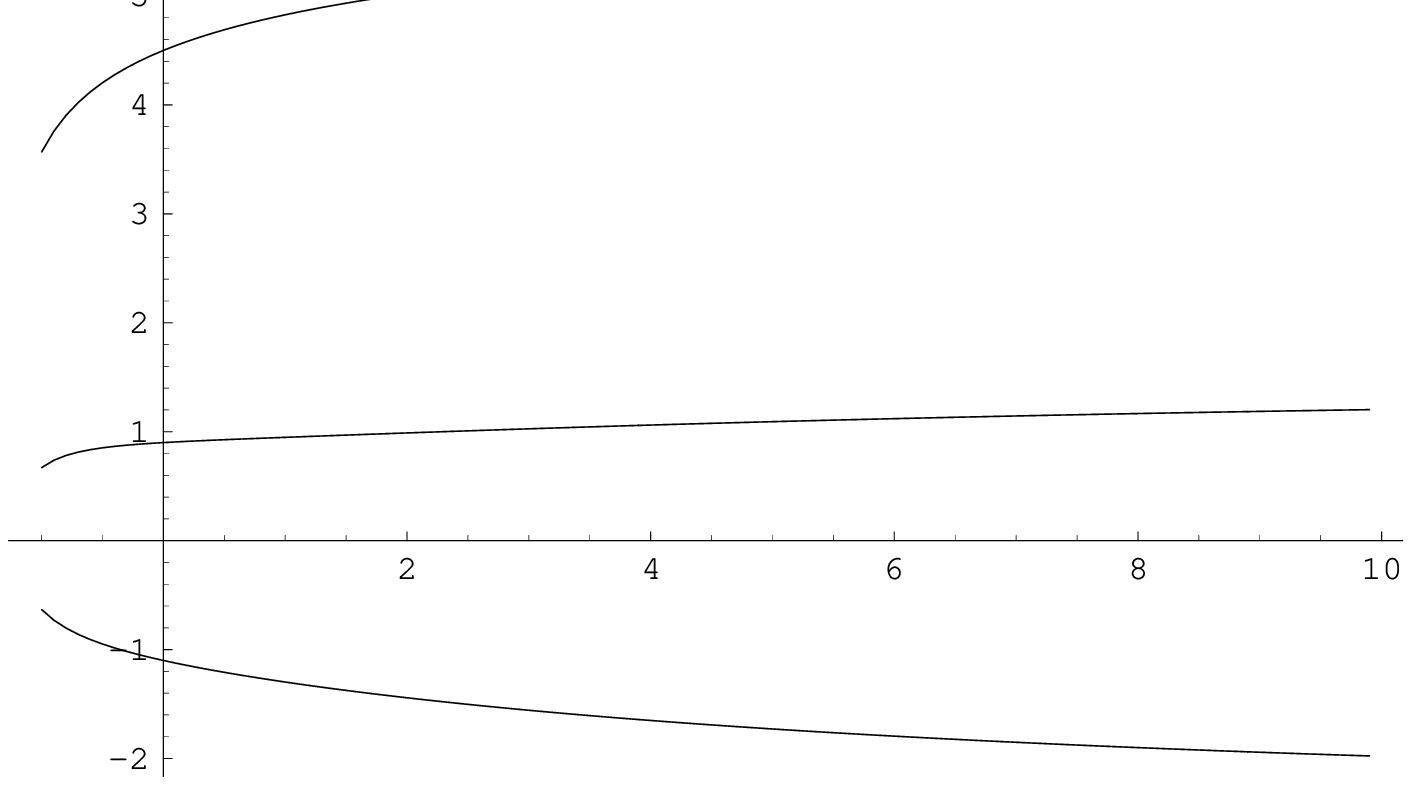,width=5cm}  
} }  
\vspace{1mm}  
\caption{Behavior of the critical dimension  
$\Delta[5,p]$ for $d=3$ with $p=1,3,5$ (from below to above) {\it  
vs} $\rho_1$ for $\rho_2=0$---{\it left}, {\it vs}  
$\rho\equiv\rho_1=\rho_2$---{\it center},{\it vs} $\rho_2$ for  
$\rho_1=0$---{\it right}.}  
\end{figure}  
  
\begin{figure}  
\centerline{  
\hbox{  
\epsfig{file=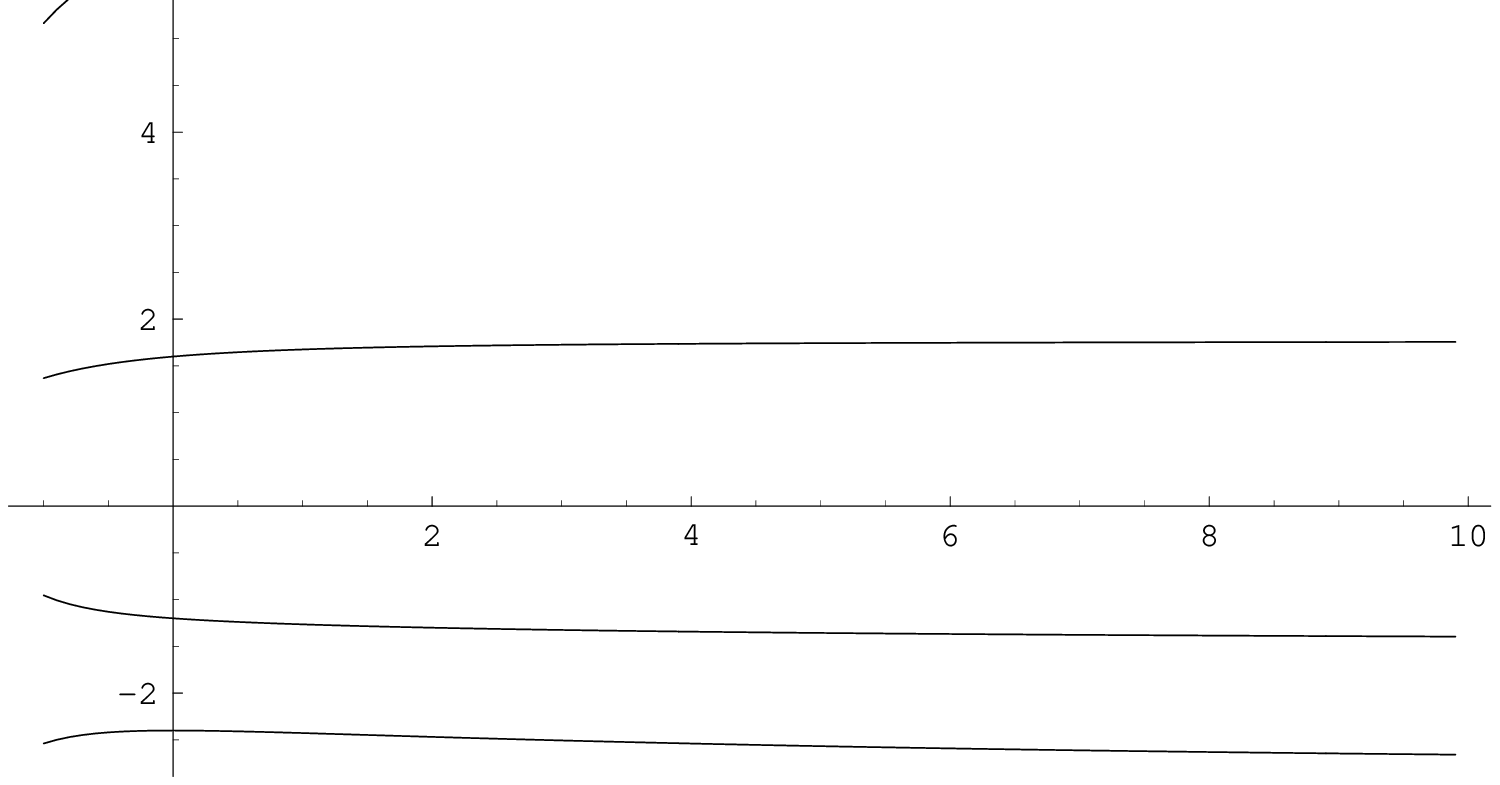,width=5cm}\hskip .5 cm  
\epsfig{file=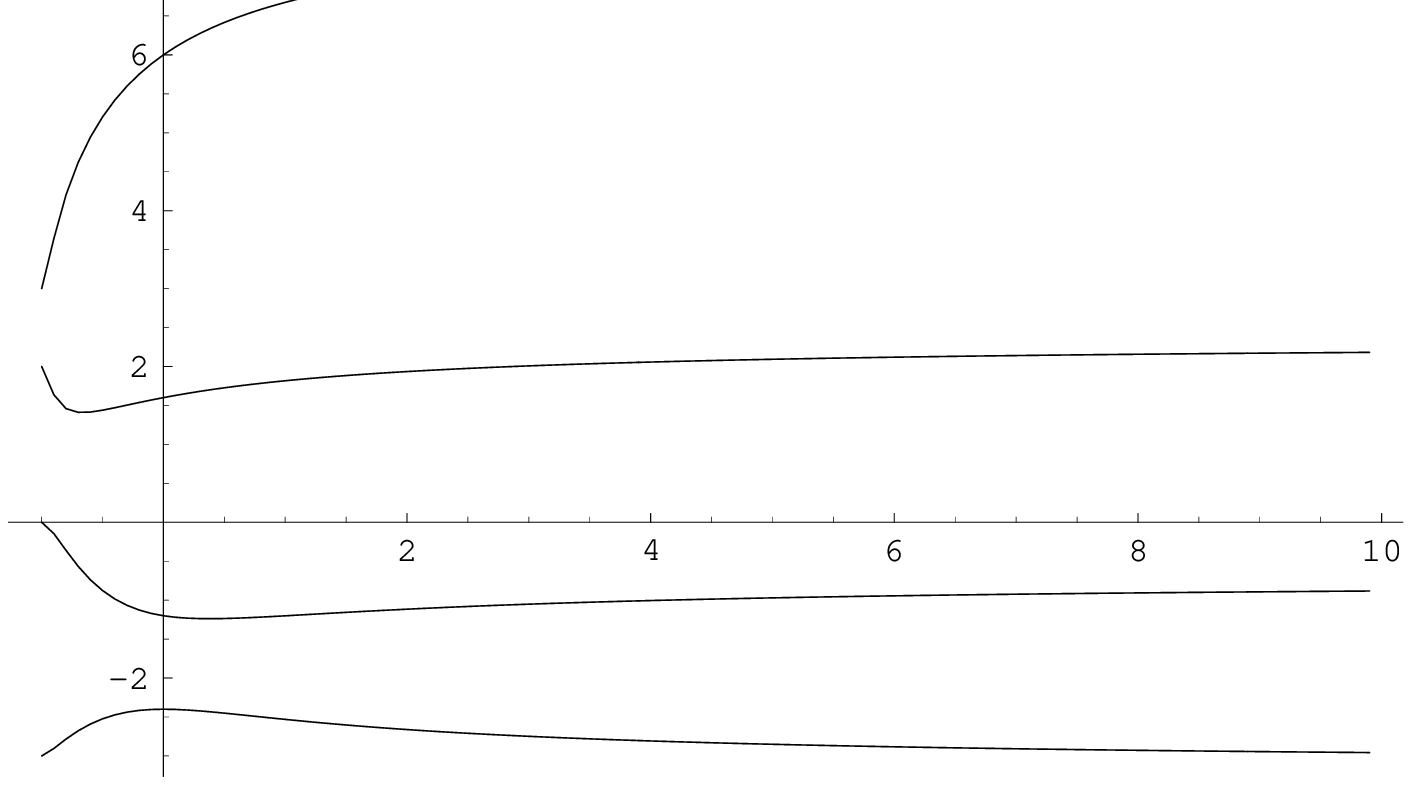,width=5cm}\hskip .5 cm  
\epsfig{file=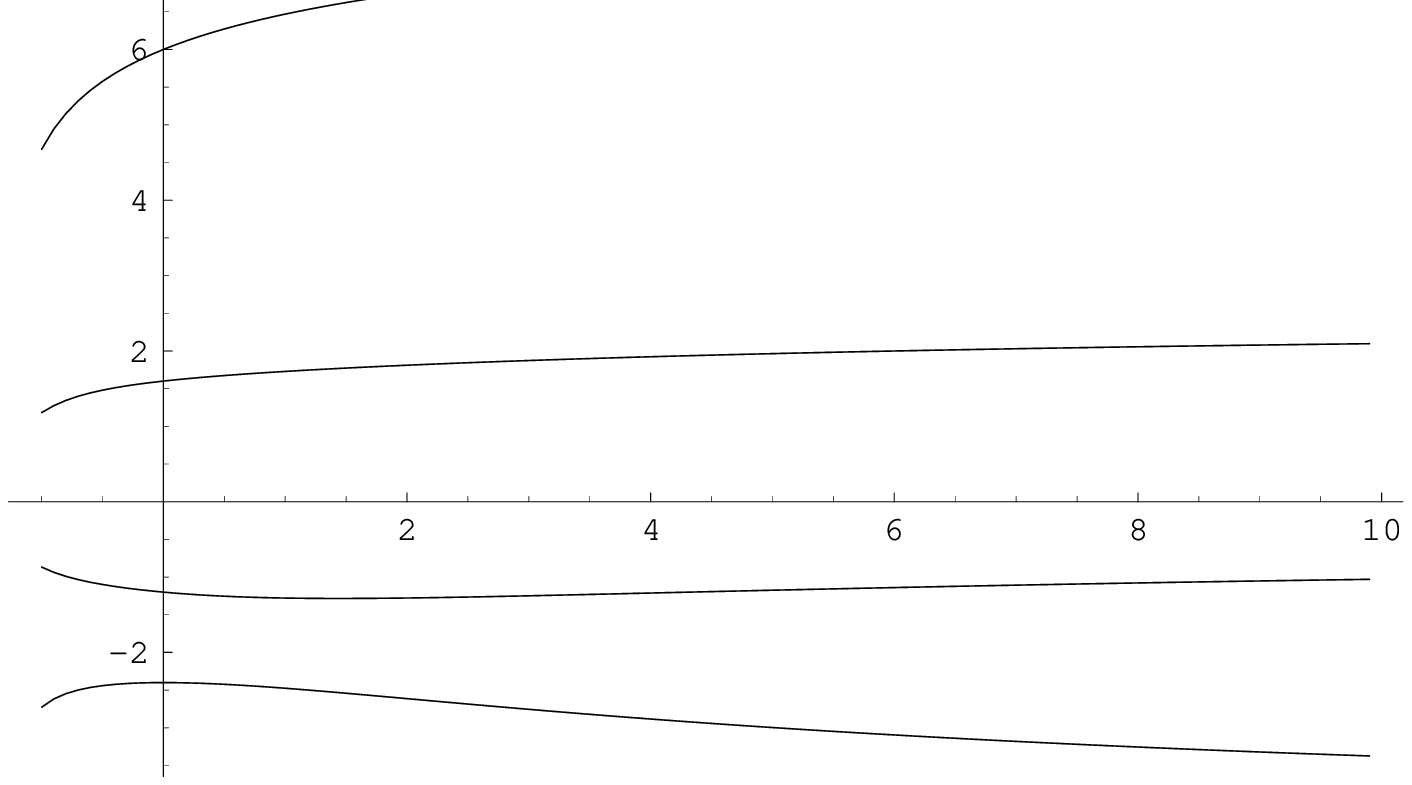,width=5cm}  
} }  
\vspace{1mm}  
\caption{Behavior of the critical dimension  
$\Delta[6,p]$ for $d=3$ with $p=0,2,4,6$ (from below to above)  
{\it vs} $\rho_1$ for $\rho_2=0$---{\it left}, {\it vs}  
$\rho\equiv\rho_1=\rho_2$---{\it center},{\it vs} $\rho_2$ for  
$\rho_1=0$---{\it right}.}  
\end{figure}

\begin{figure}  
\centerline{  
\hbox{  
\epsfig{file=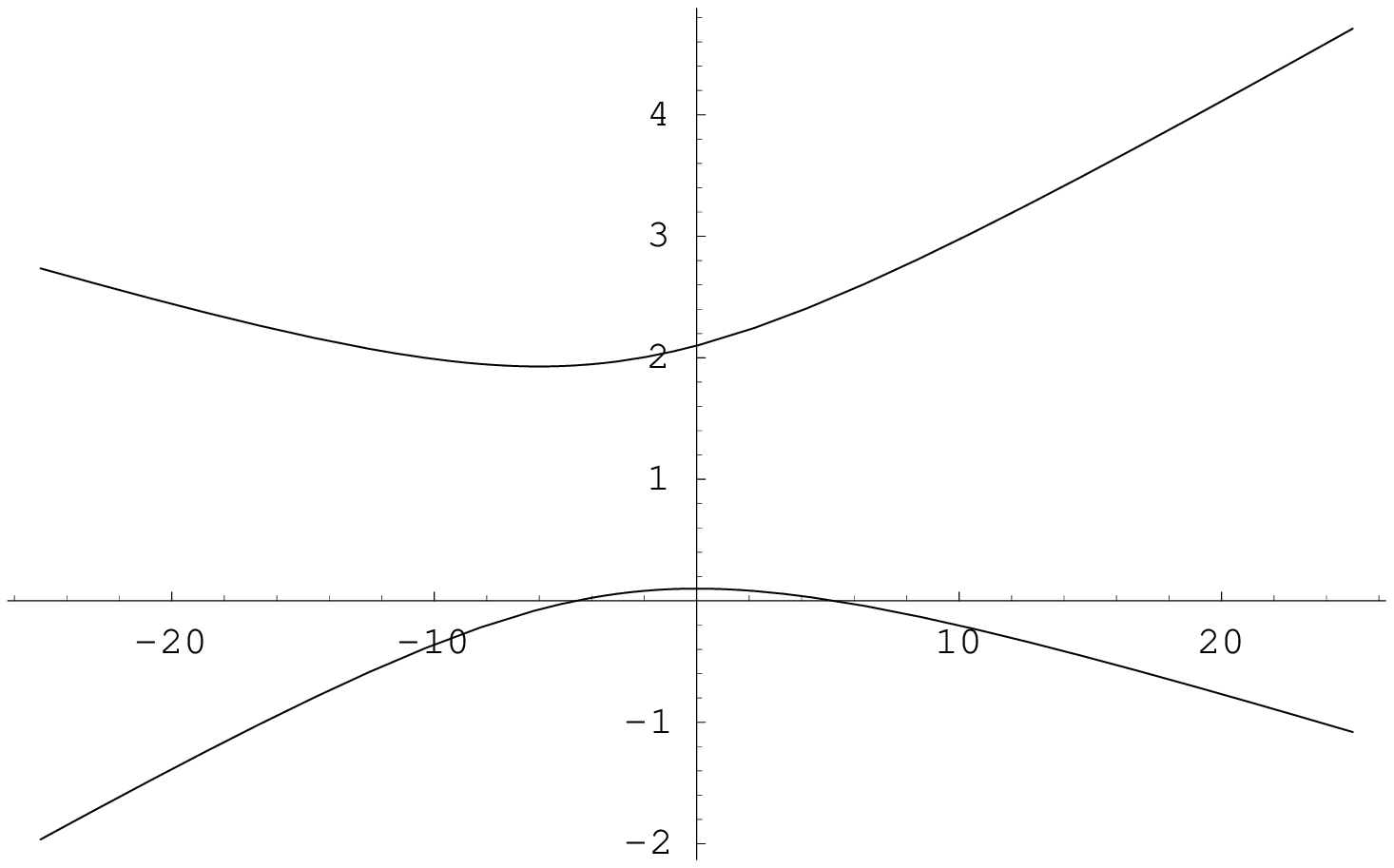,width=5cm}\hskip .5 cm  
\epsfig{file=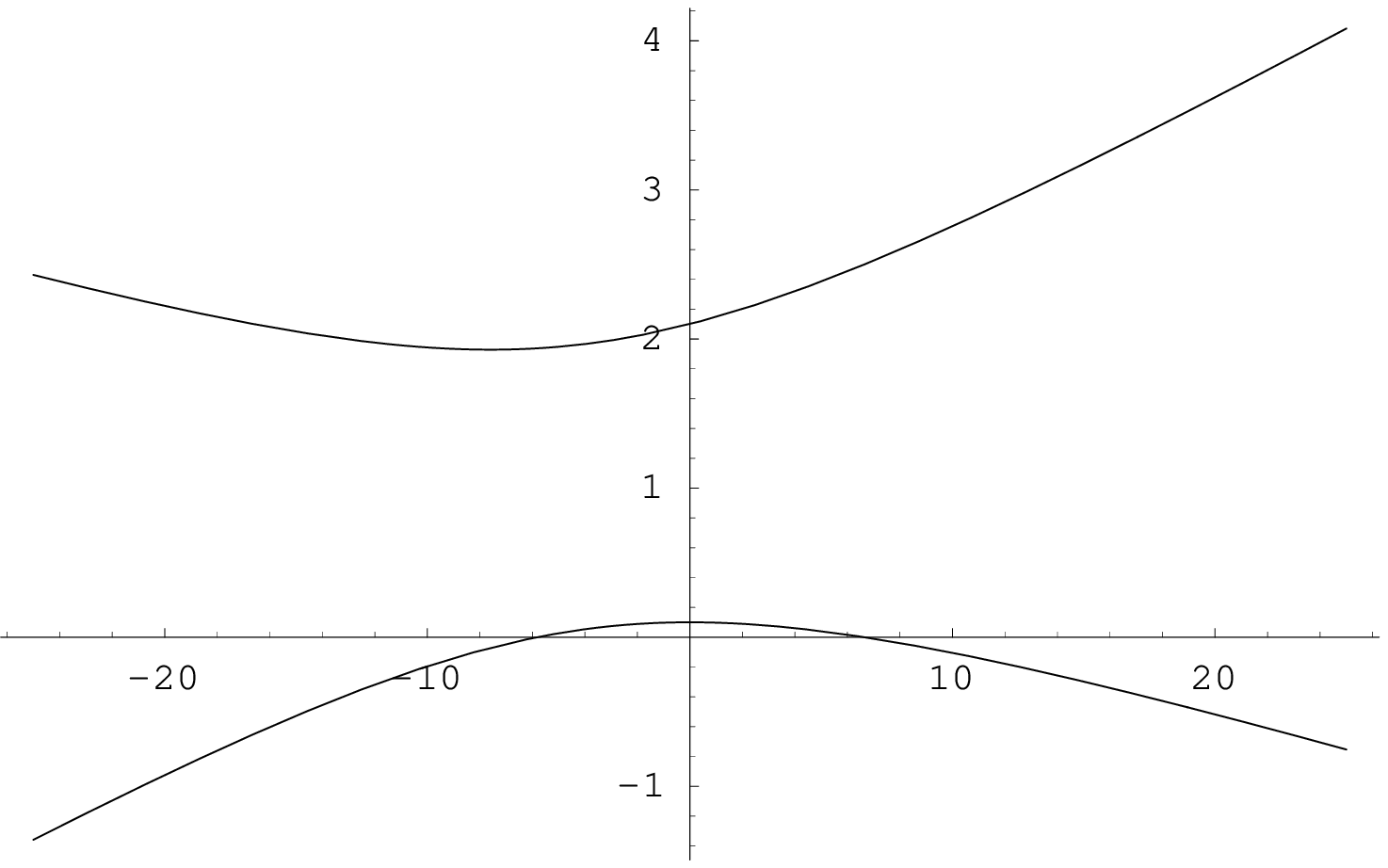,width=5cm}\hskip .5 cm  
\epsfig{file=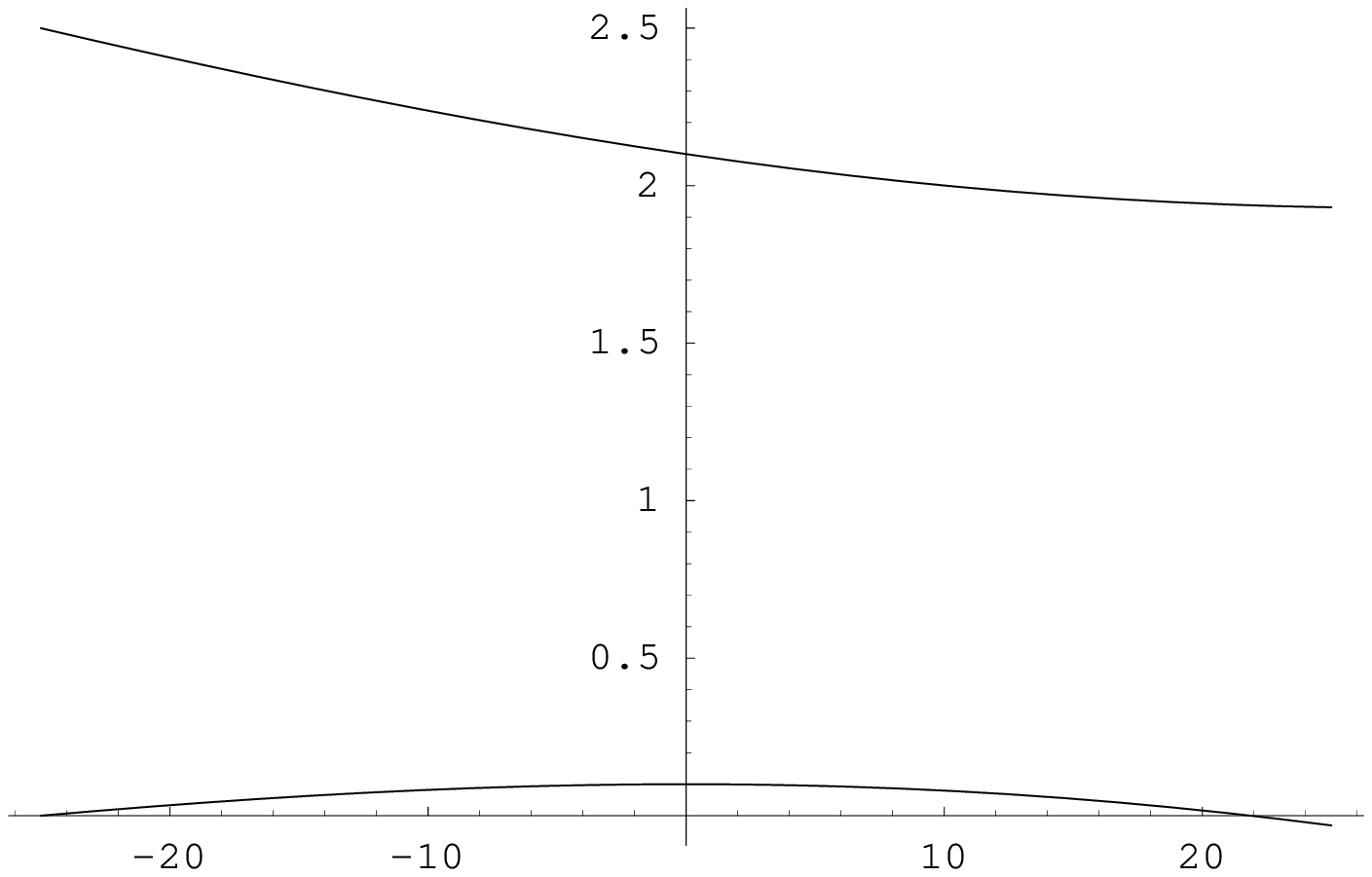,width=5cm}  
} }  
\vspace{1mm}  
\caption{Behavior of the critical dimension  
$\Delta[3,p]$ for $d=3$ with $p=1,3$ (from below to above) {\it  
vs} $a_2$ for $b_3=0$---{\it left}, {\it vs} $a_2=b_3$---{\it  
center},{\it vs} $b_3$ for $a_2=0$---{\it right}.}  
\end{figure}  
  
\begin{figure}  
\centerline{  
\hbox{  
\epsfig{file=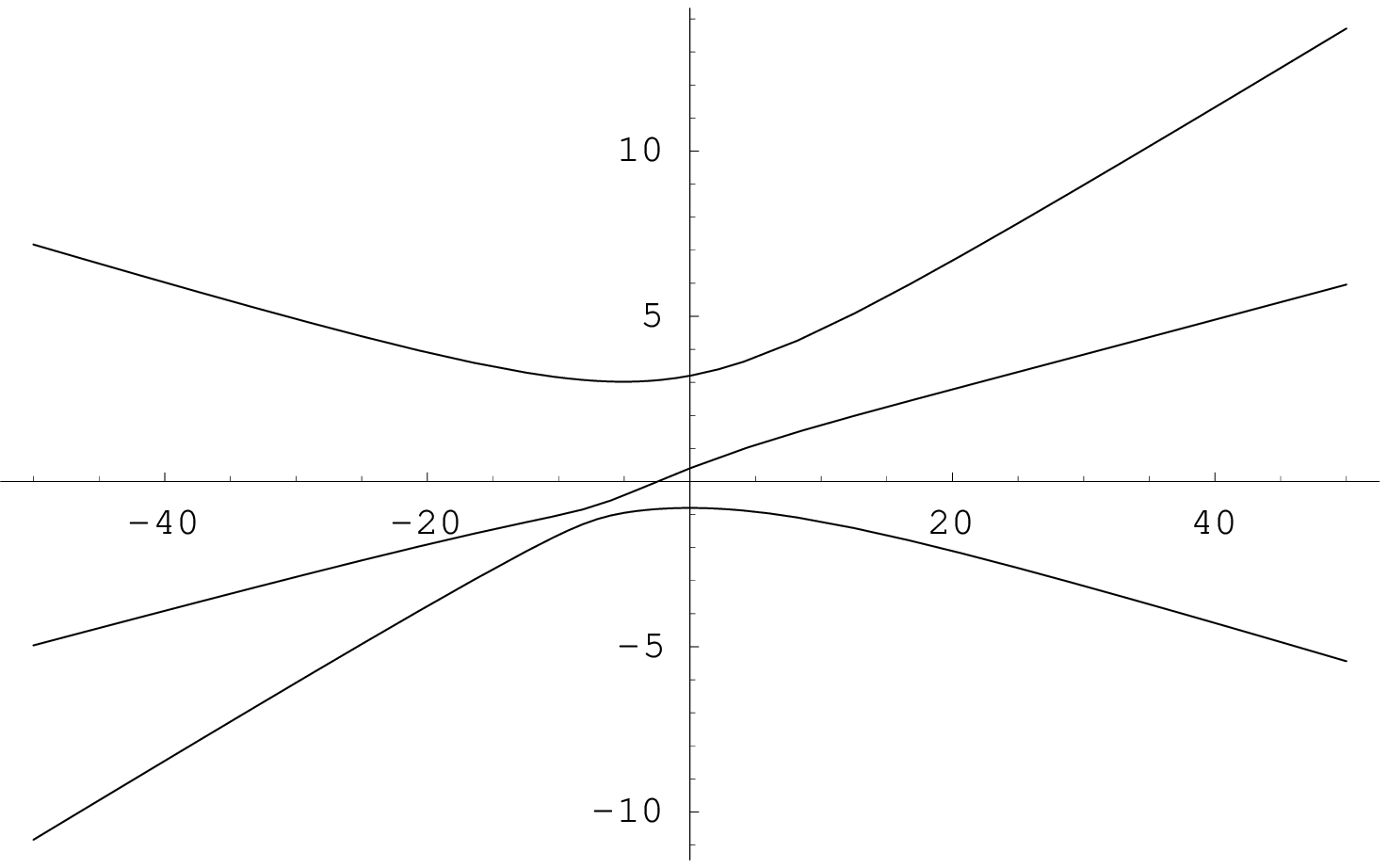,width=5cm}\hskip .5 cm  
\epsfig{file=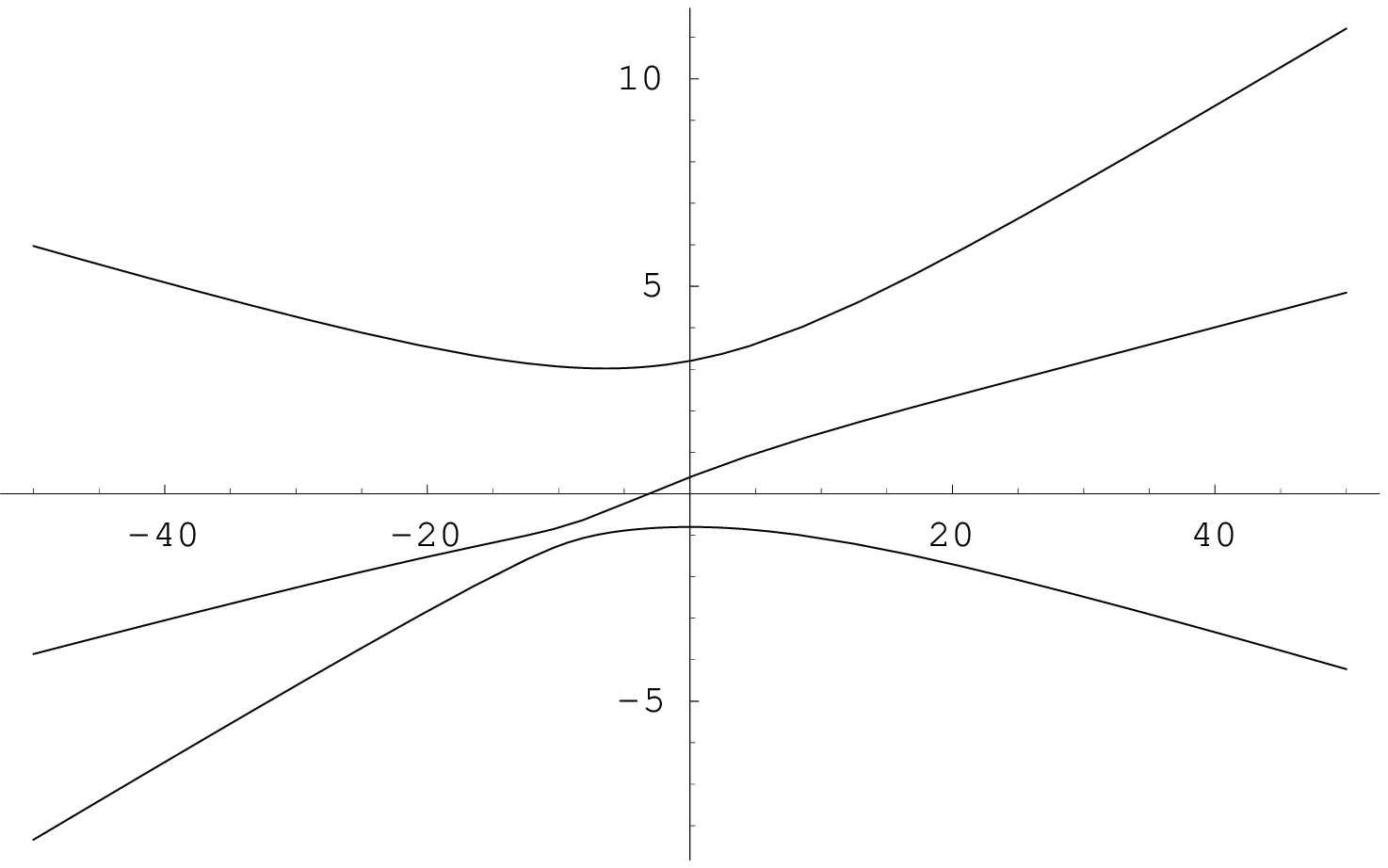,width=5cm}\hskip .5 cm  
\epsfig{file=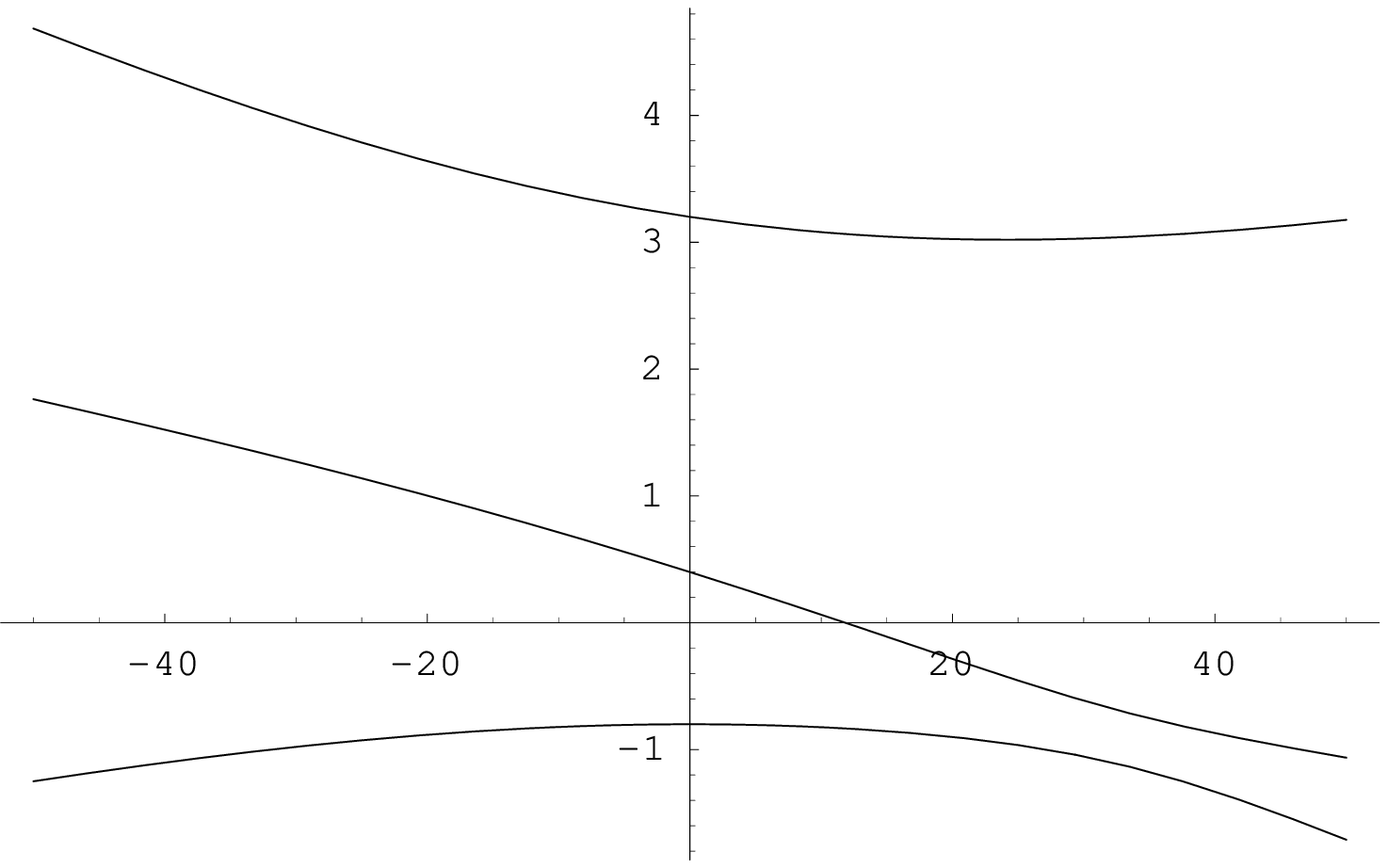,width=5cm}  
} }  
\vspace{1mm}  
\caption{Behavior of the critical dimension  
$\Delta[4,p]$ for $d=3$ with $p=0,2,4$ (from below to above) {\it  
vs} $a_2$ for $b_3=0$---{\it left}, {\it vs} $a_2=b_3$---{\it  
center},{\it vs} $b_3$ for $a_2=0$---{\it right}.}  
\end{figure}  
  
\begin{figure}  
\centerline{  
\hbox{  
\epsfig{file=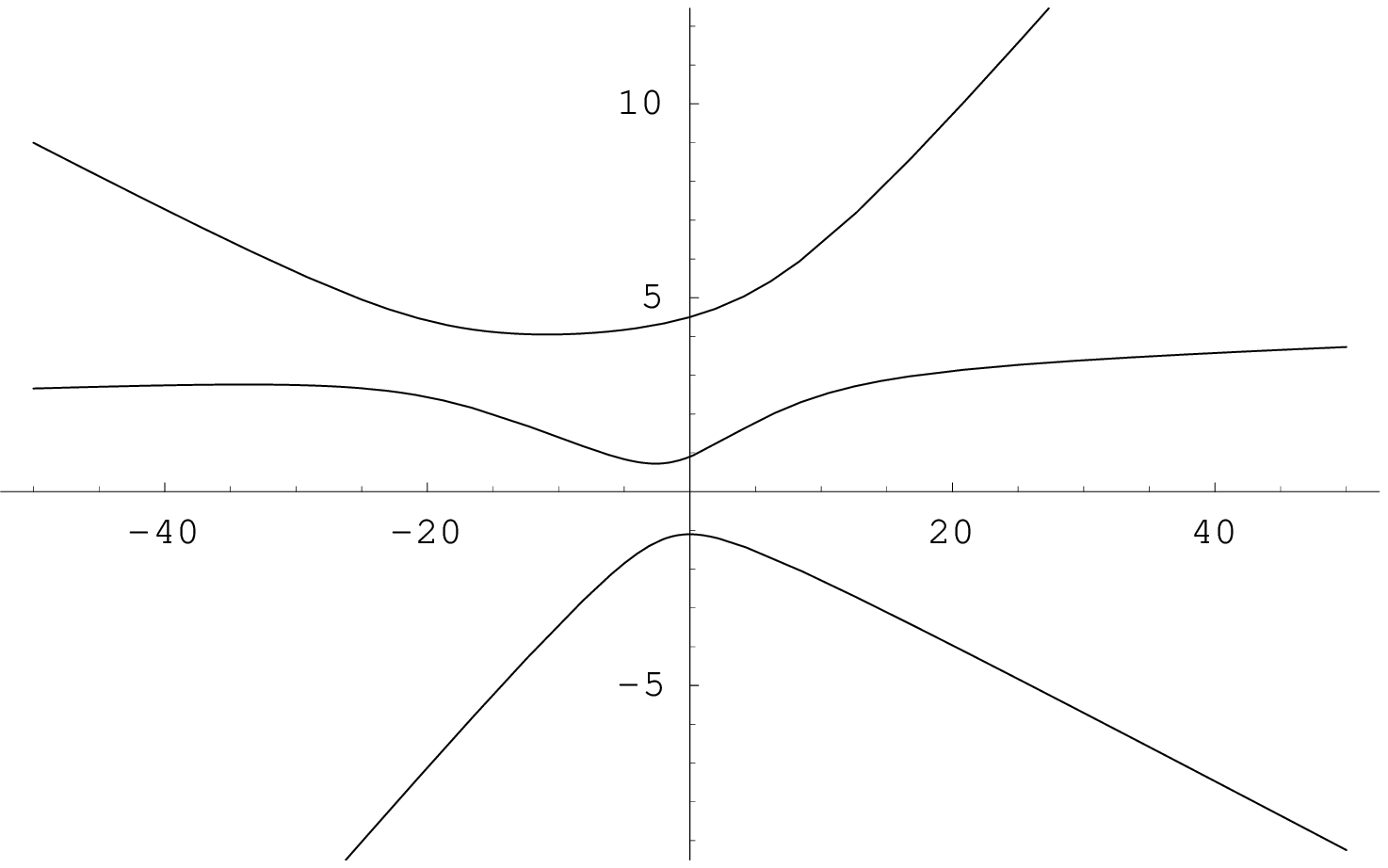,width=5cm}\hskip .5 cm  
\epsfig{file=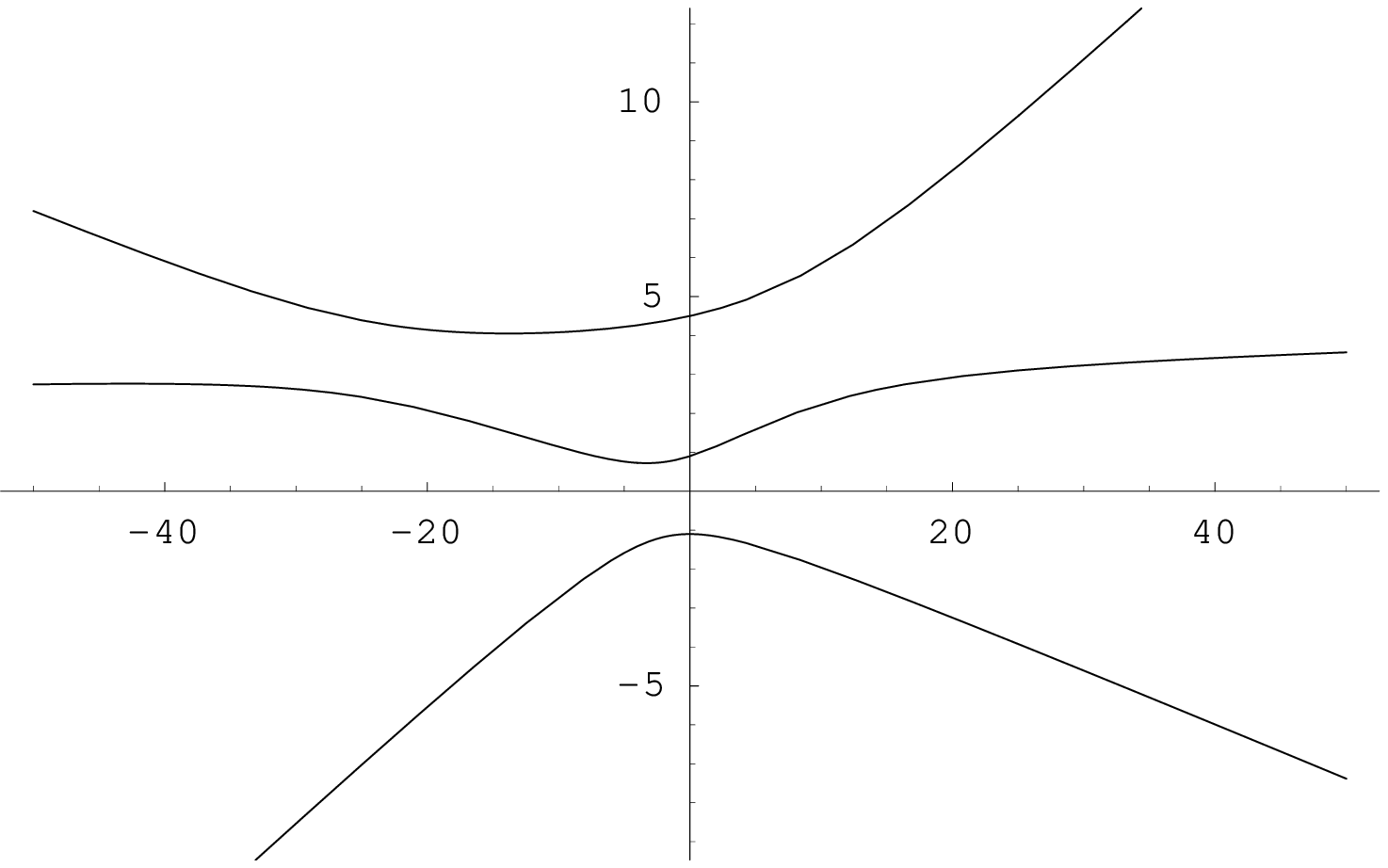,width=5cm}\hskip .5 cm  
\epsfig{file=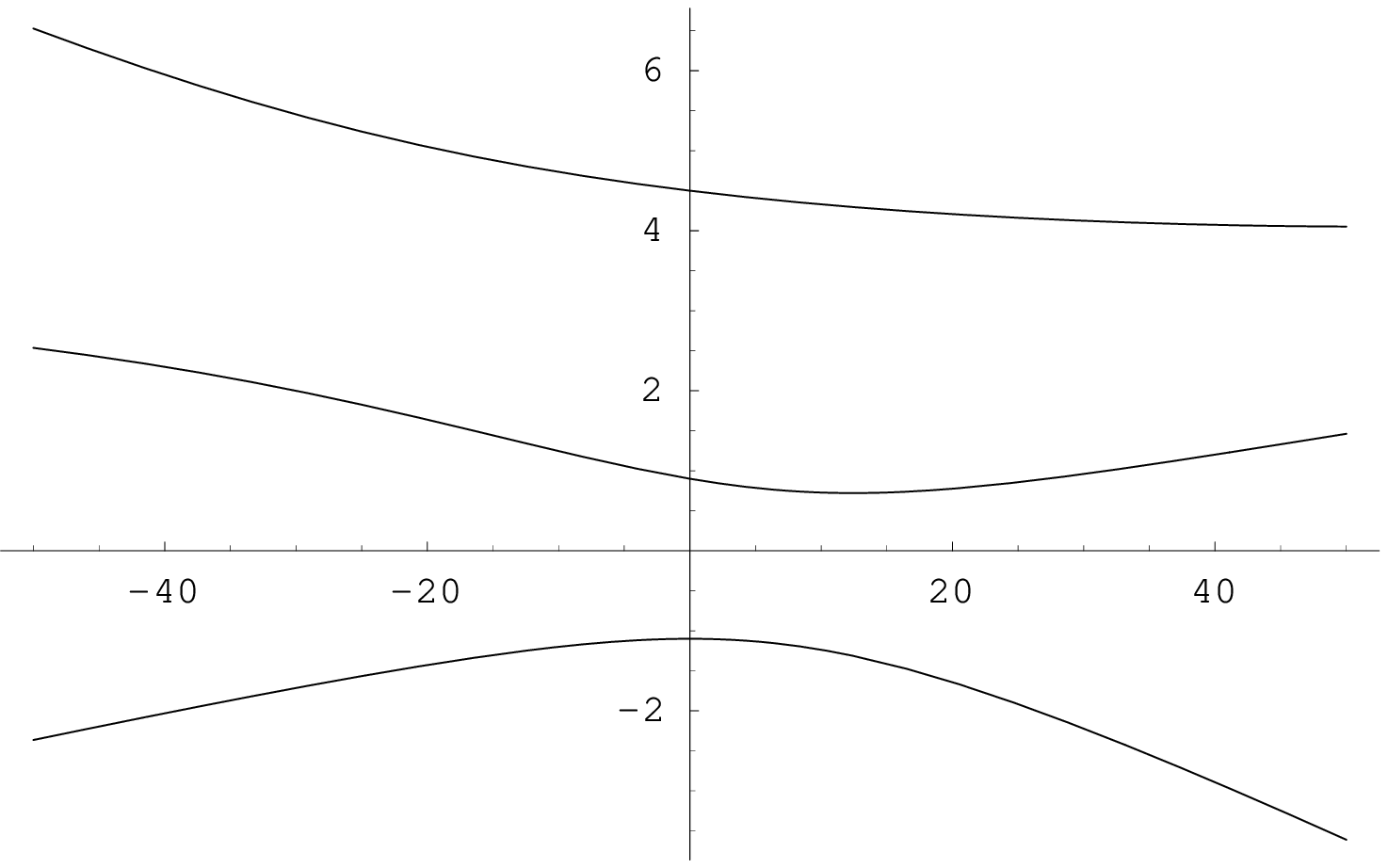,width=5cm}  
} }  
\vspace{1mm}  
\caption{Behavior of the critical dimension  
$\Delta[5,p]$ for $d=3$ with $p=1,3,5$ (from below to above) {\it  
vs} $a_2$ for $b_3=0$---{\it left}, {\it vs} $a_2=b_3$---{\it  
center},{\it vs} $b_3$ for $a_2=0$---{\it right}.}  
\end{figure}  
  
\begin{figure}  
\centerline{  
\hbox{  
\epsfig{file=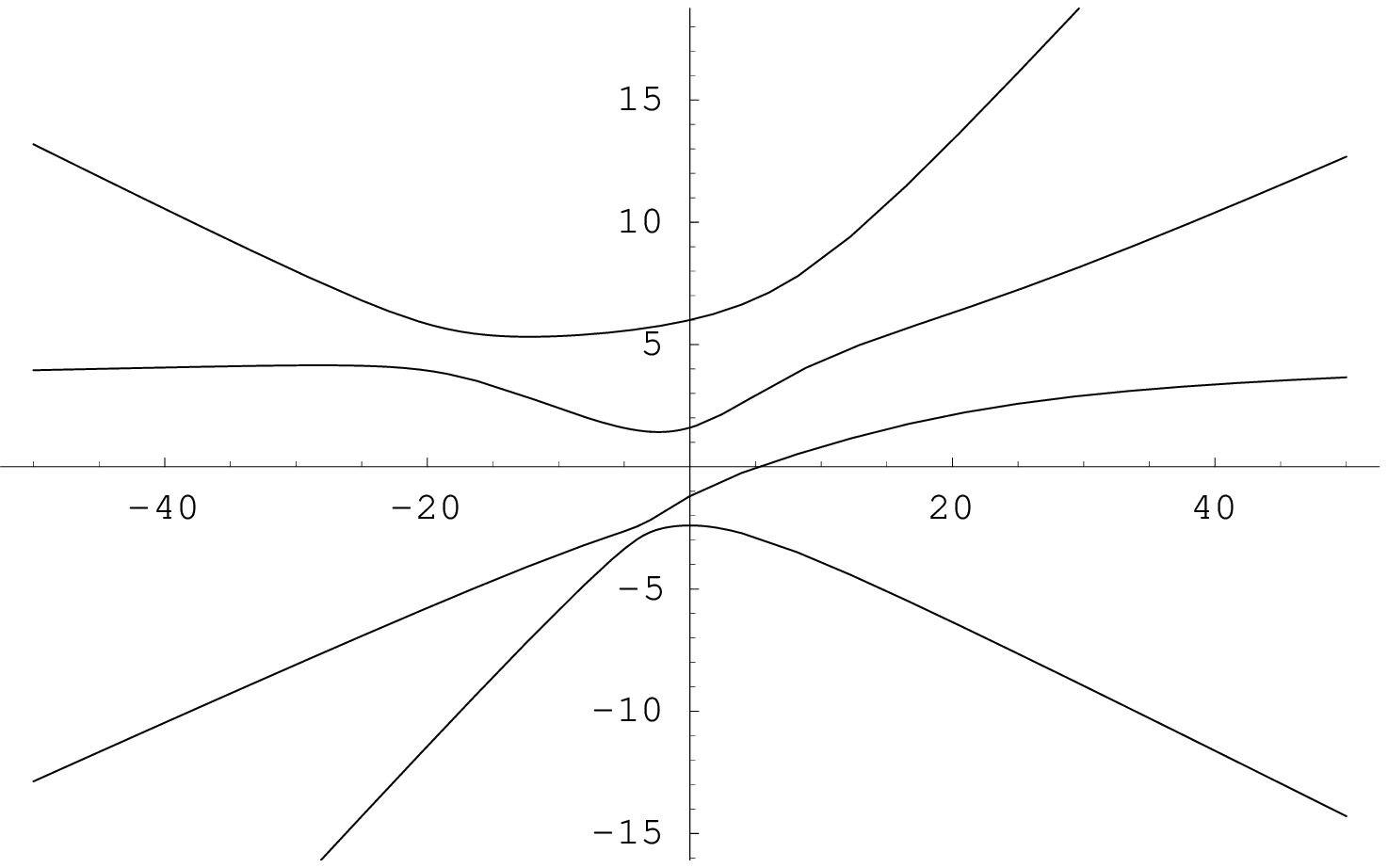,width=5cm}\hskip .5 cm  
\epsfig{file=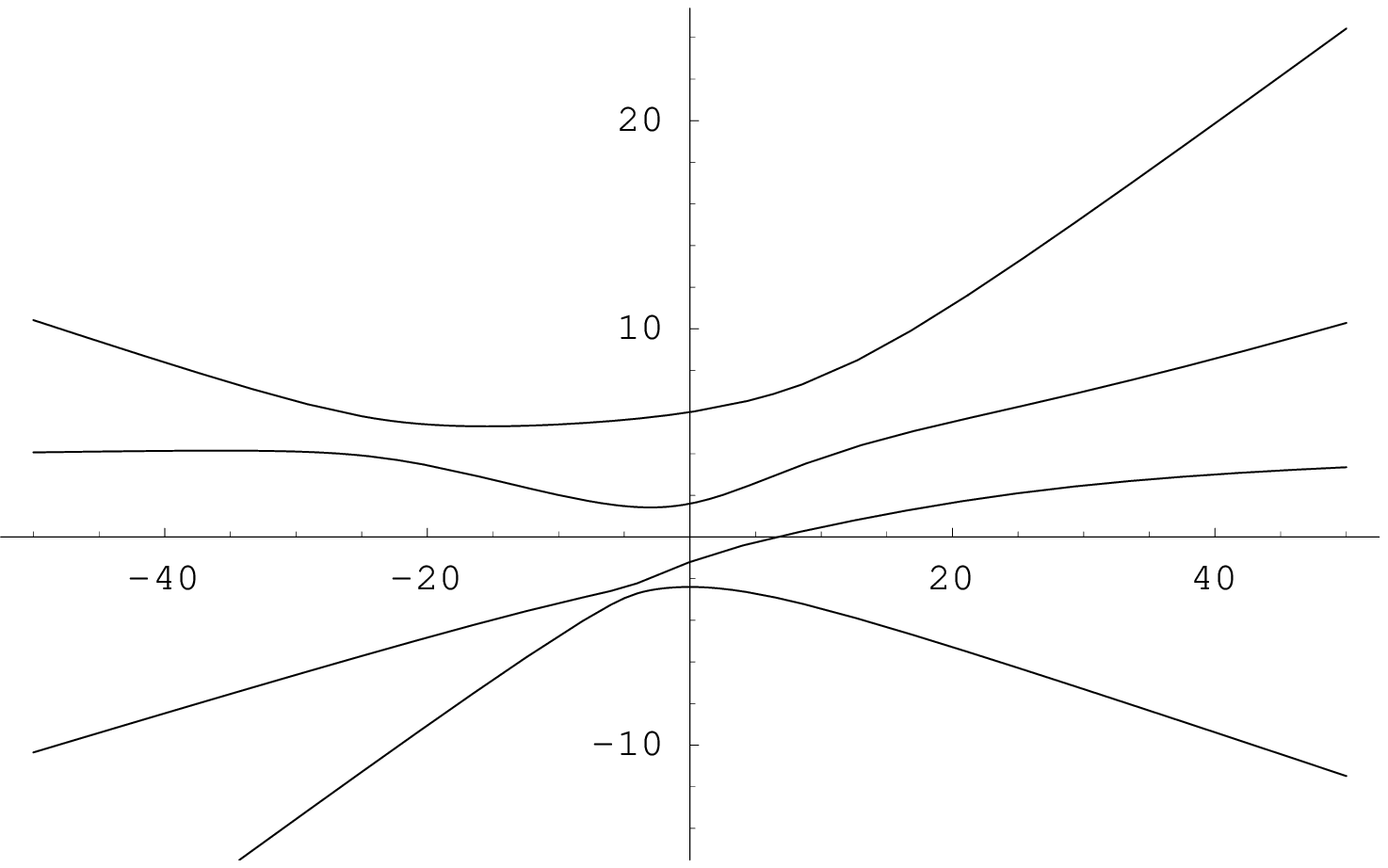,width=5cm}\hskip .5 cm  
\epsfig{file=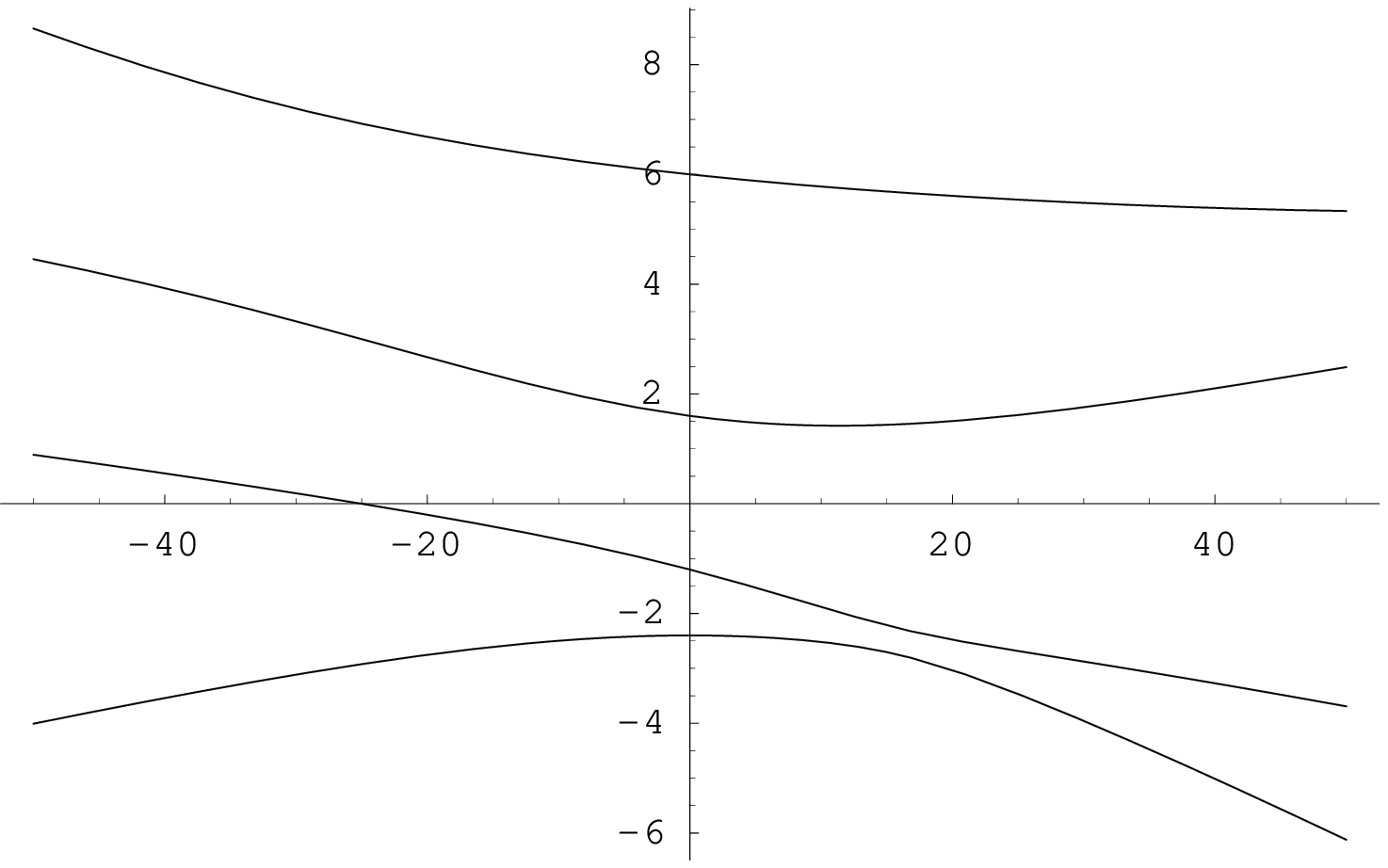,width=5cm}  
} }  
\vspace{1mm}  
\caption{Behavior of the critical dimension  
$\Delta[6,p]$ for $d=3$ with $p=0,2,4$ (from below to above) {\it  
vs} $a_2$ for $b_3=0$---{\it left}, {\it vs} $a_2=b_3$---{\it  
center},{\it vs} $b_3$ for $a_2=0$---{\it right}.}  
\end{figure}  
  
\begin{figure}  
 \centerline{\epsfig{file=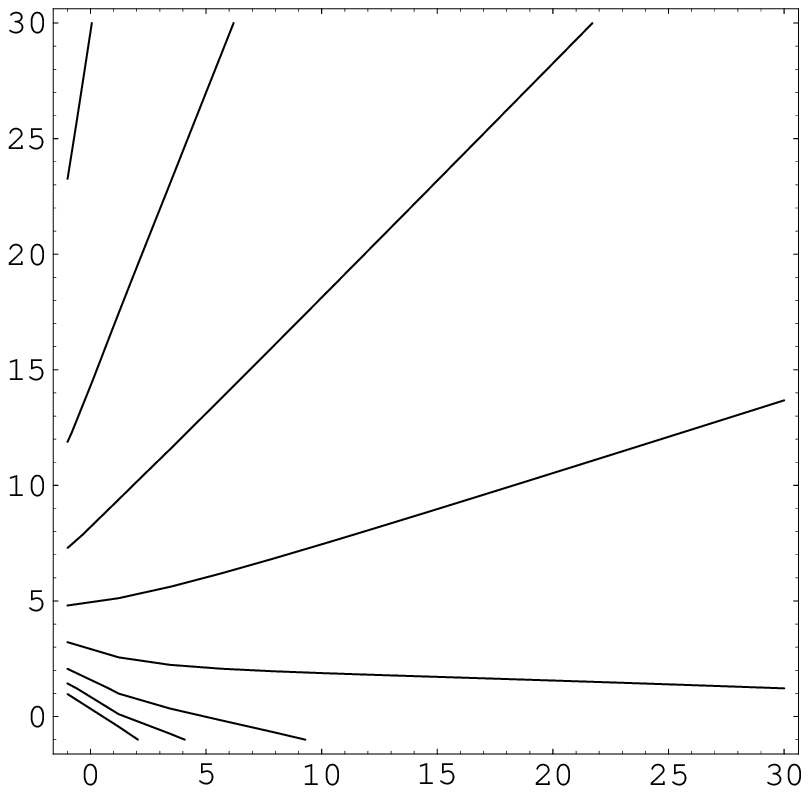,width=8cm}}  
 \vspace{1mm}  
\caption{Levels of the dimension $\Delta[3,1]$ for $d=3$ on the plane  
$\rho_1$--$\rho_2$. Value changes from~$-0.3$ (top) to~$0.1$  
(bottom) with step~$0.05$.}  
\end{figure}  
  

\begin{references}  
  
\bibitem{Monin} A. S. Monin and A. M. Yaglom, {\it Statistical Fluid  
Mechanics\/} (MIT Press, Cambridge, MA, 1975), Vol.~2.  
  
\bibitem{Legacy} U. Frisch, {\it Turbulence: The Legacy of  
      A.~N.~Kolmogorov} (Cambridge University Press, 1995).  
  
  
\bibitem{An} R. Antonia, E. Hopfinger, Y. Gagne, and F. Anselmet, Phys.  
Rev. {\bf A 30}, 2704 (1984); K. R. Sreenivasan, Proc. Roy. Soc. London A  
 {\bf 434}, 165 (1991); K. R. Sreenivasan and R. A. Antonia,  
 Annu. Rev. Fluid. Mech. {\bf 29}, 435 (1997).  
  
  
\bibitem{synth}  
  M. Holzer and E. D. Siggia, Phys. Fluids {\bf 6}, 1820 (1994);  
  A. Pumir, Phys. Fluids {\bf 6}, 2118 (1994);  
  C. Tong and Z. Warhaft, Phys. Fluids {\bf 6}, 2165 (1994).  
  
\bibitem{Kraich1} R. H. Kraichnan, Phys. Fluids {\bf 11}, 945 (1968);  
                  J. Fluid. Mech. {\bf 64}, 737 (1974); Phys. Rev. Lett.  
               {\bf 72}, 1016 (1994); {\it ibid.} {\bf 78}, 4922 (1997).  
  
\bibitem{Falk1}  M. Chertkov, G. Falkovich, I. Kolokolov, and V. Lebedev,  
                 Phys. Rev. E {\bf  52}, 4924 (1995).  
  
\bibitem{Falk2}  M. Chertkov and  G. Falkovich, Phys. Rev. Lett.  
                 {\bf 76}, 2706 (1996).  
  
\bibitem{GK}  
K. Gaw\c{e}dzki and A. Kupiainen, Phys. Rev. Lett. {\bf 75}, 3834 (1995);  
 D.~Bernard, K.~Gaw\c{e}dzki, and A.~Kupiainen,  
Phys. Rev. E {\bf  54}, 2564  (1996).  
  
\bibitem{Pumir} A. Pumir, Europhys. Lett. {\bf 34}, 25 (1996);  
  {\it ibid.} {\bf 37}, 529 (1997); Phys. Rev. E {\bf  57}, 2914 (1998).  
  
\bibitem{Siggia} B. I. Shraiman and E. D. Siggia, Phys. Rev. Lett.  
   {\bf 77}, 2463 (1996); A.~Pumir, B.~I.~Shraiman, and E.~D.~ Siggia,  
   Phys. Rev. E {\bf  55}, R1263  (1997).  
  
\bibitem{Eyink} G. Eyink, Phys. Rev. E {\bf  54}, 1497 (1996).  
  
\bibitem{RG} L. Ts. Adzhemyan, N. V. Antonov, and A. N. Vasil'ev,  
             Phys. Rev. E {\bf 58}, 1823 (1998).  
  
\bibitem{RG1} L. Ts. Adzhemyan and N. V. Antonov,  
              Phys. Rev. E {\bf 58}, 7381 (1998).  
  
\bibitem{RG3} N. V. Antonov, Phys. Rev. E {\bf 60}, 6691 (1999);  
{\it chao-dyn/9907018} (to appear in Physica D).  
  
\bibitem{KJW} K. J. Wiese, {\it chao-dyn/9911005}.  
  
\bibitem{CLMV99} A.~Celani, A.~Lanotte, A.~Mazzino, and M.~Vergassola,  
Phys. Rev. Lett. {\bf 84}, 2385 (2000).  
  
\bibitem{Lanotte}  A. Lanotte and A. Mazzino,  
Phys. Rev. E {\bf 60}, R3483 (1999).  
  
\bibitem{Lanotte2}  
N.~V.~Antonov, A.~Lanotte, and A.~Mazzino, {\it nlin.CD/0001039}  
(to appear in Phys. Rev. E).  
  
  
\bibitem{ABP}  
I. Arad, V. L'vov, E. Podivilov, and I. Procaccia, {\it chao-dyn/9907017};  
I.~Arad, L.~Biferale, and I.~Procaccia, {\it  chao-dyn/9909020}.  
  
\bibitem{Borue} S. G. Saddoughi and S. V. Veeravalli, J. Fluid. Mech.  
     {\bf 268}, 333 (1994);  
V.~Borue and S.~A.~Orszag, J. Fluid. Mech. {\bf 306}, 293 (1996);  
  
\bibitem{Arad1} I. Arad, B. Dhruva, S. Kurien, V. S. L'vov, I. Procaccia,  
         and K. R. Sreenivasan, Phys. Rev. Lett, {\bf 81}, 5330 (1998);  
         I. Arad, L. Biferale, I. Mazzitelli, and I. Procaccia,  
         Phys. Rev. Lett, {\bf 82}, 5040 (1999);  
         S. Kurien, V. S. L'vov, I. Procaccia,  
         and K. R. Sreenivasan, {\it chao-dyn/9906038.}  
  
\bibitem{Zinn} J. Zinn-Justin, {\it Quantum Field Theory and  
                Critical Phenomena} (Clarendon, Oxford, 1989).  
  
\bibitem{book3} A. N. Vasil'ev,  {\it Quantum-Field Renormalization  
  Group in the Theory of Critical Phenomena and Stochastic Dynamics}  
  (St~Petersburg Institute of Nuclear Physics, St~Petersburg, 1998)  
  [in Russian; English translation: Gordon \& Breach, in preparation].  
  
\bibitem{UFN} L. Ts. Adzhemyan, N. V. Antonov, and A. N. Vasil'ev, Usp.  
Fiz. Nauk, {\bf 166}, 1257 (1996) [Phys. Usp. {\bf 39}, 1193 (1996)].  
  
\bibitem{turbo} L. Ts. Adzhemyan, N. V. Antonov, and A. N. Vasiliev, {\it  
 The Field Theoretic Renormalization Group in Fully Developed Turbulence}  
      (Gordon \& Breach, London, 1999).  
  
\bibitem{JETP} L. Ts. Adzhemyan, N. V. Antonov, and A. N. Vasil'ev,  
                     Zh. {\'E}ksp. Teor. Fiz. {\bf 95}, 1272 (1989)  
                            [Sov. Phys. JETP {\bf 68}, 733 (1989)].  
  
\bibitem{DL}  B. Duplantier and A. Ludwig, Phys. Rev. Lett. {\bf 66},  
        247 (1991); G. L. Eyink, Phys. Lett. A {\bf 172}, 355 (1993).  
  
\bibitem{Burg} A. M. Polyakov, Phys. Rev. E {\bf 52}, 6183 (1995).  
  
\bibitem{Burg1}  M. L\"{a}ssig, J. Phys. C {\bf 10}, 9905 (1998);  
      Phys. Rev. Lett. {\bf 84}, 2618 (2000).  
  
\bibitem{Barton} R. Rubinstein and J. M. Barton,  
                 Phys. Fluids {\bf 30}, 2986 (1987).  
  
\bibitem{Carati} D. Carati and L. Brenig,  
                 Phys. Rev. A {\bf 40}, 5193 (1989).  
  
\bibitem{Denis}  L. Ts. Adzhemyan, M. Hnatich, D. Horvath, and  
        M. Stehlik,  Int. J. Mod. Phys. B {\bf 9}, 3401 (1995).  
  
\bibitem{Kim}  T. L. Kim and A. V. Serdukov, Theor. Math. Phys.  
                {\bf 105}, 412  (1995).  
  
\bibitem{Busa} J. Bu\v{s}a, M. Hnatich, J. Honkonen, and D. Horvath,  
                Phys. Rev. E  {\bf  55}, 381 (1997).  
  
\bibitem{FF}  
 J. D. Fournier and U. Frisch, Phys. Rev. A {\bf 19}, 1000 (1983).  
  
\bibitem{AV} N. V. Antonov and A. N. Vasil'ev,  
             Theor. Math. Phys. {\bf 110}, 97  (1997).  
  
\bibitem{Grad} I. S. Gradshtejn and I. M. Ryzhik, {\it Tables of  
    integrals, series and products} (Academic Press, New York, 1965).  
  
\bibitem{Triple} L. Ts. Adzhemyan, S. V. Borisenok, and  
      V. I. Girina, Theor. Math. Phys. {\bf 105}, 1556  (1995).  
  
\bibitem{SA} K. R. Sreenivasan, B. Dhruva, and I. San Gil,  
 {\it chao-dyn/9906041.}  
  
\end{references}
\end{document}